\documentclass[fleqn,usenatbib]{mnras}
\usepackage{newtxtext,newtxmath}
\usepackage[T1]{fontenc}
\DeclareRobustCommand{\VAN}[3]{#2}
\let\VANthebibliography\thebibliography
\def\thebibliography{\DeclareRobustCommand{\VAN}[3]{##3}\VANthebibliography}

\usepackage{graphicx}	
\usepackage{amsmath}	
\usepackage{nomencl}
\usepackage{threeparttable}
\usepackage{xcolor}

\def \cblu {\textcolor{blue}}

\title[AGN Feedback in Massive Galaxies]{XMAGNET : Kinetic, Thermal and Magnetic AGN Feedback in Massive Galaxies at Halo Masses $\sim 10^{13.5}$ M$_\odot$}

\author[Deovrat Prasad et al.]{
Deovrat Prasad,$^{1}$\thanks{E-mail: deovratd@cardiff.ac.uk}
Philipp Grete,$^{2}$
Brian W. O'Shea,$^{3,4}$
Forrest W. Glines,$^{3}$
G. Mark Voit,$^{3}$
\newauthor
Freeke van de Voort,$^{1}$
Martin Fournier,$^{2}$
Benjamin D. Wibking,$^{3}$
\\
$^{1}$School of Physics and Astronomy, Cardiff University, UK\\
$^{2}$Universität Hamburg, Hamburger Sternwarte, Gojenbergsweg 112, 21029 Hamburg, Germany\\
$^{3}$Department of Physics and Astronomy, Michigan State University, MI, US\\
$^{4}$Department of Computational Mathematics, Science, and Engineering, Michigan State University, MI, US
}
\begin{document}
\label{firstpage}
\pagerange{\pageref{firstpage}--\pageref{lastpage}}
\maketitle

\begin{abstract}
The interplay between radiative cooling of the circumgalactic medium (CGM) and feedback heating governs the evolution of the universe’s most massive galaxies.
This paper presents simulations of feedback processes in massive galaxies showing how kinetic, thermal, and magnetic active galactic nuclei (AGN) feedback interacts with the CGM under different environmental conditions. We find that in massive galaxies with shallower central gravitational potential and higher CGM pressure (multiphase galaxy; MPG) pure kinetic AGN feedback is most efficient in preventing CGM cooling from becoming catastrophic while maintaining the CGM entropy within the observed range. For the same galaxy, partitioning AGN energy injection into kinetic ($75\%$) and thermal ($25\%$) energy results in an entropy bump within 
$r\lesssim15$ kpc while also having a larger amount of cold gas extending out to $r\sim80$ kpc. A magnetohydrodynamic MPG run with seed magnetic field in the CGM (1~$\mu$G) and partial magnetised AGN feedback ($1\%$ of total AGN power) also shows a higher entropy (within $r<15$ kpc) and cold gas mass, albeit the cold gas remains constrained within $r\lesssim30$ kpc. For a similarly massive galaxy with deeper potential well and low CGM pressure (single phase galaxy; SPG) our simulations show that for both hydro and MHD runs with partial thermal AGN energy, the feedback mechanism remains tightly self-regulating with centrally concentrated cooling (within $r<1$ kpc). Our simulations of a similar mass galaxy with a deeper potential well and higher CGM pressure (SPG-Cool)
show that our AGN feedback mechanism cannot get rid of the high CGM density and pressure and its long term evolution is similar to the multiphase galaxy.
\end{abstract}

\section{Introduction}
\label{sec:introduction}
Galaxies can be supplied cold gas for star formation in multiple ways including cold streams (\citealt{keres2005}), stellar mass loss (\citealt{mathews2003}), and {\it in-situ} cooling of the circumgalactic medium (CGM) due to thermal instability (\citealt{Field1965,white1991}). The cold gas being fed through cold streams encounters hot CGM, which likely prevents it from reaching the core and thus also from fueling star formation in quiescent galaxies (\citealt{heitsch2009}). The
gas being shed by dying stars provides another channel to supply gas for star formation. Energetics show that supernova explosions are capable of driving the stellar ejecta from the galaxy to the CGM, thus limiting star formation (\citealt{voit15L}). However, if the CGM pressure is large, supernovae cannot do so leaving that gas available for star formation within the galaxy (\citealt{voit2020}). The cooling time ($t_{\rm cool}$) of the CGM surrounding massive galaxies is of the order of 100s Myr within the central $r<30$ kpc, much shorter than the galaxy's lifetime (\citealt{hogan2017}). 
As such, the CGM is expected to radiatively cool due to thermal instability and feed the cold gas into the galaxy, fueling star formation. If left uninterrupted, CGM cooling would become catastrophic, making massive galaxies appear blue due to very high star formation rates (\citealt{fabian1994}). However, observations of massive elliptical galaxies show them to be much more quiescent with respect to star formation (\citealt{combes2007}), necessitating the need for some feedback mechanism to quench star formation. 

The trick to any feedback mechanism quenching the star formation
in massive elliptical galaxies is to provide enough energy to the CGM to compensate for radiative cooling without further exciting thermal instabilities which lead to the overproduction of cold gas (\citealt{pizz2005,mccourt12,sharma12}). X-ray observations show that AGN feedback in massive elliptical galaxies appears to achieve these goals without greatly increasing the central entropy (\citealt{birzan2004, kormendy2013, voit15N}). Within the central few kiloparsecs (kpc) of many nearby massive ellipticals, entropy levels in the ambient medium are observed to go below 5 keV cm$^2$ (\citealt{werner2012,werner2014, babyk2018}), corresponding to a central cooling time $t_{\rm cool} \lesssim 100$~Myr. In order to remain in such a state, the feedback loop that limits cooling must be able to tune itself. Otherwise, AGN feedback would overheat the ambient galactic atmosphere in its vicinity. Hydrodynamic numerical simulations show that the feedback loop indeed tunes itself on a time scale of less than 100 Myr in massive elliptical galaxies (\citealt{Gaspari2012_ellipticals, wang2019}).  

The issue with the feedback processes in massive galaxies becomes complicated with observations showing that the amount and spatial distribution of the cold gas in massive galaxies depends on the galaxy's stellar velocity dispersion (\citealt{Wake2012, Bluck2016, Bluck2020, Terrazas2016}). Massive elliptical galaxies that are not central galaxies of galaxy clusters with central velocity dispersion $\sigma_v < 240$ km s$^{-1}$, such as NGC5044, NGC4636, or NGC5813, typically show a significant amount of extended multiphase gas and star formation in addition to high AGN activities and, as such, are called multiphase galaxies. 
On the other hand, massive elliptical galaxies that are not central galaxies of galaxy clusters with $\sigma_v>240$ km s$^{-1}$, such as NGC 4472, NGC4261, or NGC4649, show cooling to be concentrated within the central $r\sim2$ kpc with little multiphase gas and star formation despite having similar AGN activity as galaxies with $\sigma_v < 240$ km s$^{-1}$ (\citealt{voit15L, voit2020}) and are called single phase galaxies. In our previous works, we explored this close interplay between radiative cooling, stellar and AGN feedback and environmental factors like stellar velocity dispersion and CGM pressure in such single phase galaxies (SPG) and multiphase galaxies (MPG) through hydrodynamic simulations (\citealt{prasad2020, Prasad2022}).

Most of the numerical works exploring the intricate nature of feedback processes in massive galaxies, including ours, have relied on hydrodynamic simulations. Hydrodynamic models of AGN feedback in massive halos suffer from their own issues such as the formation of galactic-sized cold gas discs (\citealt{Prasad15,li15}) or physically questionable partitioning of AGN feedback between kinetic and thermal energy. In \citet{prasad2020}, the massive galaxies show elevated entropy within the central $r<5$ kpc compared to observations. This is largely because momentum-heavy jets tend to cause large atmospheric circulation that reconfigures the CGM, with  high entropy gas from larger radii settling close to the center (\citealt{Prasad2022}). Furthermore, magnetic fields can be an important factor as they provide forces that allow for collimation in the AGN jets in addition to contributing to the pressure support of the CGM, thus allowing the jets to travel longer distances before thermalizing. They are also expected to play a critical role at shaping the morphology of the cold phase, making it more filamentary \citep{Ehlert2023,Das2024,Fournier2024}.
This can be critical for the evolution of massive galaxies as AGN feedback, supernova sweeping, and CGM pressure remain tightly coupled in halos with mass $M_{200}\sim 10^{13}$ M$_\odot$ (\citealt{wang2019, prasad2020,mohapatra2025}). As such, there is a need for a magnetohydrodynamic (MHD) numerical study of the evolution of massive galaxies.

\citealt{grete2025} introduced the XMAGNET\footnote{See \url{https://xmagnet-simulations.github.io} for further material and videos.} (eXascale simulations of Magnetized AGN feedback focusing on Energetics and Turbulence) suite of simulations designed to explore the role of magnetised AGN feedback in massive galaxies, groups, and clusters. It discussed the role of magnetised and kinetic AGN feedback mechanisms in {\it Perseus}-type cool-core galaxy clusters ($M_{200}\sim 6.6\times 10^{15} \, M_\odot$). In this work, we focus on the role of AGN feedback in lower mass galactic halos with $M_{200}\sim 10^{13.5} \, M_\odot$, in which AGN feedback using the same model couples with CGM pressure and Type Ia supernovae (SNIa) heating in the presence of magnetic fields. The paper proceeds as follows : Section \ref{sec:methods} introduces the numerical and analysis methods for our simulations, Section \ref{sec:results} presents the main results of our simulations, Section \ref{sec:disc} looks more closely at how AGN feedback couples with CGM, its larger implications and compares this work with similar studies, and finally in Section \ref{sec:conc}, we present the conclusions based on our findings. 

\section{Methods}
\label{sec:methods}
Overall, the simulations follow setup described in detail in \citet{grete2025}.
This section briefly discusses the key numerical methods adopted in the open source, performance-portable AthenaPK\footnote{AthenaPK is openly developed at \url{https://github.com/parthenon-hpc-lab/athenapk}.} code in carrying out the magnetised AGN feedback simulations and analysis methods.
AthenaPK is based on the adaptive mesh refinement framework Parthenon (\citealt{grete2022}) and Kokkos \citep{Trott2022} to run on any (GPU) architecture.
All simulations presented in this paper employ an overall second-order accurate, shock-capturing, finite volume scheme consisting of RK2 
time integration, piecewise-linear reconstruction, and a HLLD (MHD) 
or HLLC (hydro) Riemann solver. In the MHD case, we use the hyperbolic divergence cleaning method for $\nabla \cdot {\bf B} = 0$ (\citealt{Dedner2002}).  Optically thin radiative cooling is treated using the exact integration method introduced by \citet{Townsend2009}. 

\begin{table*}
\caption{List of runs.} 
\resizebox{0.90 \textwidth}{!}{
\begin{tabular}{c c c c c c c c c c c c}
\hline
Runs & M$_{200}^\ddag$ & c$_{200}^{\ddag}$ & R$_{200}^\ddag$ & K$_0^\dag$ & K$_{100}^\dag$ & $\alpha_K^\dag$ & M$_{\rm BH}$ & M$_*^{\beta}$ & $r_H^\beta$ & B field & Thermal\\
    & ($10^{13}$/M$\odot$) &  & (kpc) &    &      &      & ($10^8$/M$_\odot$) & ($10^{11}$/M$_\odot$) & (kpc) & $(\mu G$) & feedback\\
\hline
 MPG-MHD & 4.4 & 9.5 & 730 & 1.3 & 150 & 1.05 & 4.6 & 1.2 & 1.2 & 1.0 & yes\\ 
 \hline
 MPG-hydro & 4.4 & 9.5  & 730 & 1.3 & 150 & 1.05 & 4.6 & 1.2 & 1.2 & no & yes\\ 
 \hline
 MPG-hydro-kinetic & 4.4 & 9.5  & 730 & 1.3 & 150 & 1.05 & 4.6 & 1.2 & 1.2 & no & no\\
 \hline
 SPG-MHD & 4 & 7.5  & 700  & 1.5 & 400 & 1.05 & 26 & 2.0 & 1.6 & 1.0 & yes \\
 \hline
 SPG-hydro & 4 & 7.5  & 700 & 1.5 & 400 & 1.05 &26 & 2.0 & 1.6 & no & yes\\
 \hline 
 SPG-Cool-MHD &  4 & 7.5  & 700 & 1.5 & 200 & 1.05 & 26 & 2.0 & 1.6 & 1.0 & yes\\
 \hline
 SPG-Cool-hydro &  4 & 7.5  & 700 & 1.5 & 200 & 1.05 & 26 & 2.0 & 1.6 & no & yes\\
 \hline
 \end{tabular} }
 \begin{tablenotes}
      \item  $\ddag$ - Parameters for the NFW profile : 
      $\Phi_{\rm NFW}$ = $-\frac{G M_{200}}{r} \frac{\ln(1+ c_{200}r/r_{200})}{[\ln(1+c_{200}) - c_{200}/(1+c_{200})]}$
      \item $\dag$ - Parameters are for baryon entropy profile, $K = K_0 + K_{100} (\frac{r}{100 kpc})^{\alpha_K}$ (\citealt{cavagnolo09})
      \item $\beta$ - Parameters for the BCG's Hernquist profile : $\Phi_{\rm BCG} = \frac{G M}{r + r_H}$
      \item SMBH mass, $M_{\rm BH}$, is taken from \citet{kormendy2013} for SPG and from \citet{David2009} for the MPG halo.
      \item All the runs have been run for 2 Gyr except for MPG-hydro-kinetic (1.6 Gyr) and SPG-Cool-MHD (1.7 Gyr)
 \end{tablenotes}
\label{Tab:Runs}
\end{table*}

\subsection{Simulation Grid}
\label{sec:grid}
The simulation is set up in a (6.4 Mpc)$^3$ cubic box, covered by $512^3$ cells in the static root grid. We enforce 7
levels of refinement within $([-25,25] \mathrm{kpc})^3$ (where the root grid is the 0th level). Thus, the central (50 kpc)$^3$ of the setup domain is covered with a uniform grid of $512^3$ cells with cell side length of $\Delta x \approx 100$ pc. 
Outside of this central region the resolution is progressively coarsened by factors of 2 (with intervening regions of constant resolution either 256 or 512 cells wide) to ensure that the transition in resolution is smooth and all phenomena of interest are sufficiently resolved.

\subsection{Gravitational Potential }
\label{sec:grav}
The gravitational potential confining each simulated galactic atmosphere consists of three components: a dark-matter potential that follows an NFW profile (\citealt{nav1997}), a stellar potential that follows a Hernquist profile (\citealt{hern1990}), and a supermassive black hole (SMBH) potential approximated with a Paczyinski-Witta profile (\citealt{pacwita1980}). 

Table \ref{Tab:Runs} lists the runs discussed in this paper along with the parameters used for initial setup. In the SPG (halo with deeper central potential, $\sigma_v>240$ km s$^{-1}$, and lower CGM pressure) and SPG with cooler core (Halo with deeper central potential, $\sigma_v>240$ km s$^{-1}$, and higher CGM pressure) simulations, a central galaxy with stellar mass $M_* = 2 \times 10^{11} \, M_\odot$ and central velocity dispersion $\sigma_v \approx 280 \, {\rm km \, s^{-1}}$ is embedded in a halo of mass $M_{\rm halo} = 10^{13.6} \, M_\odot$ ($c_{200}=7.5$) and has a central black hole mass $M_{\rm BH} = 2.6 \times 10^9 \, M_\odot$. In the MPG simulations (halo with shallower central potential, $\sigma_v<240$ km s$^{-1}$, and higher CGM pressure), a central galaxy with stellar mass $M_* = 1.2 \times 10^{11} \, M_\odot$ and central velocity dispersion $\sigma_v \approx 230 \, {\rm km \, s^{-1}}$ is embedded in a halo of mass $M_{\rm halo} = 10^{13.6} \, M_\odot$ ($c_{200}=7.5$) and has a central black hole mass $M_{\rm BH} = 4.6 \times 10^8 \, M_\odot$. 

\subsection{Initial Baryon Profile}
\label{sec:init}
\begin{figure*}
\centering
 \includegraphics[width=0.45\textwidth]{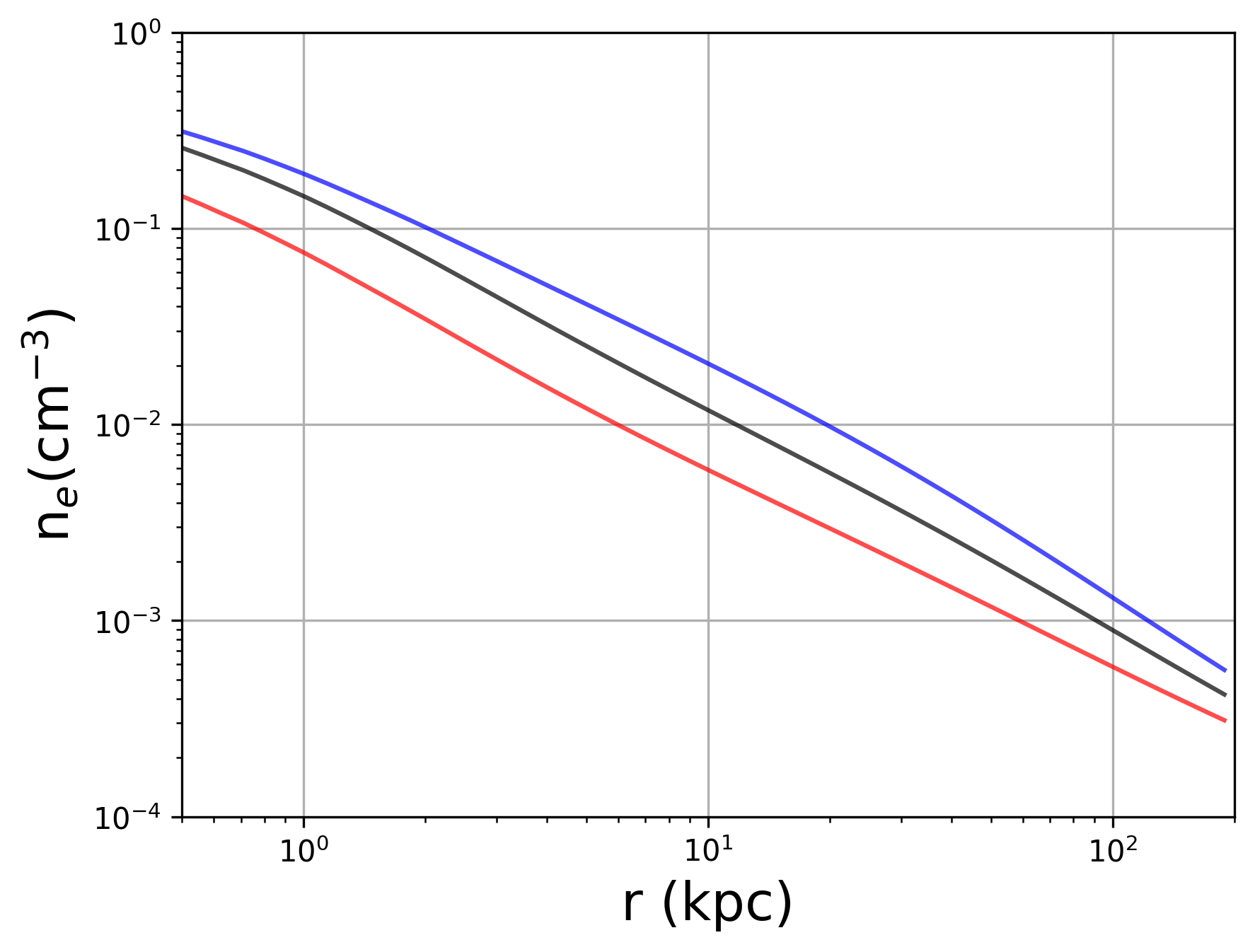}
 \includegraphics[width=0.45\textwidth]{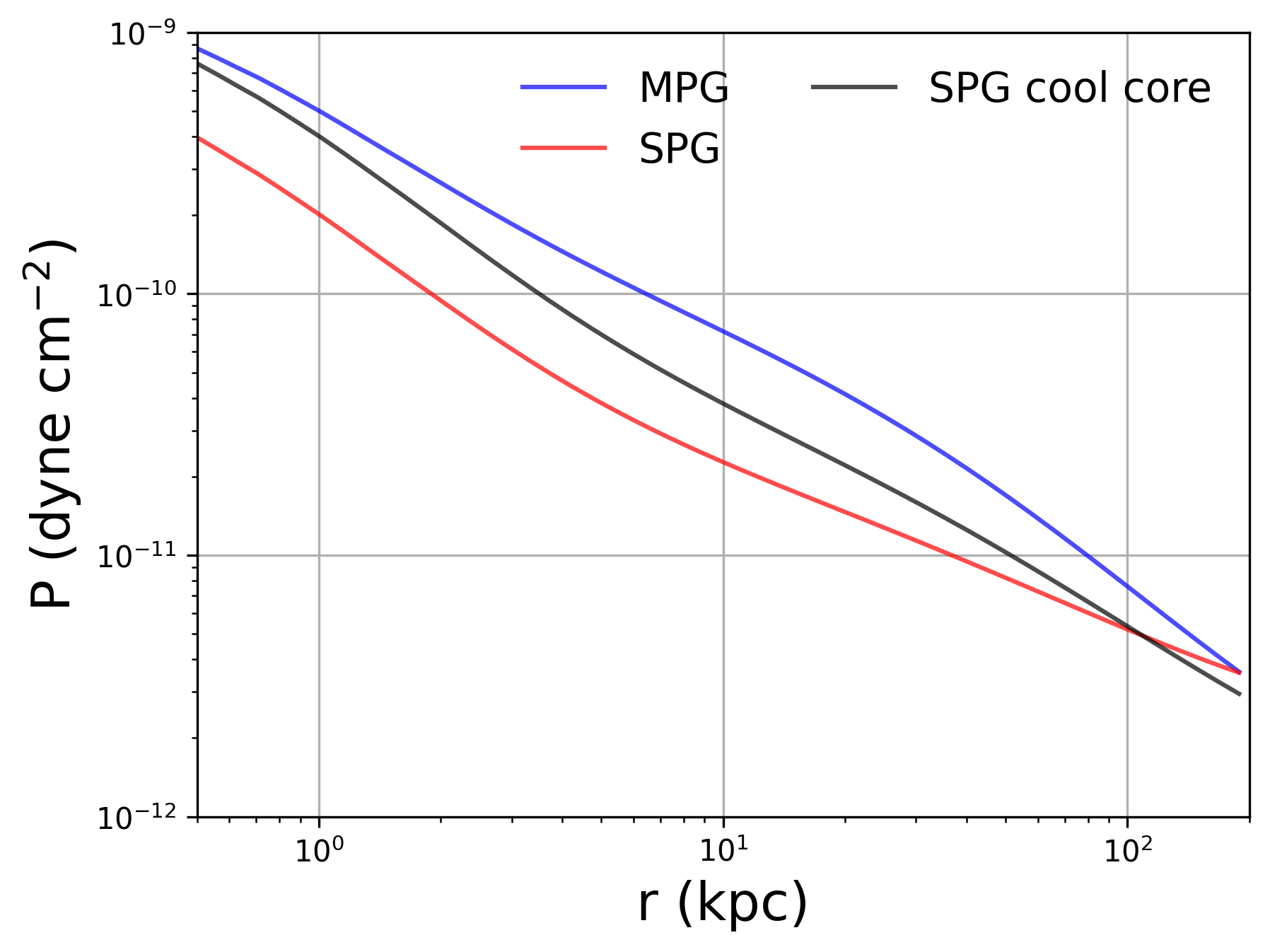}
 \includegraphics[width=0.45\textwidth]{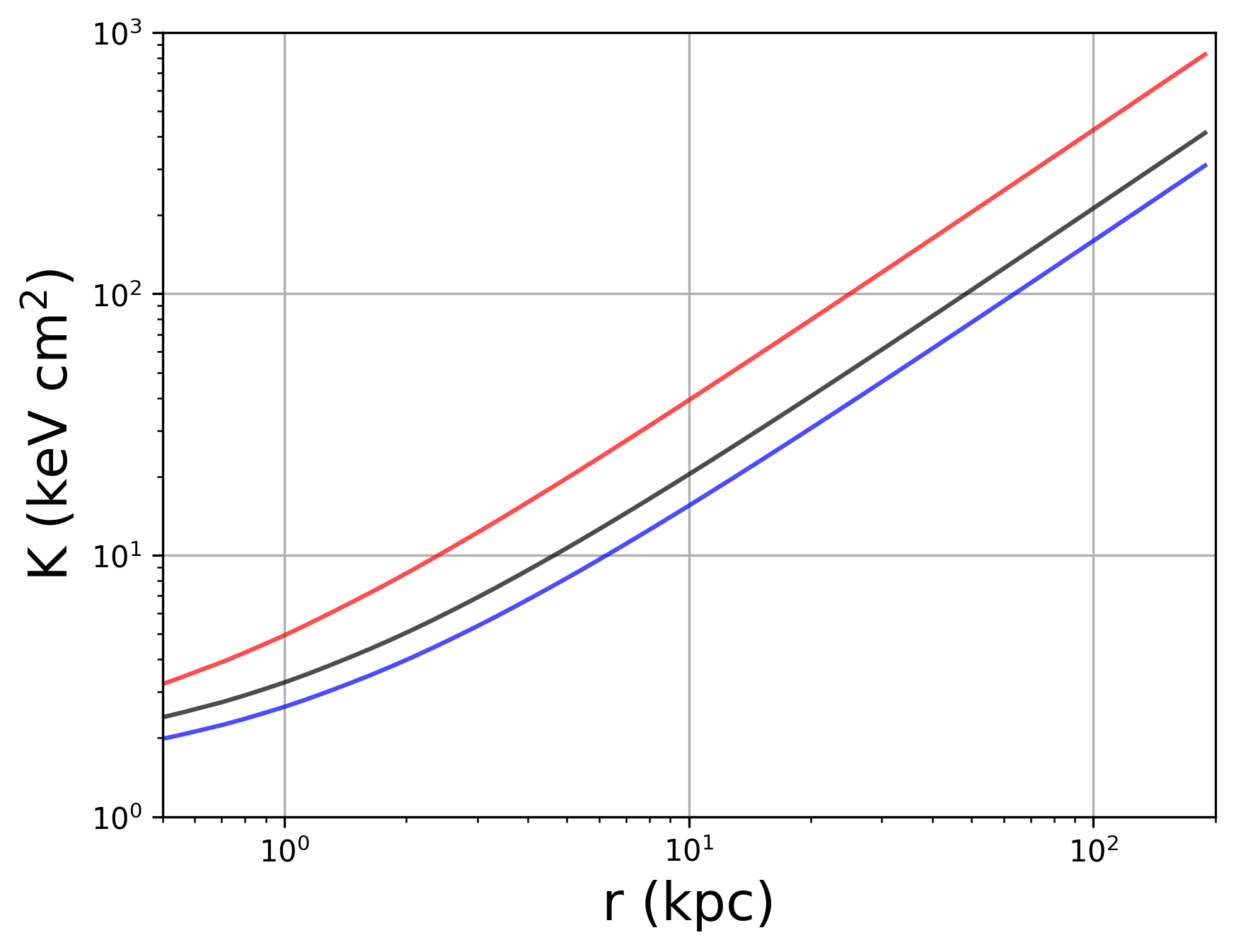} 
 \includegraphics[width=0.45\textwidth]{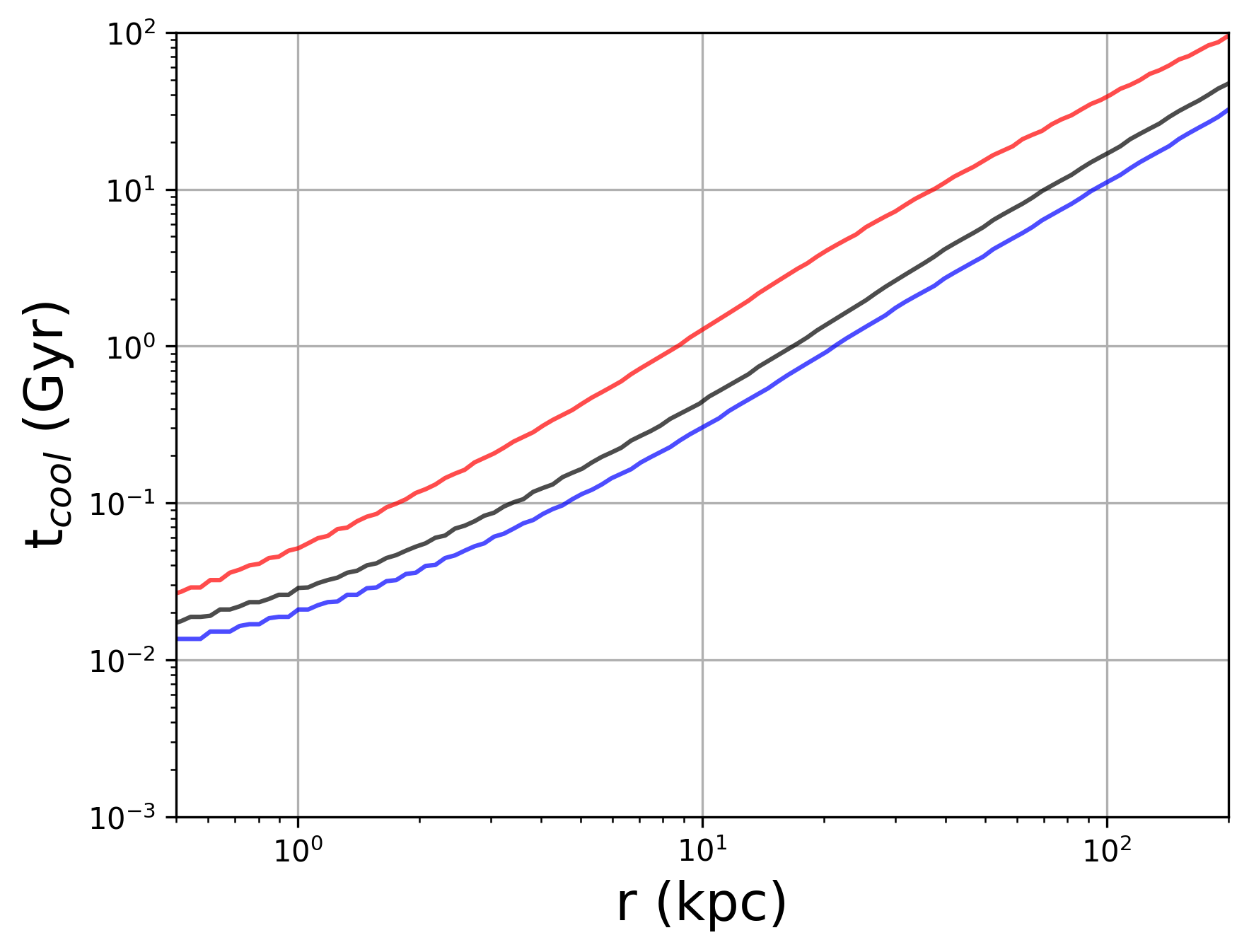}
 \caption{ Radial electron density ({\it top left panel}), pressure ({\it top right panel}), entropy ({\it bottom left panel}) and cooling time
 ({\it bottom right panel}) profile for the MPG (blue line), SPG (red line) and SPG-Cool (black line) runs at $t=0$. The baryon profile of the SPG-Cool halo is closer to the MPG halo.  
}
 \label{fig:init_prof}
\end{figure*}
As stated in Section \ref{sec:grav} we initialize the baryon profile in hydrostatic equilibrium within the gravitational potential well given by the NFW+Hernquist+SMBH profile. Figure \ref{fig:init_prof} shows the initial radial electron density (top panel), entropy (middle panel) and cooling time ($t_{\rm cool}$; lower panel) profile for the single phase galaxy (SPG; blue line), multiphase galaxy (MPG; red line) 
and galaxy with SPG potential with shallower entropy and higher CGM density and pressure (SPG-cool core; black line). For the SPG halo we have chosen NGC4472 as the analogue (\citealt{werner2014}), and for the MPG halo NGC5044 has been chosen as the analogue (\citealt{werner2014}).
The SPG with cooler core has been initialised to study the interplay between AGN feedback, SNIa-driven sweeping of gas out of the interstellar medium (and thus supression of star formation),  
and CGM pressure in a situation where the central stellar velocity dispersion is $\sigma_v > 240$ km s$^{-1}$ (like the SPG case) but also has a much higher CGM pressure (like the MPG case) to start with. This will allow us to test whether AGN+SNIa feedback alone is able to reconfigure the CGM entropy profile to be similar to the SPG-type system -- in other words, if it can be reconfigured by feedback alone to have the lower CGM density and pressure that observations tell us are appropriate for the central potential. (If this is not the case, this likely implies that either the AGN feedback has to be active during the halo assembly stage to keep the CGM pressure low or a merger between comparably massive cosmological halos is needed to reconfigure the CGM.) Figure~\ref{fig:init_prof} shows that the baryon profile for the SPG with cooler core is closer to that of the multiphase galaxy with higher CGM pressure and cooling time, $t_{\rm cool}\lesssim 0.5$ Gyr within the central ($r<10$ kpc). The parameter of the runs are listed in table \ref{Tab:Runs}.

In order to break the symmetry in the initial baryon profile, we initialise the baryons with velocity perturbations as in \citet{grete2025}. The perturbations are generated in spectral space, based on 40 wave modes chosen randomly with a characteristic scale between 50 kpc and 200 kpc. The amplitudes of the velocity perturbations are set by an inverse parabolic shape with a peak characteristic length scale of 30 kpc and scaled to a root mean squared velocity of 40 km/s.

\subsection{Dynamical Equations}
\label{sec:equation}
Galaxy evolution with radiative cooling, star formation and stellar feedback, including SNIa feedback and AGN feedback, is done through the following magnetohydrodynamic (MHD) equations:
\begin{equation}
\frac{\partial \rho}{\partial t} + \nabla \cdot (\rho {\bf u}) = \dot{\rho}_{\rm AGN} + \dot{\rho}_{\rm SNIa} - \dot{\rho}_{\star} - \dot{\rho}_{\rm acc}
\end{equation}
\begin{equation}
\frac{\partial{\rho {\bf u}}}{\partial t} + \nabla \cdot [\rho {\bf u} {\bf u} - \frac{{\bf B}{\bf B}}{4\pi}] + \nabla P = \rho {\bf g} + \dot{\mu}_{\rm AGN},
\end{equation}
\begin{equation}
\begin{split}
\frac{\partial E}{\partial t} + \nabla \cdot [ (E^\star+P^\star){\bf u} - \frac{{\bf B({\bf B}\cdot{\bf u})}}{4\pi}] = \rho {\bf u}\cdot{\bf g} + \dot{E}_{\rm SNIa} + \dot{E}_{\star} \\
+ \dot{E}_{\rm AGN} - n_e n_i \Lambda(T)
\end{split}
\end{equation}
\begin{equation}
\frac{\partial {\bf B}}{\partial t} - \nabla \times ({\bf u} \times {\bf B}) = 0,
\end{equation}
\begin{equation}
E = \frac{\rho{\bf u}\cdot{\bf u}}{2} + \frac{P}{\gamma -1} + \frac{B^2}{2}
\end{equation}
where $\mu_{\rm AGN}$ is the momentum jet material, $\Lambda(T)$ is the cooling function, and P, {\bf B}, {\bf u} and $E$ are the pressure, magnetic field, fluid velocity, and total energy density respectively. $E^\star$ is the sum of gas kinetic and internal energy density and $P^\star$ is the sum of gas and magnetic pressure. ${\dot \rho_{\rm AGN}}$, ${\dot \rho_{\rm SNIa}}$, ${\dot \rho_{\star}}$, and ${\dot \rho_{\rm acc}}$ represent the density addition/depletion due to AGN feedback, SNIa feedback, star formation and accretion onto the central supermassive black hole. ${\dot E_{\rm AGN}}$, ${\dot E_{\rm SNIa}}$, and ${\dot E_{\star}}$ represent the power due to AGN, SNIa, and stellar feedback. For all of our simulations we assume $\gamma=5/3$ for the ideal gas. All other physical processes, including radiative cooling, star formation and stellar feedback, and AGN feedback are implemented in an operator split way. We expand upon the details of each of these implemented physics modules in the following sections. 

\subsection{Initial magnetic fields and Radiative Cooling}
\label{sec:cooling}

All of the magnetised runs have been initialised with a weak magnetic field of 1 $\mu$G. The initial magnetic fields are also perturbed in a manner identical to that of the initial velocity.

The plasma composition of the ICM is assumed to be $25\%$ helium with the remaining baryonic mass being in hydrogen and electrons.
The temperature, $T$, is defined as
\begin{equation}
T = \frac{\mu m_p P}{k_B \rho}
\end{equation}
where $m_p$ is the mass of a proton, $k_B$ is Boltzmann’s constant, and $\mu$ is the mean weight. The plasma cooling rate ($n_e n_i \Lambda(T)$) is based on tables from \citet{schure2009} assuming $1 Z_\odot$ metallicity for all the runs, which is appropriate for cosmological halos in the mass range under consideration. The CGM cooling time $t_{\rm cool}$, is defined as the ratio of total internal energy to the cooling rate i.e. $t_{\rm cool}= n k_B T/ [n_e n_i \Lambda(T)]$, where $\Lambda$ is the cooling function. To account for the net cooling of the CGM, we calculate X-ray luminosity ($L_X$) using pyXSIM (\citealt{Zuhone2016}) for all our simulations. For all our $L_X$ calculations we keep the temperature range between 0.5-7 keV.

\subsection{Stellar Feedback}
\label{sec:stellar}
To account for star formation and stellar feedback, we deplete gas from cells with density $n \gtrsim 50$ cm$^{-3}$ and temperature lower than $2\times10^4$ K within $r=25$ kpc of the galaxy center. From such cells, we remove gas mass, $\Delta {\rm M_{dep}}=50$ cm$^{-3} \times \bar{\mu} \times$ Vol$_{\rm cell}$ instanteneously, where $\bar{\mu}$ is the mean molecular weight and Vol$_{\rm cell}$ is the cell volume. To account for stellar feedback, we deposit an equivalent thermal energy ($\sim \epsilon_{*}  \Delta$M$_{\rm dep}$ c$^2$) by assuming a feedback efficiency of gas rest mass into thermal feedback energy of $\epsilon_* = 5\times10^{-6}$.   

In addition, to account for feedback from Type Ia supernova events in the galaxy, we use a spherically symmetric kernel depositing mass and thermal energy in the domain based on the assumed stellar density profile. The energy ($e_{\rm SNIa}$) and mass ($\rho_{\rm SNIa}$) density deposited in the domain is given as:
\begin{equation}
\dot{e}_{\rm SN
Ia} = \eta E_{\rm SNIa} \rho_\star , {\rm and}
\end{equation}
\begin{equation}
\dot{\rho}_{\rm SNIa} = \alpha \rho_\star 
\end{equation}
where $\eta = 3 \times 10^{-14}$ SNIa yr$^{-1}$ M$_\odot^{-1}$ is the SNIa rate in the local universe, $E_{\rm SNIa} = 10^{51}$ ergs is the energy released per SNIa event, $\alpha = 10^{-19}$ s$^{-1}$ is the mass ejection rate by the old stellar population (\citealt{voit15L}) and $\rho_\star$ is the stellar density given by the {\it Hernquist} profile. 

\subsection{Cold gas accretion}
\label{sec:agn}
Cold gas ($T<5\times10^4$ K) within the accretion zone ($r_{\rm acc} < 1$ kpc) triggers AGN feedback. The accretion rate is calculated as
\begin{equation}
\dot{M}_{\rm acc} = \int_{r<r_{\rm acc}} \rho_{\rm cold}/t_{\rm acc} dV 
\end{equation}
where $t_{\rm acc}$ is the cold gas depletion time scale. For all the runs in this paper a fixed value of $t_{\rm acc}=10$ Myr has been used. The accreted gas is removed from the cold cells ($T<5\times10^4$ K) within the accretion zone. 

\subsection{AGN feedback}
\label{sec:agnfb}
The AGN feedback is introduced into the simulation using a zone 
centered on the supermassive black hole (SMBH) at the center of the domain. We assume the SMBH is located at the center of the galaxy.  

The total AGN power, $\dot{E}_{\rm AGN}$, is then set to
\begin{equation}
\dot{E}_{\rm AGN} = \epsilon_{\rm AGN}  \dot{M}_{\rm acc} c^2
\end{equation}
where $\epsilon_{\rm AGN} = 10^{-3}$ is the accretion efficiency for all the runs and $c$ is the speed of light. The total AGN power is then partitioned into thermal heating, kinetic jet, and magnetic power as follows:
\begin{equation}
\dot{E}_{\rm AGN} = \dot{E}_{\rm th} + \dot{E}_{\rm kin} + \dot{E}_{\rm mag} = (f_{\rm th} + f_{\rm kin} + f_{\rm mag}) \dot{E}_{\rm AGN}
\end{equation}
where, $f_{\rm th}, f_{\rm kin}$, and $f_{\rm mag}$ are the thermal, kinetic and magnetic fraction of the total AGN power respectively. For all the simulations with magnetised jets $f_{\rm th} = 0.25$, $f_{\rm kin}=0.74$, and $f_{\rm mag} = 0.01$, while for the hydrodynamic runs $f_{\rm th} = 0.25$ and $f_{\rm kin}=0.75$. For the pure kinetic AGN feedback, MPG-hydro-kinetic run, the kinetic energy fraction $f_K=1$.  

For thermal AGN feedback, we inject mass and thermal energy into the domain volumetrically within a sphere centered around the supermassive black hole. The equations for the spherical energy and mass deposition are:
\begin{equation}
    \label{eq1}
\begin{split}
\dot{e}_T(r) & = \frac{3 \dot{E}_{\rm th}}{4\pi R_T^3}\ \ \ {\rm if}\  r\leq R_T \\
             & = 0\ \ \ \ {\rm otherwise}
\end{split}
\end{equation}
\begin{equation}
    \label{eq2}
\begin{split}
\dot{\rho}_T(r) & = \frac{3 f_{\rm th} \dot{M}_{\rm acc}}{4\pi R_T^3},\ \ \ {\rm if}\  r\leq R_T \\
 & = 0,\ \ \ \ {\rm otherwise}
\end{split}
\end{equation}
where $R_T=1$ kpc is the thermal feedback radius. 

For the kinetic AGN feedback, mass and energy are injected through discs at a specified distance above and below the supermassive black hole (SMBH) along the z-axis. The discs have radius $R_D = 0.5$ kpc and a thickness $H_D=2\Delta_x = 0.2$ kpc, so that the jet source region is resolved by several grid points across its radius and two in depth. The discs are offset from the central SMBH by $R_{\rm offset} = 1$ kpc.  
The total kinetic energy injection $\dot{E}_{\rm kin}$ is put in the domain through the jet source region, with jet velocity $v_{\rm jet}$  given by:
\begin{equation}
\label{eq:vjet}
v_{\rm jet}^2 = 2[\epsilon c^2 - (1-\epsilon) T_{\rm jet}k_B/((\gamma -1 )\mu m_p)]
\end{equation}
where $T_{\rm jet} = 10^8$ K is the initial jet temperature, $\epsilon=10^{-3}$ is the accretion efficiency, $\mu$ is the mean molecular weight, $k_B$ is the Boltzmann constant and $m_p$ is the proton mass. Equation \ref{eq:vjet} takes into account the split of total kinetic energy of the jet into kinetic and internal energy of the injected jet mass in the jet source region. The density, $\rho_{\rm jet}$ , of the injected jet material is given as:
\begin{equation}
\dot{\rho}_{\rm jet} = f_{\rm kin} \dot{M}_{\rm acc}/(2 \pi R_D^2 H_ D)
\end{equation}
 where the symbols have their usual meaning.

 The magnetic energy for the AGN, $E_{\rm mag}$, is introduced in the domain via a closed field loop (i.e., ``donut'') magnetic field configuration using a vector potential and normalising the magnetic field strength to match the target magetic energy $\dot{E}_B$. The choice of using vector potential is to ensure that the injected magnetic field configuration
 \begin{equation}
   \begin{split}
   B_\theta(r,\theta, h) = B_0 L_M {\rm exp}(-r^2/L_M^2)\\
   \ \ \ {\rm if}\ \ h_{0,M} \leq |h| \leq h_{0,M} + h_M \\
   = 0\ \ \ \ \ \ {\rm otherwise}\ \ \ \ \ 
   \end{split}
 \end{equation}
 remains divergence-free, which is necessary for numerical stability.  The closed magnetic field loops in our simulations are seeded within the jet source region for kinetic feedback with $h_{0,M}=1$ kpc, $h_M = 2\Delta_x \sim 0.2$ kpc, and $L_M=0.25$ kpc.
 The injected magnetic field strength is then normalised with respect to the $\dot{E}_{\rm mag}$ at each time step. Readers are referred to \citet{grete2025} for a detailed explanation of the normalisation step. 

\section{Results}
\label{sec:results}
This section describes the key results of our simulations. We first present the results of the MPG-hydro-kinetic, MPG-hydro and MPG-MHD simulations in Section \ref{sec:mpg}.   
In Section \ref{sec:hyd_spg} we discuss the results of the SPG-hydro and SPG-MHD simulations. This is followed by Section \ref{sec:spg_cool} where we explore the results of our SPG-Cool hydro and SPG-Cool MHD simulations. In Section \ref{sec:morph} we compare and contrast the jet morphology and AGN-CGM coupling as it evolves in the MHD and hydro simulations. Finally, in Section \ref{sec:config} we look at the mass and energy flows with time for different simulations.

\subsection{Multiphase Galaxy}
 \label{sec:mpg}
We simulated the multiphase galaxy (MPG) with radiative cooling, depletion of cold gas due to star formation, stellar feedback (including Type Ia SN feedback) and AGN feedback with kinetic, thermal and magnetic AGN feedback.  As discussed in Sections \ref{sec:cooling} and \ref{sec:agnfb}, in the MHD run the intragroup medium is initialised with a seed magnetic field of $1\ \mu$G along with magnetic energy being injected into the domain through magnetic AGN feedback during the galaxy evolution. 
All MPG runs show explosive AGN feedback and formation of extended cold gas clumps and filaments; the pure kinetic AGN feedback run, however, shows much higher jet power  
compared to other runs during the multiphase galaxy evolution 
In the following subsections, we look at the evolution of different quantities for all the MPG runs. 
\subsubsection{Temporal evolution}
\begin{figure}
\centering
 \includegraphics[width=0.47\textwidth]{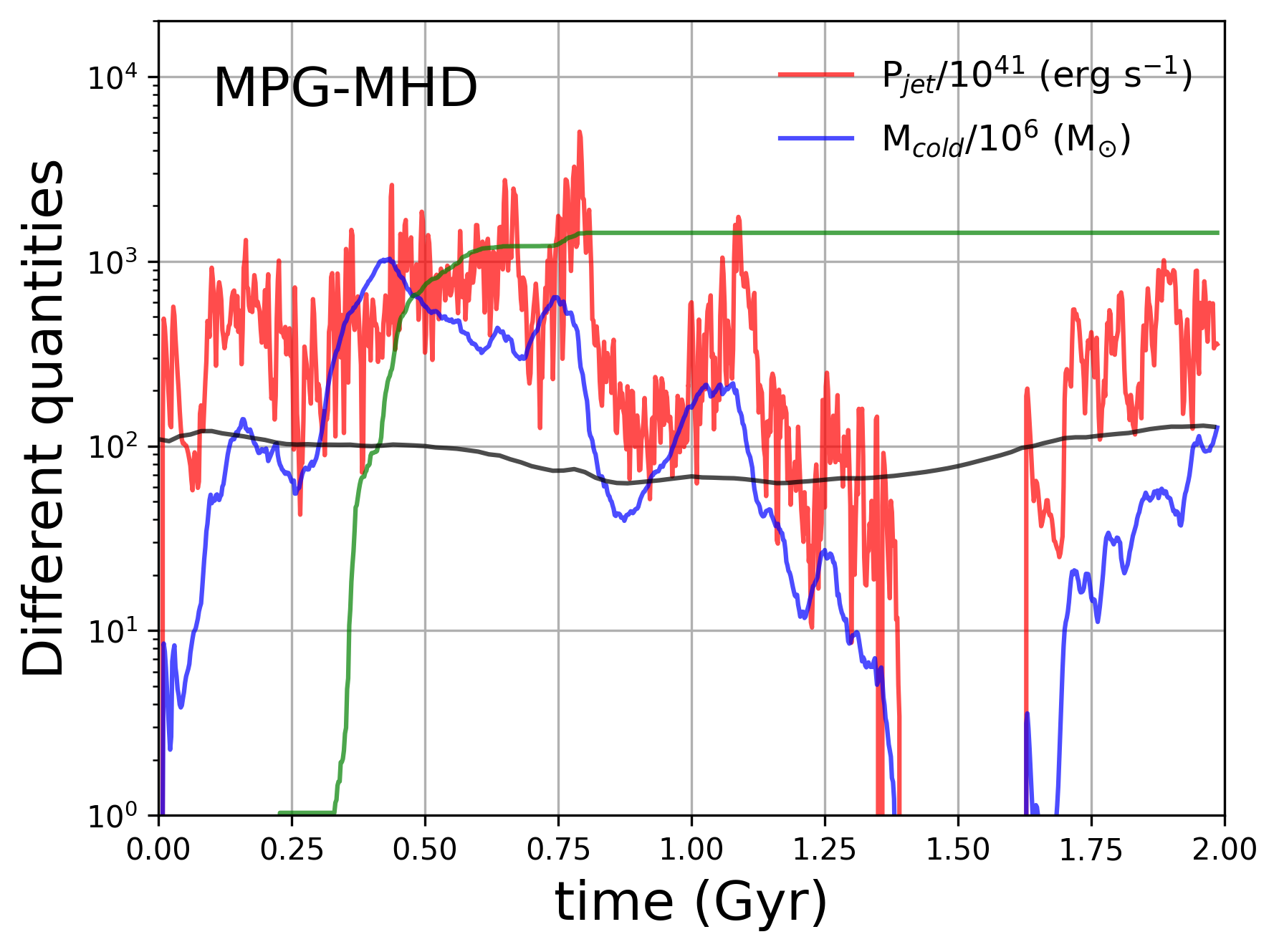}
 \includegraphics[width=0.47\textwidth]{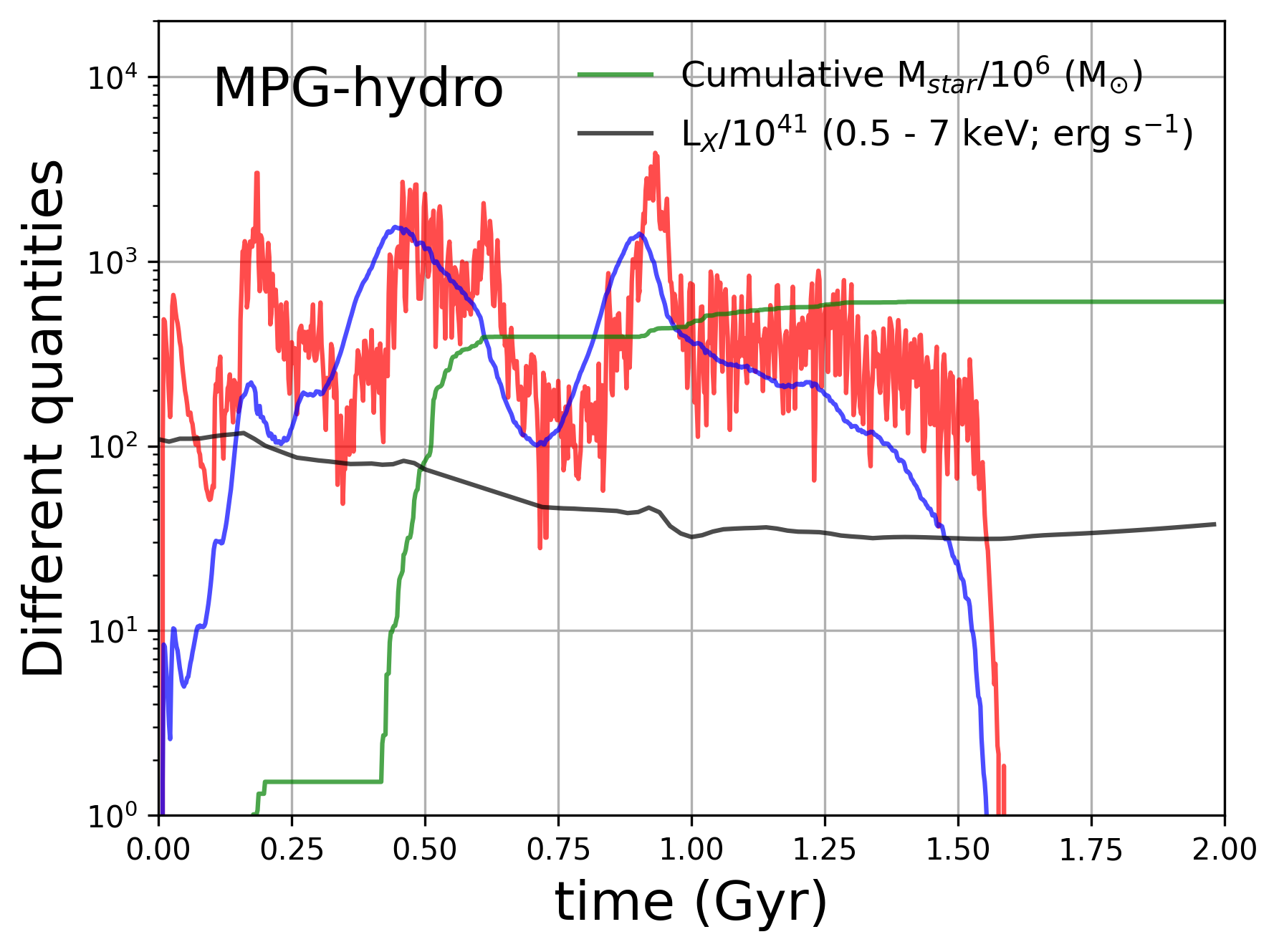}
\includegraphics[width=0.47\textwidth]{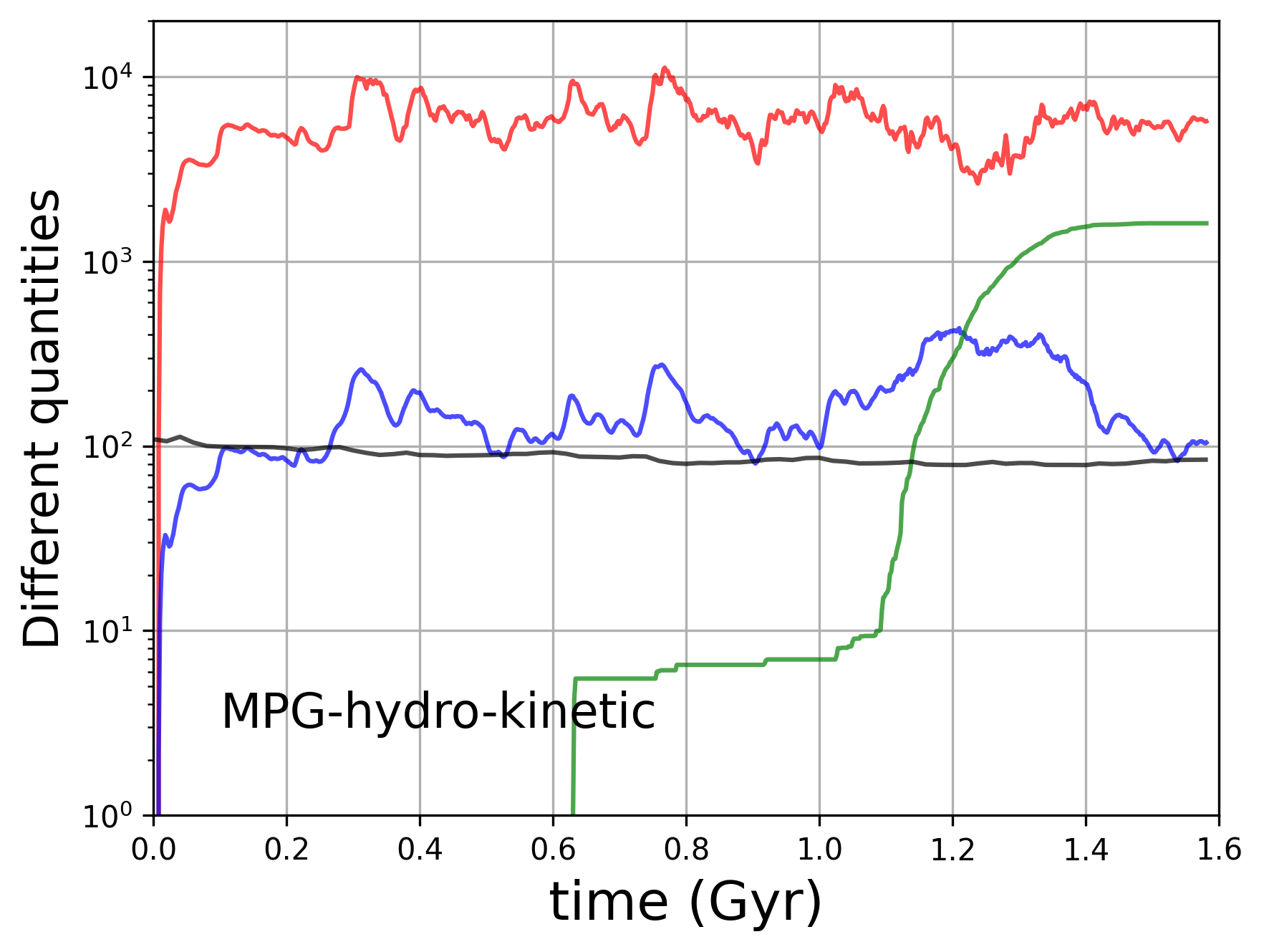}
 \caption{ Jet power ($P_{\rm jet}$; red line) and cold gas mass (blue line), total stellar mass (M$_\star$; green line) and X-ray luminosity for the $0.5-7$ keV gas within central $r\lesssim50$ kpc (black line) with time for the MPG-hydro-kinetic run (bottom panel), MPG-hydro run (middle panel) and MPG-MHD run (top panel). Note the higher $P_{\rm jet}$ ($<P_{\rm jet}>\sim 5.6\times10^{44}$ erg s$^{-1}$) and lower $M_{\rm cold}$ for the pure kinetic AGN feedback run compared to MPG-hydro ($<P_{\rm jet}>\sim 3.9\times10^{43}$ erg s$^{-1}$) and MPG-MHD ($<P_{\rm jet}>\sim 4.2\times10^{44}$ erg s$^{-1}$) runs. 
}
 \label{fig:mpg_pjet}
\end{figure}
Figure \ref{fig:mpg_pjet} shows the temporal evolution of jet power (P$_{\rm jet}$), total cold gas mass ($T<10^5$ K), X-ray  luminosity ($T_{\rm keV} = 0.5-7$ keV), and total stellar mass for the MPG-hydro-kinetic run with pure kinetic AGN feedback (bottom panel), MPG-hydro run with kinetic+thermal AGN feedback (middle panel) and MPG-MHD run with kinetic+thermal+magnetic AGN feedback (top panel). The pure kinetic AGN feedback run has an order of magnitude higher jet power ($P_{\rm jet}$) compared to kinetic+thermal and kinetic+thermal+magnetic AGN feedback runs. It also shows a persistent population of cold gas throughout the simulation run time (with $M_{\rm cold} \sim 10^8$ M$_\odot$), unlike other runs where the cold gas mass shows large fluctuations as the galaxy evolves. As a result, AGN feedback remains on for the entire simulation time of 1.6 Gyr for the MPG-hydro-kinetic run.

For the MPG-hydro run with kinetic+thermal AGN feedback and MPG-MHD run with kinetic+thermal+magnetic AGN feedback, the cold gas mass fluctuates with time with a peak cold gas mass of $M_{\rm cold} \sim 10^9\ {\rm M}_{\odot}$, resulting in a peak AGN power ${\rm P_{jet}} \sim$ few times $10^{44}$ erg s$^{-1}$. Both MPG-MHD and MPG-hydro runs show star formation with total stellar mass ($M_\star$) exceeding a few times $10^8$ M$_\odot$ ($<\dot{M}>\sim$ few times $\times0.1$ M$_\odot$). Similar to the kinetic AGN feedback run, the AGN remains active for a long duration ($\sim 1.2-1.5$ Gyr) albeit with 
a lower jet power fluctuating between $10^{42}$ erg s$^{-1}$ - $3\times10^{44}$ erg s$^{-1}$.  
The long duration jet activity leads to overheating of the galaxy core, resulting in a decline of the cold gas mass that ultimately shuts off the AGN activity.  The quiescent phase for the MPG-MHD run lasts for $\delta t \sim 0.25$ Gyr, after which cooling picks up again. This leads to the next AGN feedback cycle. On the other hand, for the MPG-hydro run the overheated galaxy core ($r<30$ kpc) does not show any further cold gas formation and, consequently, AGN activity until the end of the simulation. 
The overheating of the CGM can also be inferred from the X-ray luminosity ($L_X$;0.5 keV$< T<$ 7 keV), calculated using {\it PyXSIM} (\citealt{Zuhone2016}) for the MPG-hydro run, where $L_X$ declines from $\sim 10^{43}$ erg s$^{-1}$ to $\sim 5\times10^{42}$ erg s$^{-1}$ by $t\sim 1.25$ Gyr.

\subsubsection{Radial profiles}
\begin{figure*}
\centering
 \includegraphics[width=0.47\textwidth]{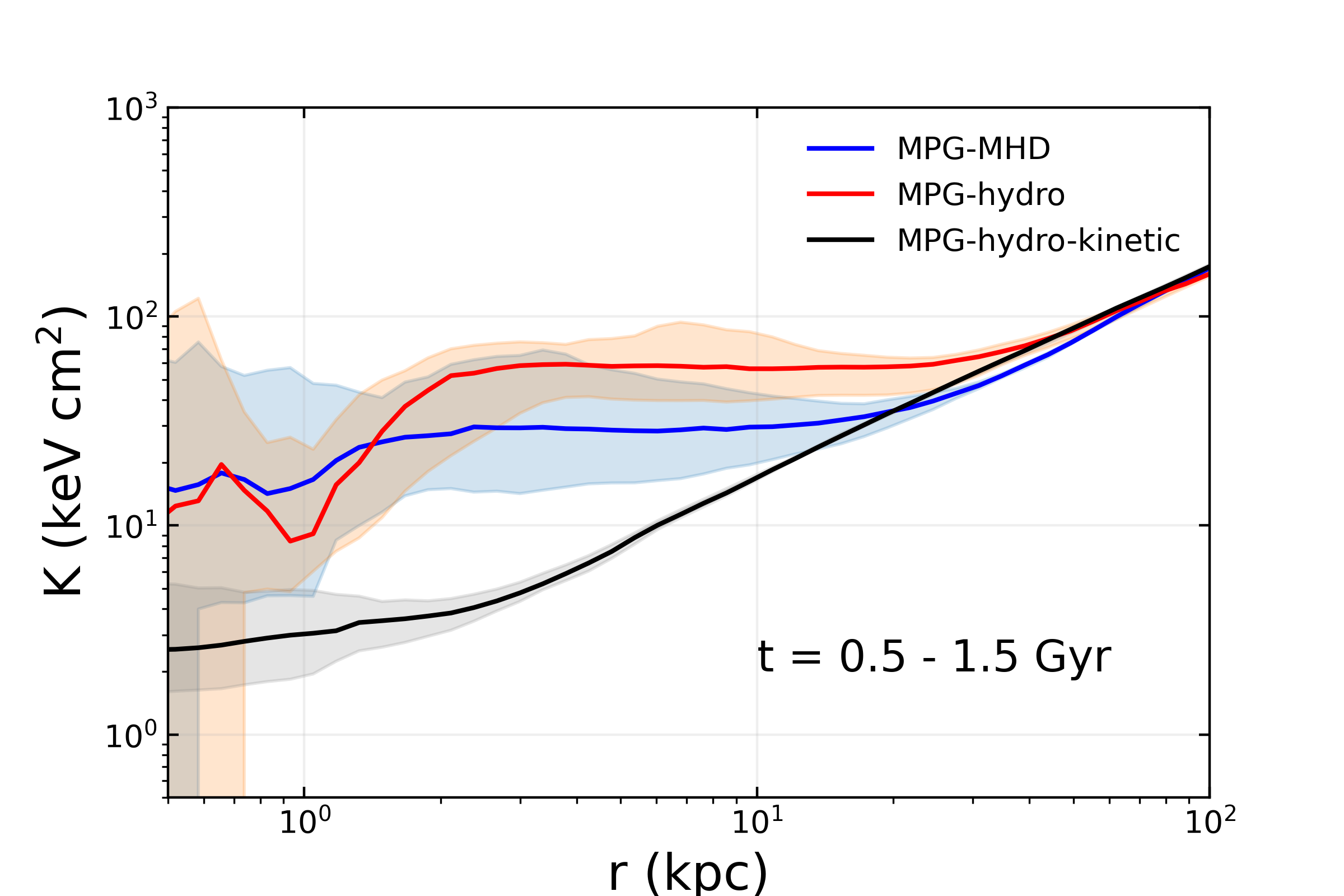}
 \includegraphics[width=0.47\textwidth]{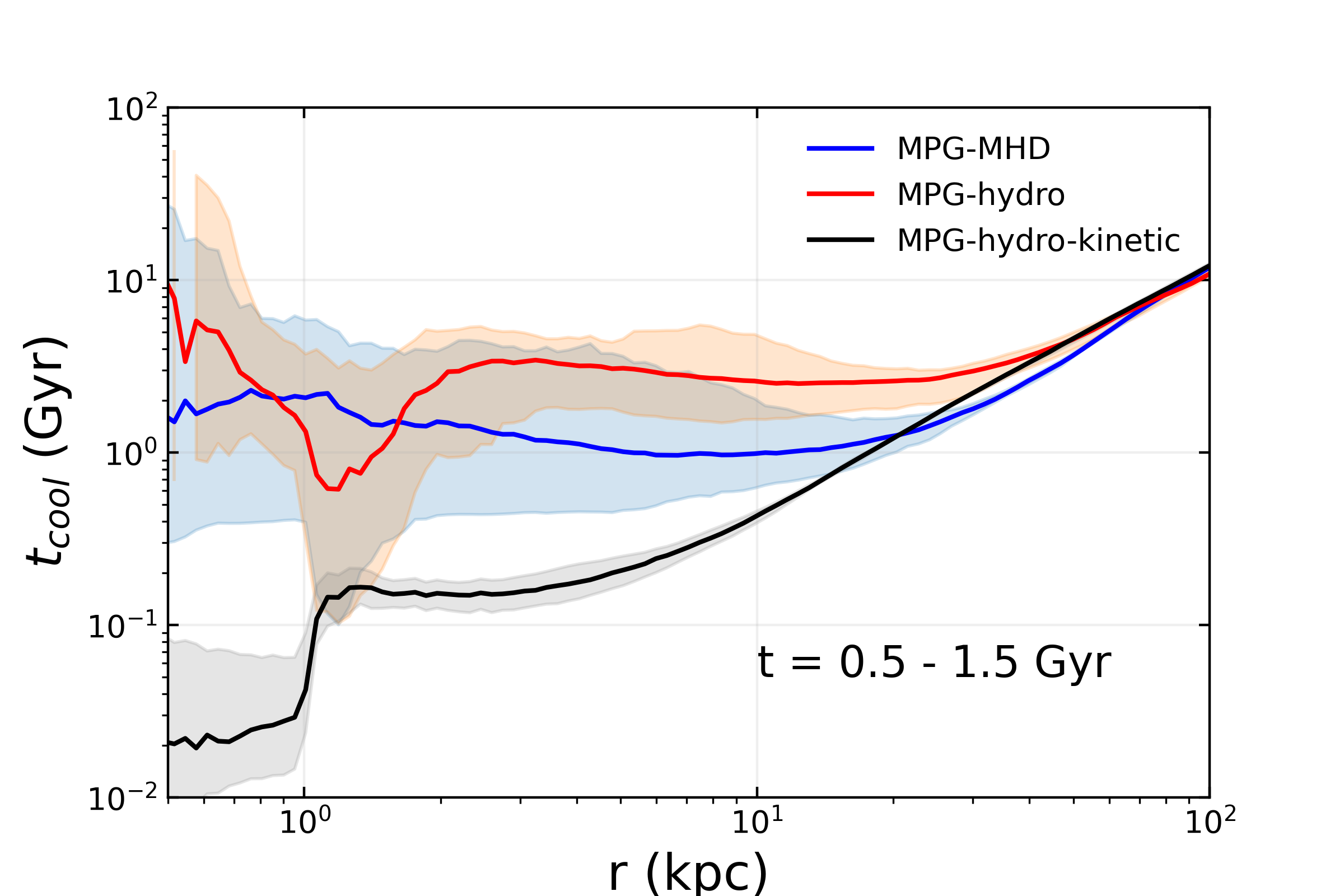}
 \includegraphics[width=0.47\textwidth]{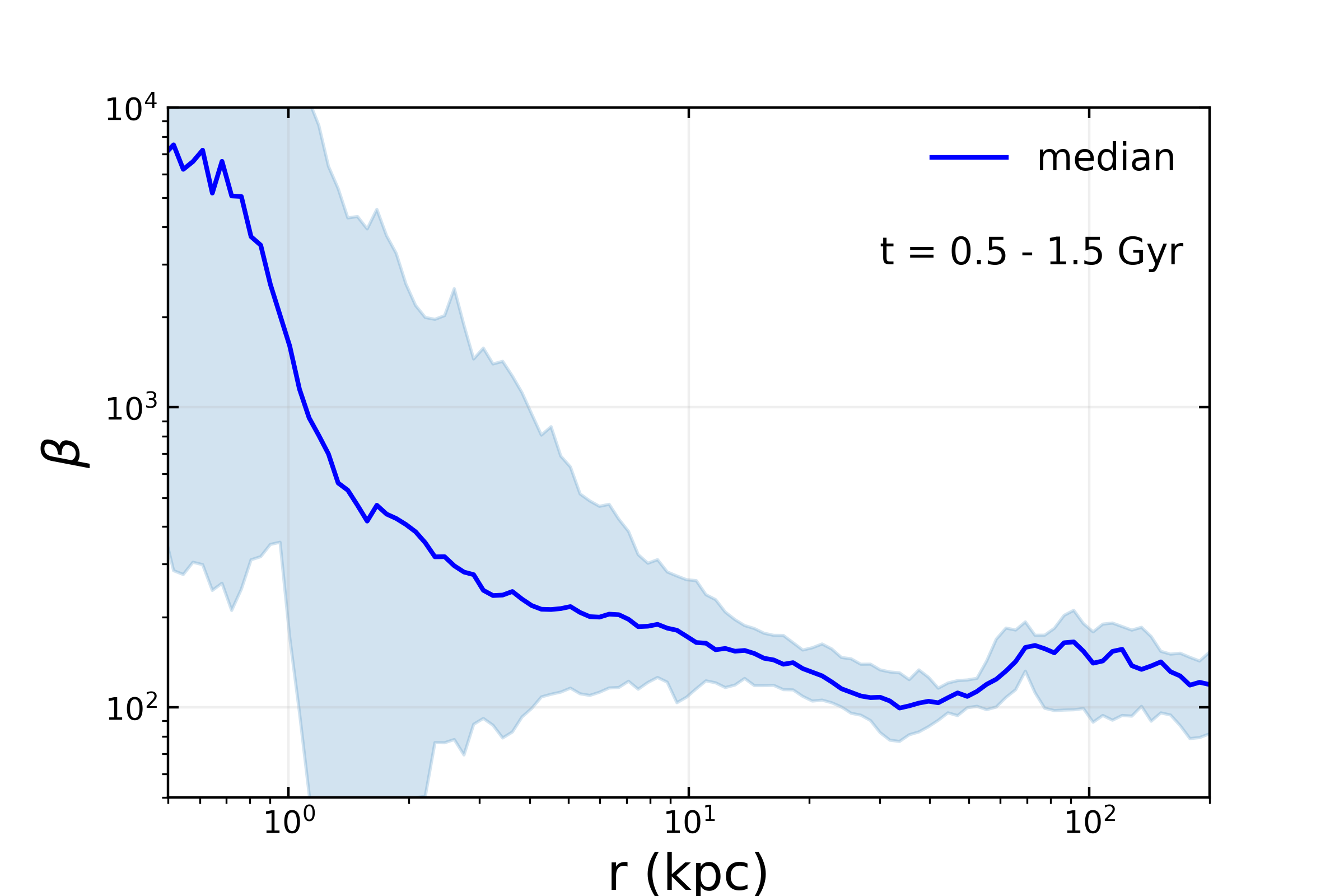}
 \includegraphics[width=3.3in,height=2.2in]{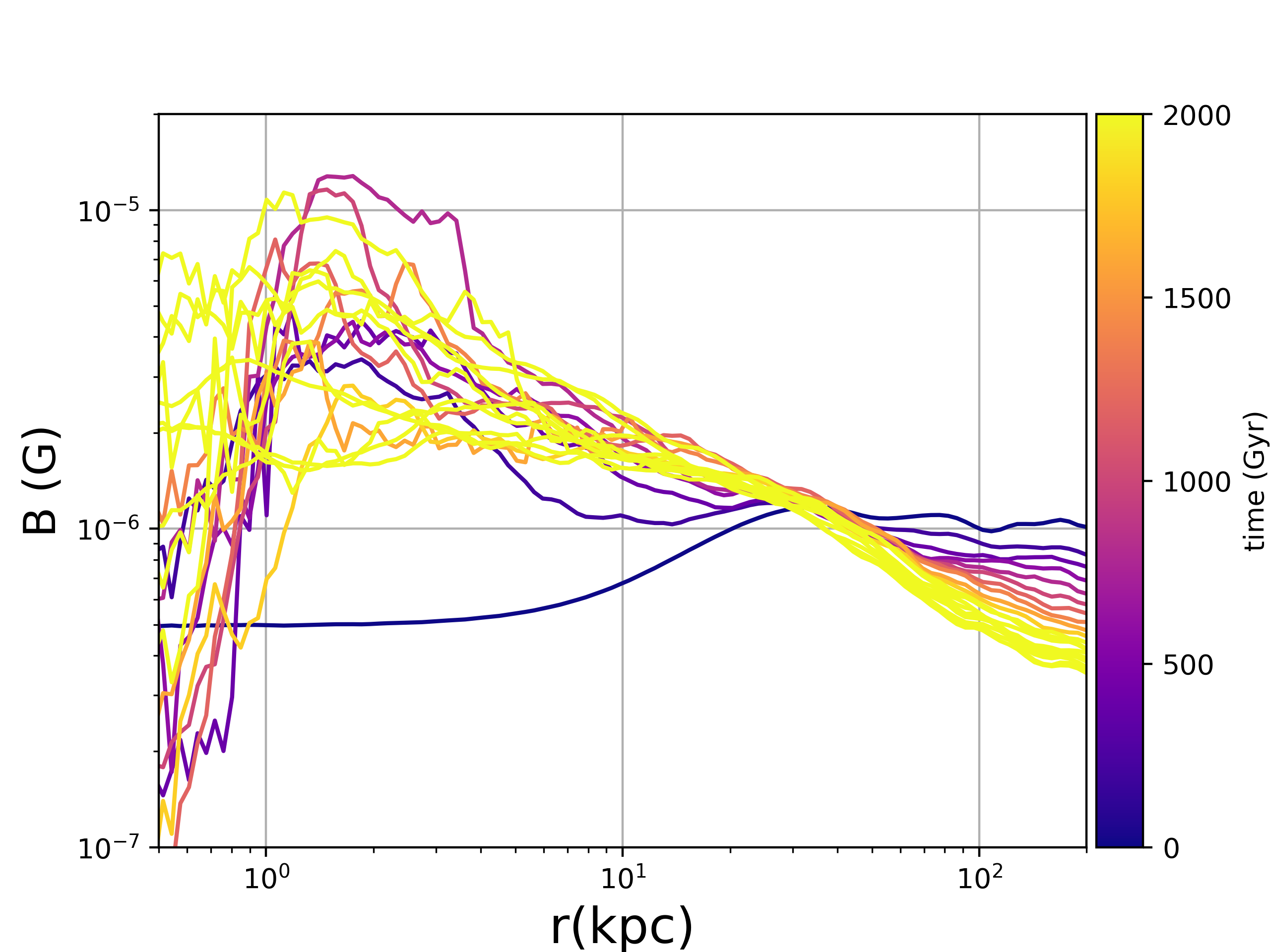}
  \caption{ {\it Top Left panel : } Median mass-weighted radial entropy profile for the  X-ray gas ($0.2$ keV $< T < 8$ keV) for MPG-MHD (blue line), MPG-hydro (red line) and MPG-hydro-kinetic (black line) runs. The faded cyan, red and grey region represent the 5th-95th percentile range of radial entropy profile at each radius between  
  $t=0.5-1.5$ Gyr for the MHD, hydro and hydro-kinetic MPG runs respectively.
{\it Top Right panel :} Median mass-weighted radial $t_{\rm cool}$ for the multiphase galaxy MPG-MHD (solid blue line), MPG-hydro (red line) and MPG-hydro-kinetic (black line) runs. {\it Bottom left panel :} Median plasma-$\beta$ ($=P/[B^2/2\mu_0]$) for the MPG-MHD run (solid blue line) with the shaded region showing the spread of plasma-$\beta$ between 5th to 95th percentile at each radii. {\it Bottom right panel :} Angle-averaged radial B-field at different times for the MPG-MHD run. The color of the lines show the time of the radial profile between t = 0-2 Gyr with a cadence of 100 Myr. 
}
 \label{fig:mpg_radial_prof}
\end{figure*}
Figure \ref{fig:mpg_radial_prof} shows the radial entropy, cooling time, magnetic field strength, and plasma $\beta$ for the multiphase galaxy (MPG). The top left panel shows the median of the radial entropy profiles of the X-ray gas ($0.2<T_{\rm keV}<8$) for the MPG-hydro-kinetic (black line), MPG-hydro (red line), and MPG-MHD (blue line) runs for $t=0.5-1.5$ Gyr. The shaded cyan, red, and grey regions show the entropy spread between 5th and 95th percentile at each radius for each simulation. For the MPG-hydro-kinetic run, the median entropy shows power law behaviour at all radii with some flattening ar $r<2$ kpc and very little spread around the median during the galaxy's evolution. The entropy profile remains within range of the observed entropy profiles of multiphase type galaxies discussed in \citet{voit2020}. On the other hand, both runs with partial thermal AGN feedback (MPG-hydro and MPG-MHD runs) show elevated entropy between 3 kpc~$\lesssim r \lesssim 15$ kpc.  The entropy profile for both of these runs shows a  
drop within $r< 3$ kpc signifying  conditions conducive to CGM cooling.
Comparing MPG-hydro and MPG-MHD runs, the plot shows that AGN activity in hydro runs causes a larger disruption within the central $r\lesssim 30$ kpc. AGN activity overheats the CGM in the hydro case, which is consistent with the shutdown of the AGN activity after $t\simeq1.5$ Gyr, as seen in Figure \ref{fig:mpg_pjet} (middle panel).  

The top right panel in Figure \ref{fig:mpg_radial_prof} shows the median of the radial $t_{\rm cool}$ profiles of all the gas for the MPG-MHD (blue line), MPG-hydro (red line) and MPG-hydro-kinetic (black line) with the shaded regions representing the spread in the $t_{\rm cool}$ profile between the 5th and 95th percentile at each radius for $t=0.5-1.5$ Gyr. Similar to the entropy profile, the pure kinetic AGN feedback run shows a very different $t_{\rm cool}$ behaviour compared to runs with partial thermal feedback. For runs with partial thermal feedback, this panel shows overheating of the CGM within $r\simeq 40$ kpc, similar to the entropy panel.

The bottom left panel shows the median of the radial plasma-$\beta$ ($\equiv n k_B T/[B^2/2\mu_0]$) 
profiles between $t=0.5-1.5$ Gyr for the MPG-MHD run. The plasma $\beta$ stays $\gtrsim100$ for $r\gtrsim10$ kpc and rises steeply to $\sim 10^4$ within $r<10$ kpc, signifying that the CGM is thermally dominated at all radii. The bottom right panel shows the radial magnetic field strength at different times for the same run.
As the simulation evolves the magnetic field strength declines at larger radii ($r>30$ kpc) and rises to a saturation stage within the central $\simeq 30$~kpc. This panel also shows that within $r \simeq 1$ kpc, the magnetic field strength again drops sharply. This might be because the strong AGN feedback evacuates gas within $r \simeq 1$ kpc, dragging the magnetic field lines along to larger radii due to flux freezing in the outflowing plasma. 

\subsubsection{SNIa heating and Radiative Cooling}
 \begin{figure*}
\centering
 \includegraphics[width=0.3\textwidth]{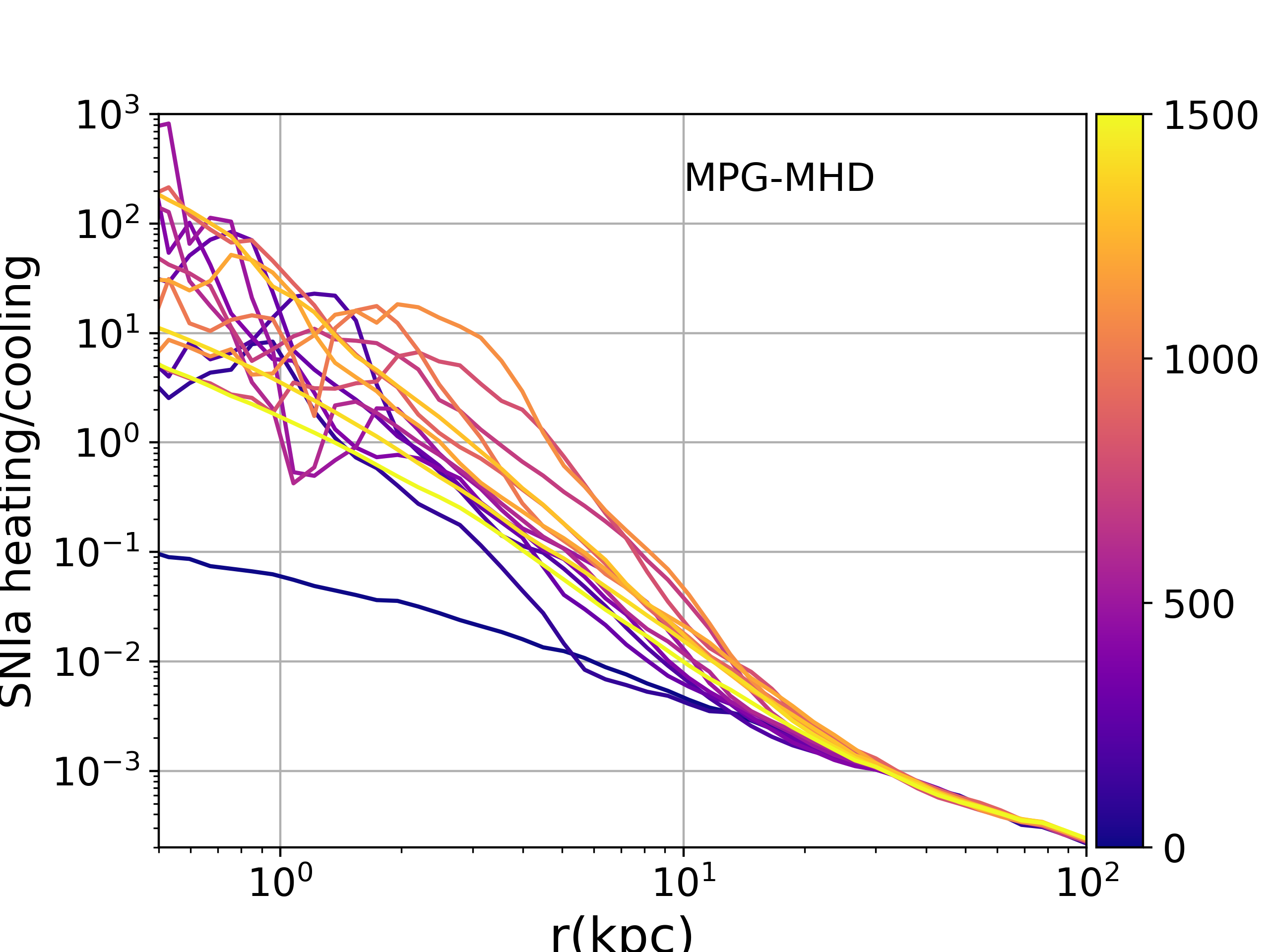}
 \includegraphics[width=0.3\textwidth]{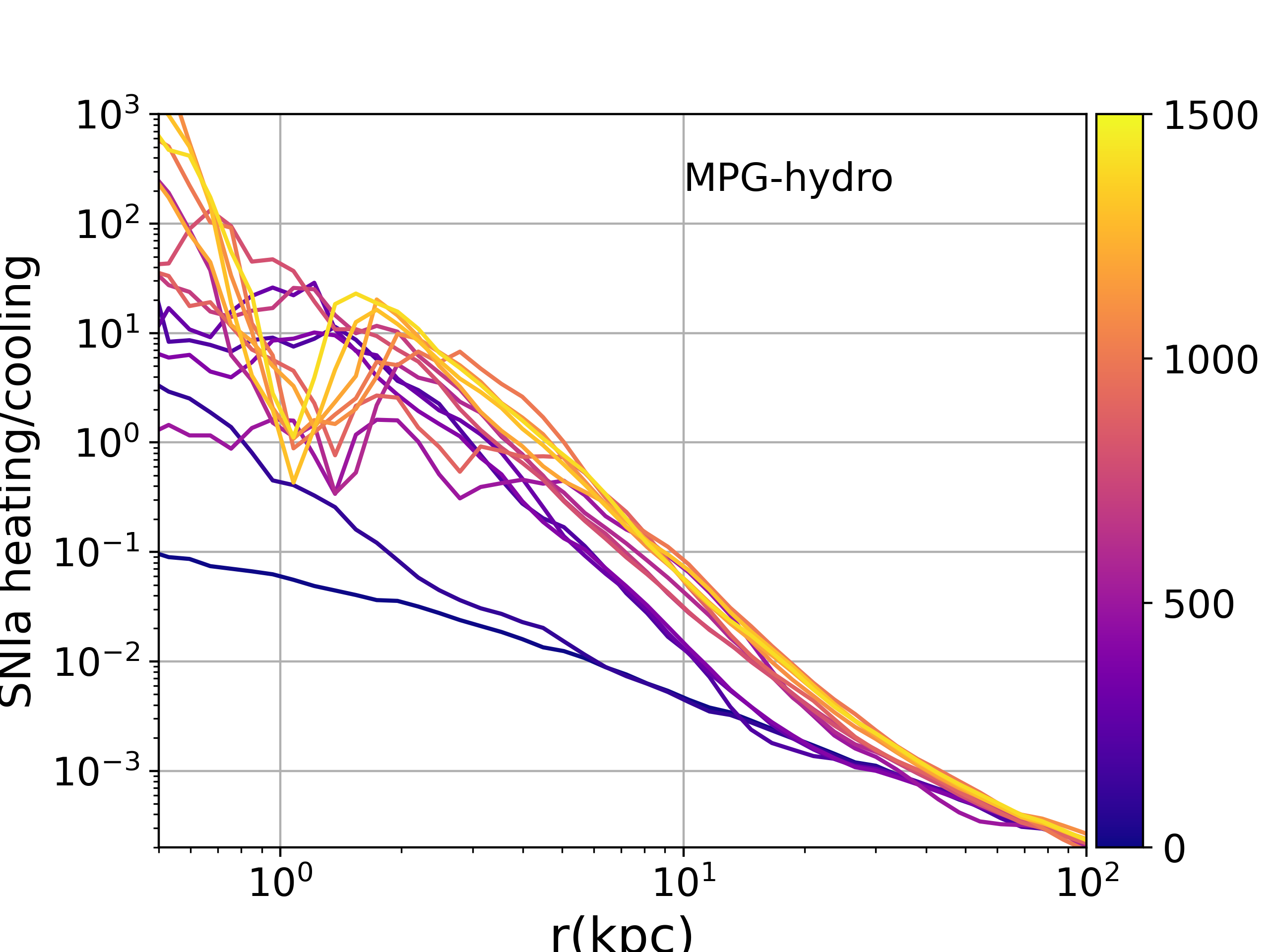}
 \includegraphics[width=0.3\textwidth]{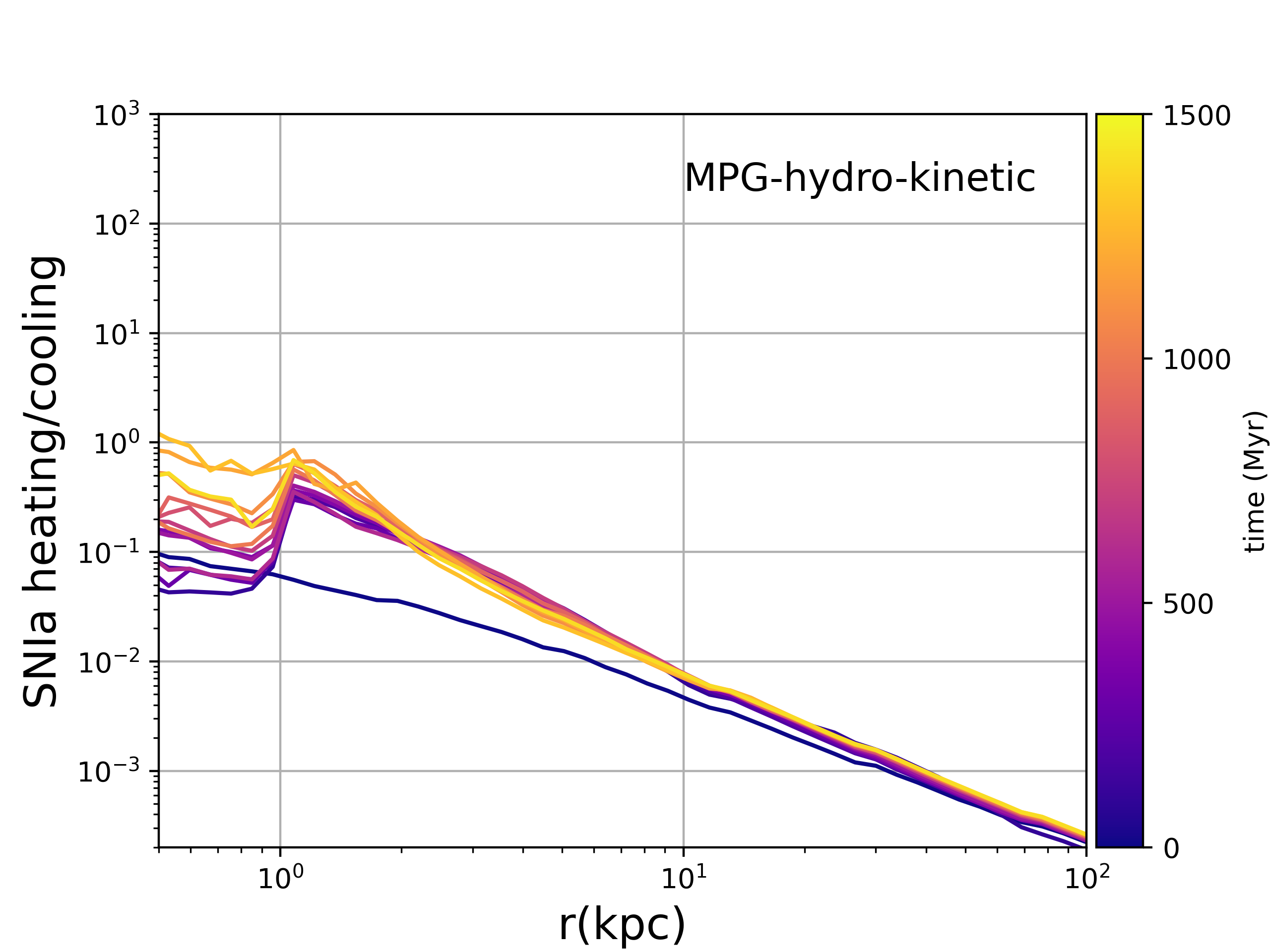}
 \caption{ SNIa heating to radiative cooling ratio for the MPG-MHD (left panel), MPG-hydro (middle panel) and MPG-hydro-kinetic (right panel) runs between t = 0 - 1.5 Gyr with a cadence of 100 Myr. The color of the lines represents the time of the galaxy evolution. For the MPG-hydro and MPG-MHD runs, SNIa heating remains dominant over cooling within the central $r\lesssim5$ kpc while for MPG-hydro-kinetic run cooling dominates over heating. 
}
 \label{fig:mpg_heatcool}
\end{figure*}

Figure \ref{fig:mpg_heatcool} shows the evolution of the ratio of SNIa heating to radiative cooling for all MPG runs. As described in Section \ref{sec:stellar}, the Type Ia supernova heating is modeled assuming the stellar density given by the {\it Hernquist} profile. To start with, optically thin radiative cooling dominates over SNIa heating at all radii for the multiphase galaxy. For the pure kinetic MPG run (right panel), cooling remains dominant over SNIa heating throughout the simulation. However, for the MPG-hydro (middle panel) and MPG-MHD (left panel) runs, SNIa heating dominates over cooling within the central $r<1$ kpc after the simulation starts because thermal AGN feedback overheats the gas close to the SMBH. Between 1 kpc~$< r < 5$~kpc cooling becomes dominant over SNIa heating to larger radii as time goes on,  
while beyond $r>5$ kpc cooling remains dominant over SNIa heating as the stellar density (and thus volumetric heating rate) declines with increasing radius.   

\subsection{Single Phase Galaxy}
\label{sec:hyd_spg}
To explore how magnetic fields affect the baryon cycle in single phase massive galaxies, we simulated the evolution of single phase galaxy (SPG) with the same physics as in the prior simulations.  
Qualitatively, both SPG-hydro and SPG-MHD runs show very similar tight coupling between AGN feedback and radiative cooling in the CGM, with centrally concentrated cooling ($r<2$ kpc). However, the SPG-MHD run shows temporally shorter AGN cycles compared to the SPG-hydro run, though both have comparable peak AGN power. In the following subsections, we look at the evolution of different quantities for both runs. 
 
\subsubsection{Temporal evolution}
 \begin{figure*}
\centering
 \includegraphics[width=0.47\textwidth]{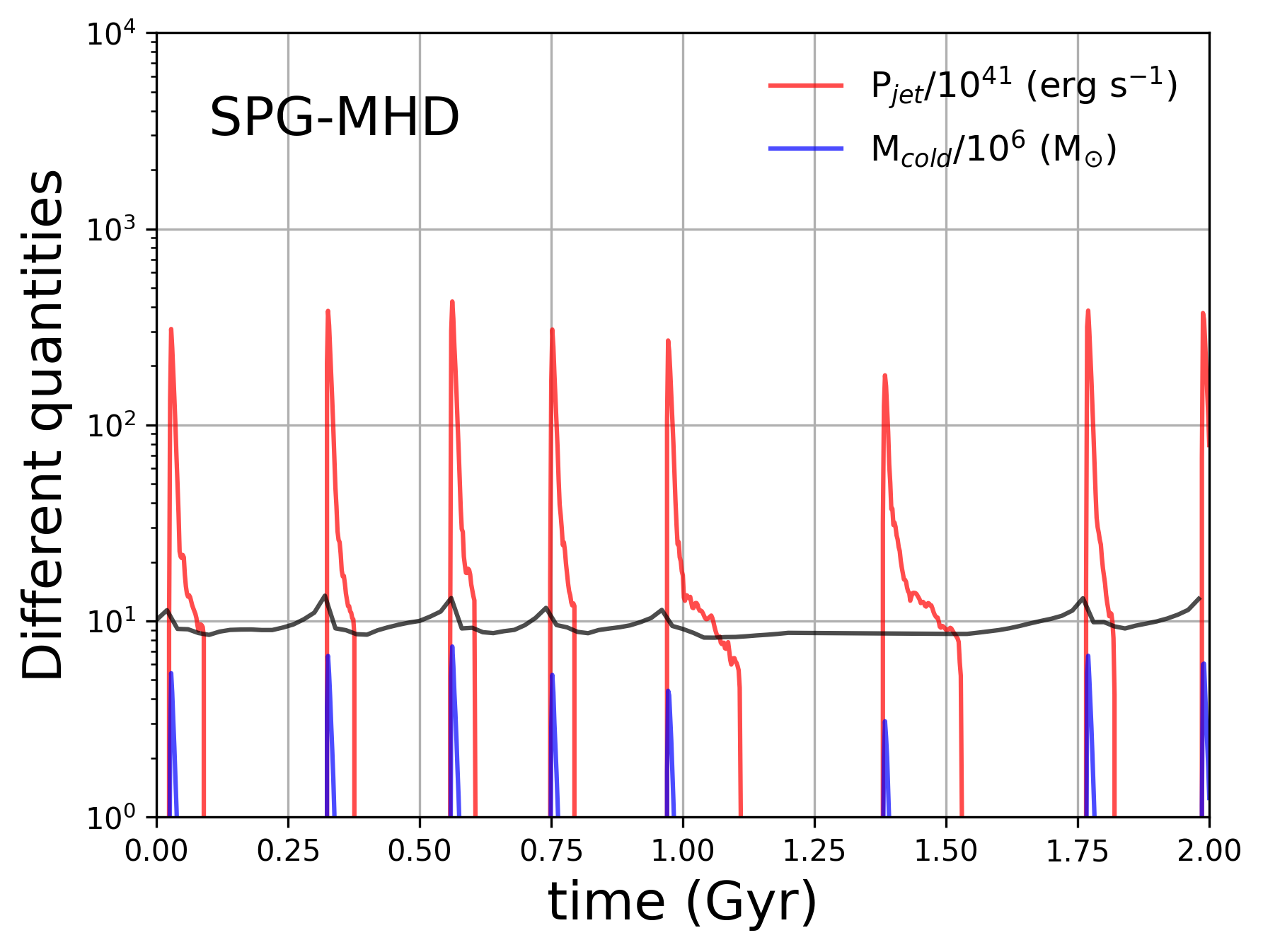}
 \includegraphics[width=0.47\textwidth]{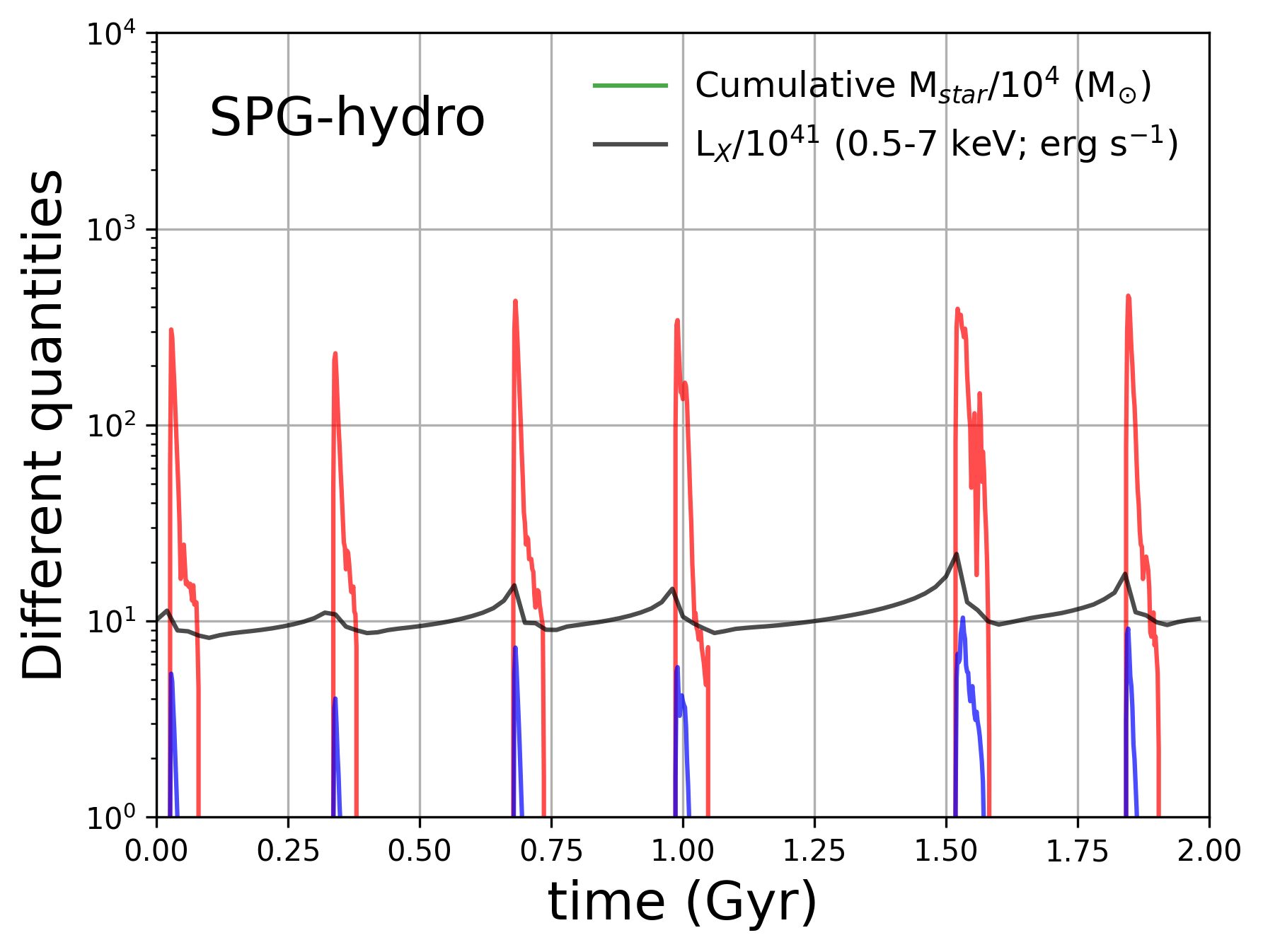} 
 \caption{ Jet power ($P_{\rm jet}$; red line) and cold gas mass (blue line), total stellar mass (M$_\star$; green line) and X-ray luminosity for the $0.5-7$ keV within the central $r = 50$ kpc (black line) with time for the SPG-MHD (left panel) and the SPG-hydro run (right panel). While both runs show similar AGN behaviour with ($<P_{\rm jet}>\sim1.5\times 10^{42}$ erg s$^{-1}$), the number of AGN cycles is higher for the SPG-MHD run compared to SPG-hydro run.  
 }
 \label{fig:fid_pjet}
\end{figure*}
Figure \ref{fig:fid_pjet} shows the evolution of the total cold gas mass ($T<10^5$ K), jet power (P$_{\rm jet}$), total stellar mass, and X-ray luminosity ($L_X$) within $r = 50$ kpc for the $0.5-7$ keV gas for the SPG-MHD (left panel) and SPG-hydro (right panel) runs. These plots show a tight correlation between jet power ($P_{\rm jet}$) and X-ray luminosity ($L_X$; 0.5 keV$< T < 7$keV),as every AGN outburst is preceded by a rise in $L_X$. For both runs, cold gas mass remains $\lesssim 10^7\ {\rm M}_{\odot}$ resulting in AGN activity with peak power (${\rm P_{jet}}\sim 10^{43}$ erg s$^{-1}$). The jet duty cycle is typically $\sim 100$ Myr for both the runs with no star formation during their evolution until at least $t=2$ Gyr. The cold gas formation remains centrally concentrated within $r = 1$ kpc with no extended cold gas filament formation (see Figure \ref{fig:rad_ext}). As such, the heating-cooling cycle remains very tightly coupled unlike the MPG runs, where extended cold gas formation leads to much longer jet events with an order of magnitude higher jet power. 

\subsubsection{Radial profiles}
\begin{figure*}
\centering
 \includegraphics[width=0.47\textwidth]{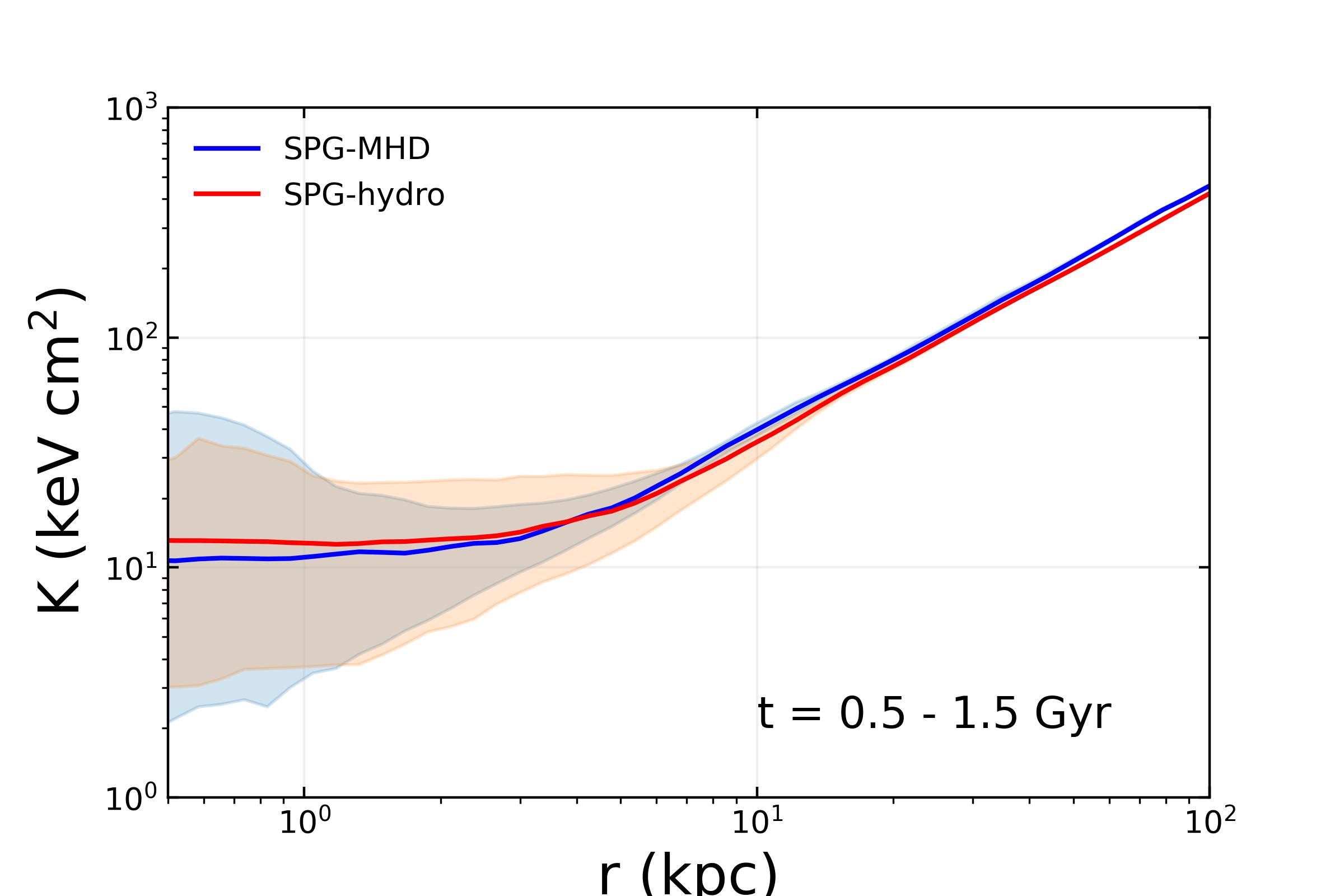}
 \includegraphics[width=0.47\textwidth]{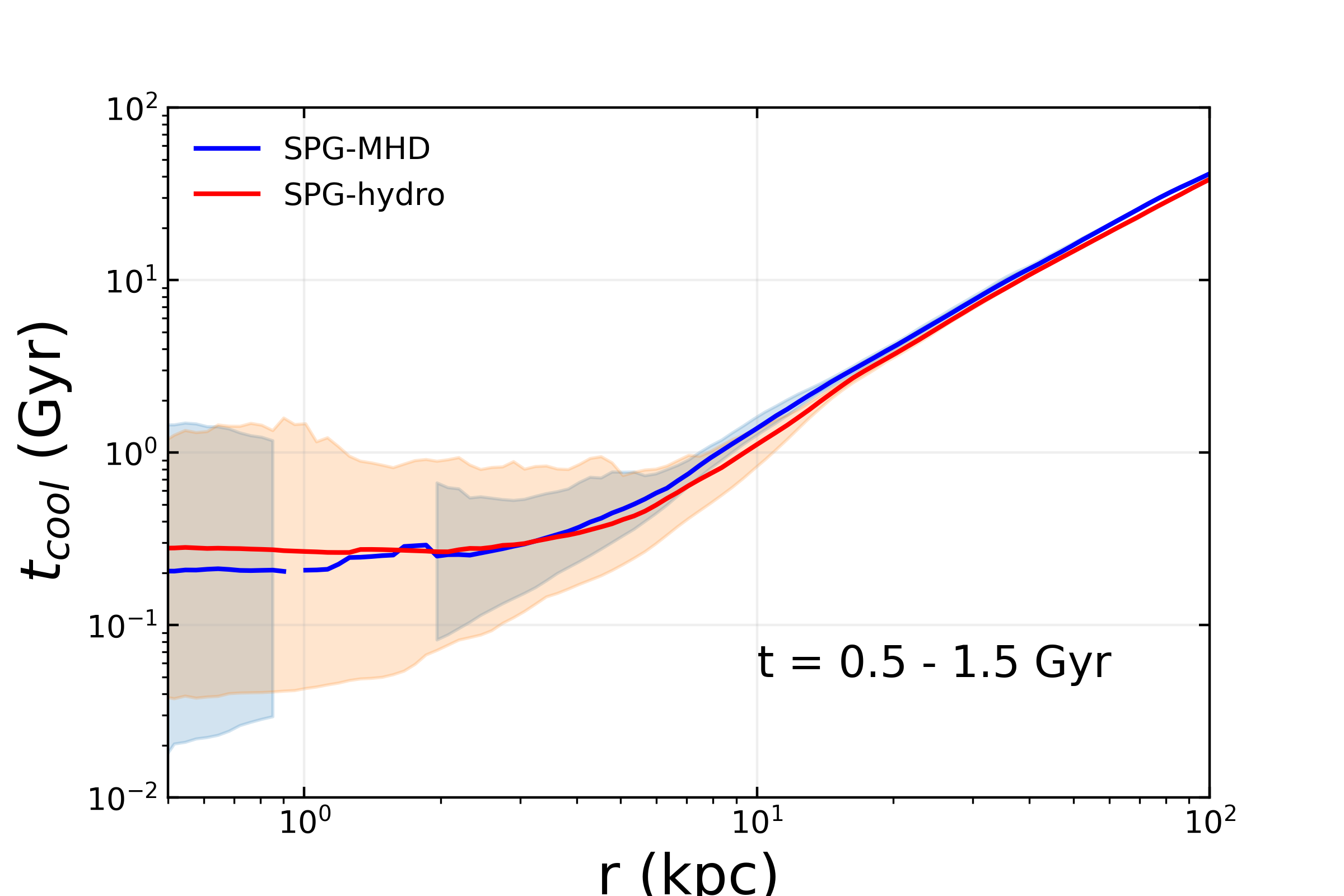}
 \includegraphics[width=0.47\textwidth]{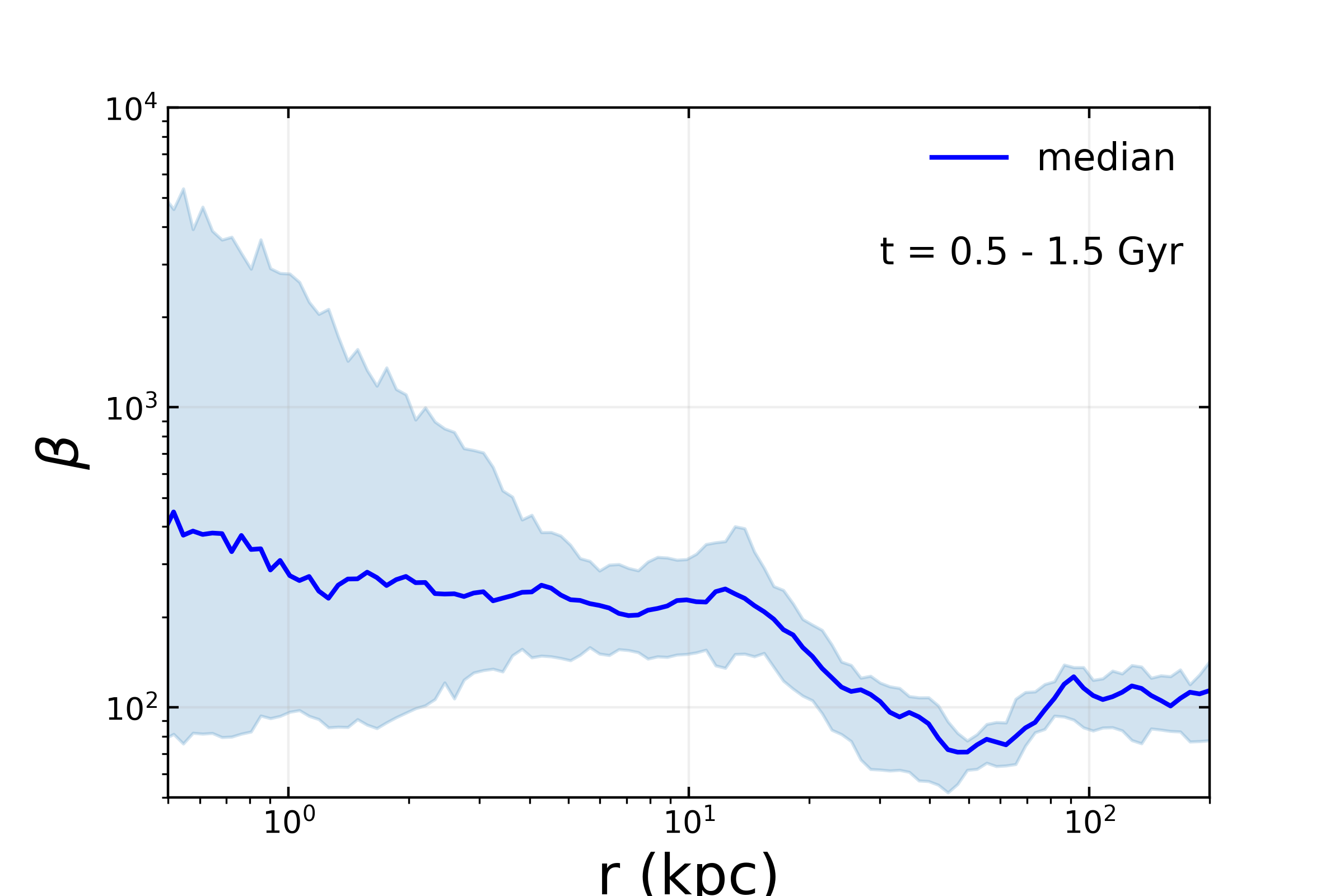}
 \includegraphics[width=3.3in,height=2.2in]{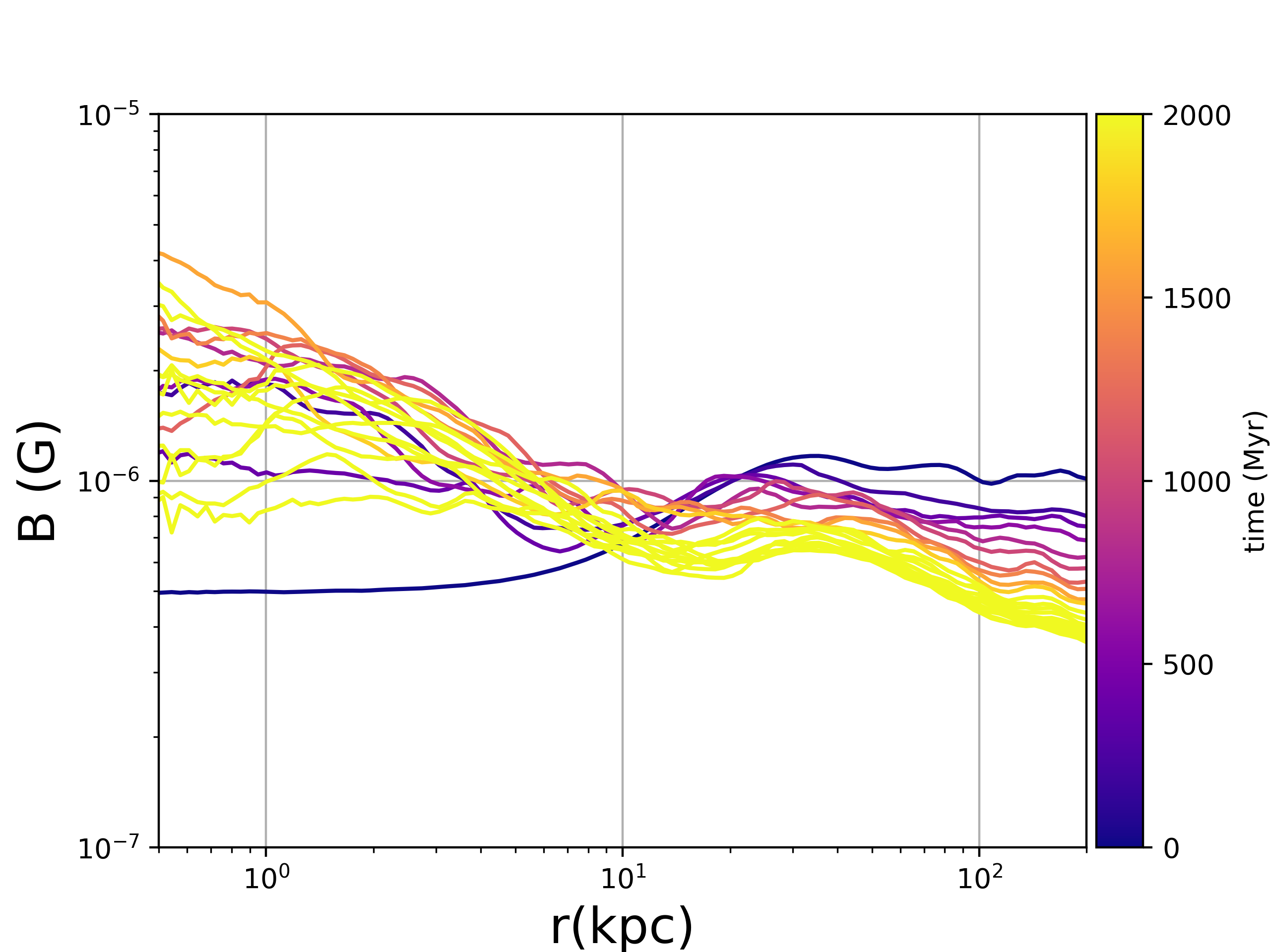}
 \caption{ {\it Left panel :} Median mass-weighted radial entropy profile for the  X- ray gas ($0.2$ keV$< T < 8$ keV) for SPG-MHD run (solid blue line) and SPG-hydro run (dashed red line). The faded cyan, red and grey region represent the 5th-95th percentile range of radial entropy profile at each radius between t=0.5-1.5 Gyr for the SPG-MHD and SPG-hydro runs respectively.
{\it Right panel :} Median mass-weighted radial $t_{\rm cool}$ for the SPG-MHD run (solid blue line) and the SPG-hydro run (dashed red line) for all the gas.
{\it Bottom left panel :} Median plasma-$\beta$ ($=P/[B^2/2\mu_0]$) for the SPG-MHD run (solid blue line) with the shaded region showing the spread of plasma-$\beta$ between 5th to 95th percentile at each radius between $t=0.5-1.5$ Gyr. {\it Bottom Right panel :} Angle-averaged B-field with time for the SPG-MHD run. The color shows the time of the B-field profile between $t = 0-2$ Gyr with a cadence of 100 Myr.}
 \label{fig:spg_radial_prof}
\end{figure*} 

Figure \ref{fig:spg_radial_prof} shows the mass-weighted radial entropy, cooling time ($t_{\rm cool} $), plasma-$\beta$ 
and magnetic fields for the single phase galaxy (SPG). The upper left panel shows the median of the radial entropy profile for the X-ray gas ($0.2<T_{\rm keV}<8$) for the SPG-MHD (blue line) and SPG-hydro (red line) runs.  
Shaded cyan and red regions represent the 5th-95th percentile range of entropy at each radius between t=0.5-1.5 Gyr for the SPG-MHD and SPG-hydro runs, respectively. The panel shows that the spread in entropy is confined within the central 5 kpc, signifying the centrally concentrated cooling-heating cycle for the SPG runs. The presence of magnetic fields does not cause any significant change to this tight interplay between CGM cooling and AGN heating. Similar behaviour is seen in the $t_{\rm cool}$ profile (includes all gas) in the upper right panel for both runs.  

The lower left panel shows the median plasma-$\beta$ profile between $t=0.5-1.5$ Gyr for the SPG-MHD run, with the cyan shaded region showing the 5th-95th percentile range of the spread in plasma $\beta$ at each radius. 
This panel shows that thermal pressure dominates over the magnetic pressure throughout the domain, with the central $r\sim 10$ kpc having even higher plasma-$\beta$ values than the group outskirts.  
The concentrated nature of the cooling-heating cycle leads to a bigger spread in the plasma $\beta$ only within the central $r\sim 3$ kpc.

The lower right panel shows the radial magnetic profile between $0-2$ Gyr with a 100 Myr cadence. Similar to the MPG-MHD run, the magnetic field strength rises within the central $r\sim10$ kpc to a saturation level as the galaxy evolves, while it declines at larger radii (r $\gtrsim 10$ kpc) with time.

\subsubsection{SNIa heating and Radiative Cooling}
\label{sec:heat-cool}
\begin{figure*}
\centering
 \includegraphics[width=0.45\textwidth]{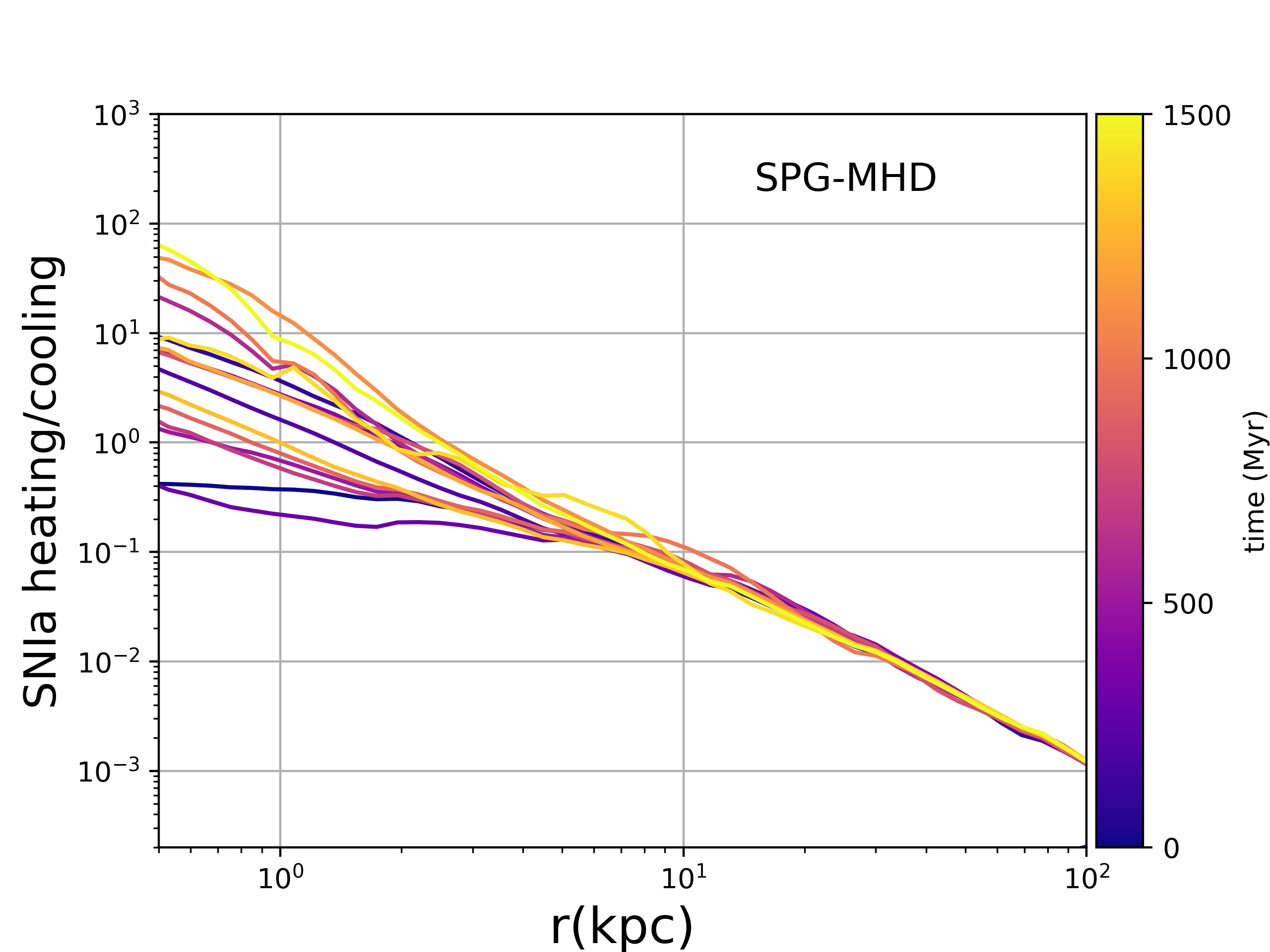}
 \includegraphics[width=0.45\textwidth]{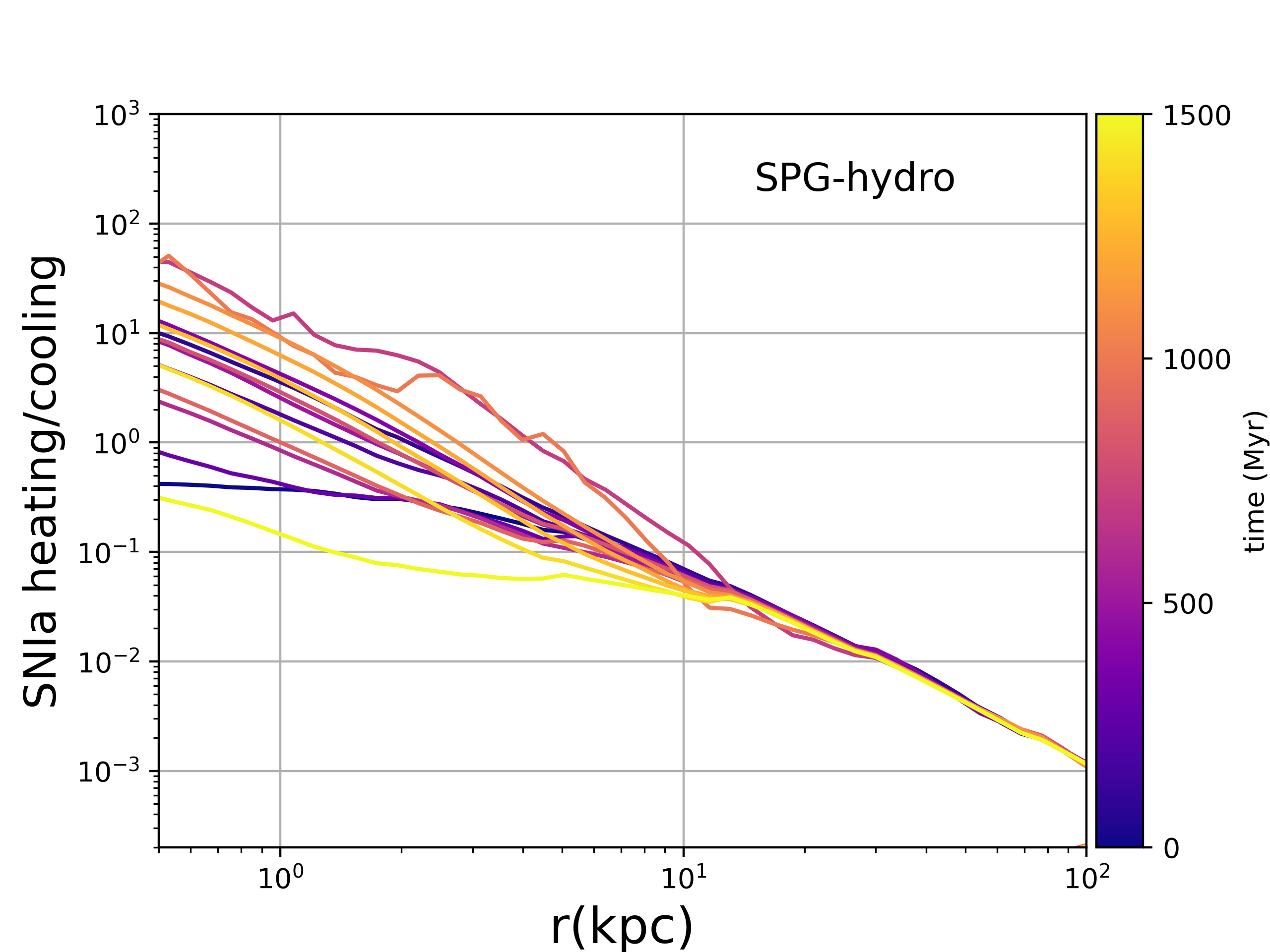} 
 \caption{Radial SNIa heating to radiative cooling ratio for the SPG-MHD (left panel) and SPG-hydro (right panel) runs. The lines on the plots are with a cadence of 100 Myr from $t=0-1.5$ Gyr. The color of the lines represent the time of the galaxy evolution. For both the runs cooling dominates over heating within $r<2$ kpc for short duration fueling AGN cycle.  
}
\label{fig:spg_heatcool}
\end{figure*}

Figure \ref{fig:spg_heatcool} shows the evolution of the ratio of SNIa heating to radiative cooling of the CGM for the SPG-MHD (left panel) and SPG-hydro (right panel) runs. At first,  cooling dominates over SNIa heating at all radii. For most of the simulation time, however, the SNIa heating dominates over radiative cooling within the central $r \sim 5$ kpc, with the ratio declining below one only in short phases coincident with cold gas formation as seen in Figure \ref{fig:fid_pjet}. This is because the thermal AGN feedback overheats the gas within central ($r \sim 2$) kpc, pushing the cooling rate lower and the SNIa heating to cooling ratio greater than 1. Similar behavior in the SNIa heating to cooling ratio is seen for both the SPG-MHD and SPG-hydro runs.

\subsection{SPG with Cooler Core}
\label{sec:spg_cool}
In the final set of numerical experiments, we evolved SPG-cool-core halos to explore the evolution of massive galaxies with a deep central gravitational potential ($\sigma_v > 240$ km s$^{-1}$) and a higher CGM pressure that is similar to a massive galaxy with a multiphase CGM. The main aim of this set of experiments is to explore whether it is possible to create a single phase CGM with lower density and longer cooling times than an MPG-type CGM solely by feedback-driven reconfiguration of the circumgalactic medium (i.e., without requiring the injection of significant amounts of energy by an external source such as a galaxy merger).  For these experiments, we evolved the SPG-Cool systems with radiative cooling, star formation, and stellar feedback (including SNIa feedback), kinetic+thermal AGN feedback (hydro run), and kinetic+thermal+magnetic AGN feedback for MHD runs. Although both the SPG-Cool-hydro and SPG-Cool-MHD runs show formation of extended cold gas, they show a significantly different evolution during the initial $t\lesssim1$ Gyr. In the following subsections we look at the evolution of different quantities for both the SPG-Cool-MHD and SPG-Cool-hydro runs. 

\subsubsection{Temporal evolution}
\begin{figure*}
\centering
 \includegraphics[width=0.47\textwidth]{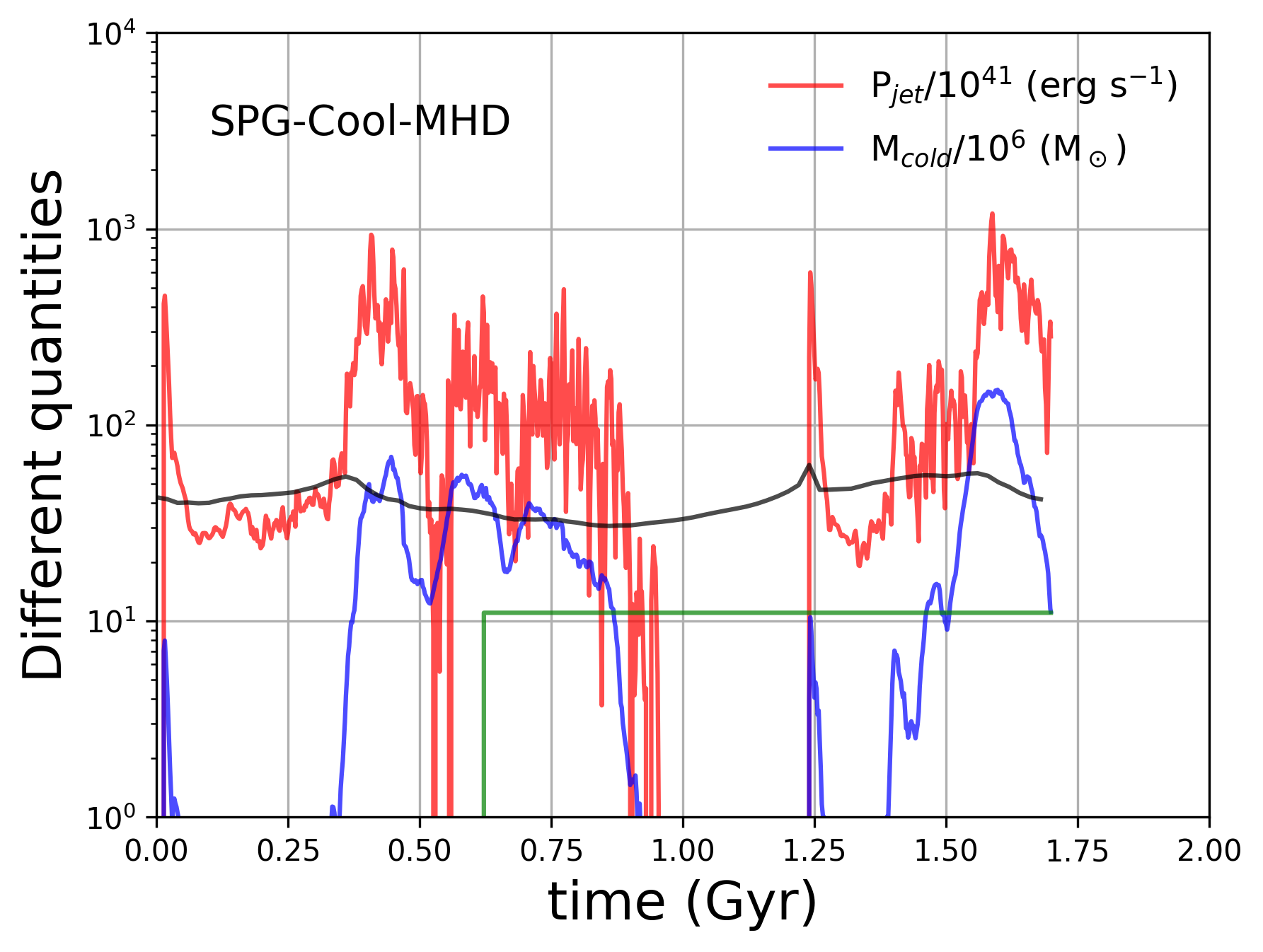}
 \includegraphics[width=0.47\textwidth]{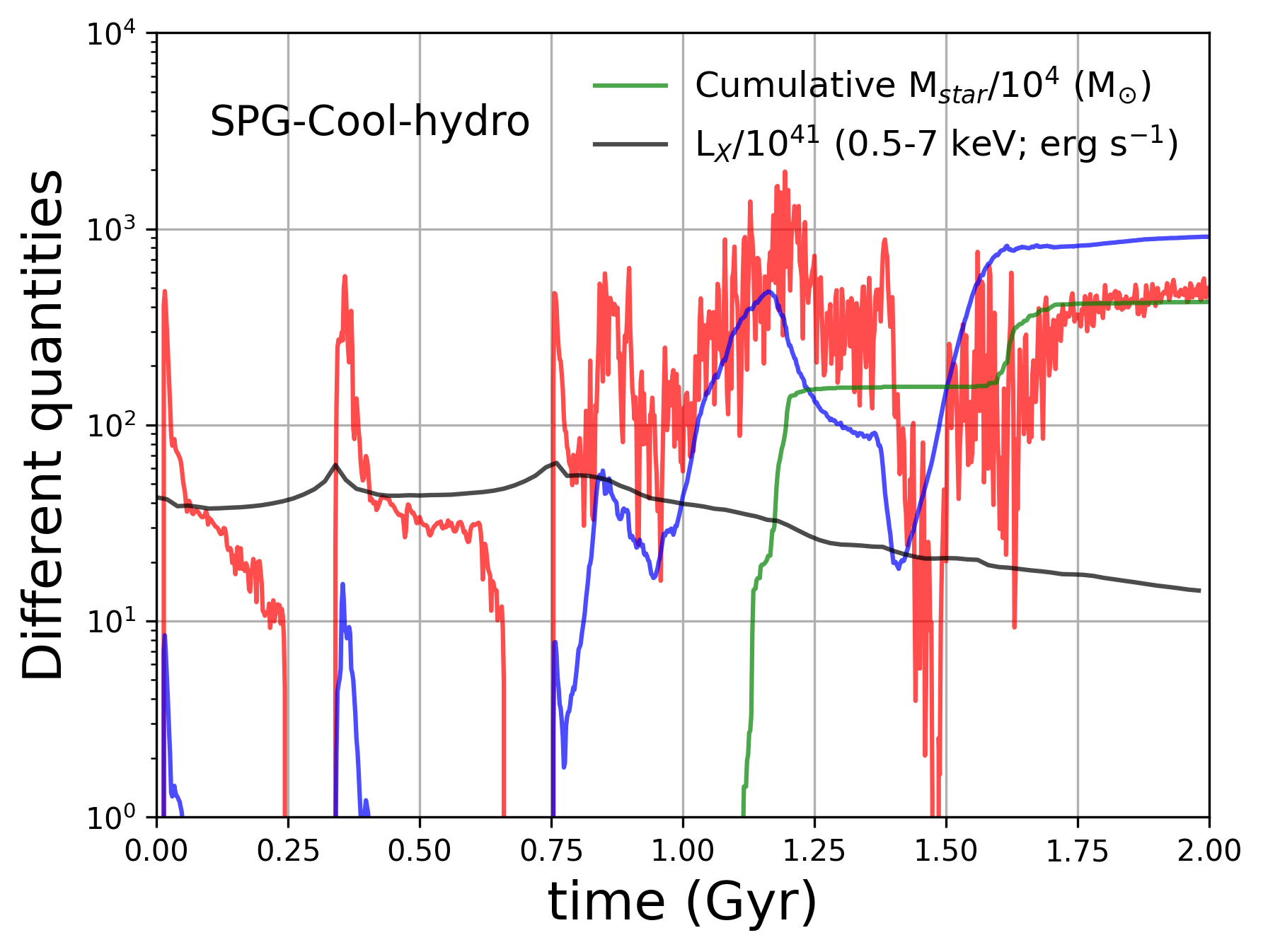} 
 \caption{  Jet power ($P_{\rm jet}$; red line) and cold gas mass (M$_{\rm cold}$;blue line), total stellar mass (M$_\star$; green line) and X-ray luminosity for the $0.5-7$ keV gas within central $r\lesssim50$ kpc (black line) with time for the SPG-Cool-MHD (left panel) and SPG-Cool-hydro (right panel) runs. While both SPG-Cool-MHD and SPG-Cool-hydro run show similar average jet power ($<P_{\rm jet}>\sim 2\times10^{43}$ erg s$^{-1}$), the strong AGN activity after $t>1$ Gyr leads to over heating/evacuation of the CGM within $r<50$ kpc for the hydro run resulting to steady decline in X-ray luminosity. 
}
 \label{fig:spg_cool_time}
\end{figure*}
Figure \ref{fig:spg_cool_time} shows the time evolution of the cold gas mass (blue lines), jet power ($P_{\rm jet}$; red lines), cumulative stellar mass (green lines), and X-ray luminosity ($0.5$ keV$<T<7$~keV) 
within the central $r = 50$ kpc for the SPG-Cool-MHD (left panel) and SPG-Cool-hydro (right panel) runs. These runs show a significant deviation in their evolution compared to the standard SPG runs. There is also a significant difference in the initial evolution ($t\lesssim1$ Gyr) between the SPG-Cool-MHD and SPG-Cool-hydro runs. SPG-Cool systems show an order of magnitude more cold gas compared to SPG runs for both the MHD and hydro cases. Cold gas also persists in the domain for several hundred Myr, unlike SPG runs where cold gas gets depleted within $t\sim50$ Myr. SPG-Cool-hydro run shows an evolution similar to single phase galaxies with jet activity very tightly correlated with X-ray luminosity for the initial $t\sim0.75$ Gyr. During this phase, centrally concentrated cooling fuels the AGN activity as in the SPG case. However, AGN feedback cannot stop extended cooling in the CGM because the density and pressure is much higher compared to the standard SPG systems, and thus the local cooling timescales are shorter. The late stage evolution ($t>0.75$ Gyr) of the SPG-Cool-hydro run is similar to that of the equivalent multiphase galaxy calculation (MPG-Hydro), with formation of extended cold gas and stars and $P_{\rm jet}$ exceeding $10^{44}$ erg~s$^{-1}$. The powerful jets overheat the CGM as the X-ray luminosity declines rapidly after $0.75$ Gyr, similar to the MPG-hydro run. On the other hand, the SPG-Cool-MHD run shows extended cold gas formation by $t\sim0.3$ Gyr with cold gas mass, $M_{\rm cold}\sim5\times10^7$ M$_\odot$, persisting for several hundred Myrs. The cold gas mass (M$_{\rm cold}$) peaks at $10^8$ M$_\odot$ at $t\sim1.55$ Gyr with a corresponding peak $P_{\rm jet}\sim10^{44}$ erg~s$^{-1}$. Unlike the hydro run, the X-ray luminosity ($L_X$) does not show any large decline and rises and falls with the feedback cycle.

\subsubsection{Radial profiles}

\begin{figure*}
\centering
 \includegraphics[width=0.47\textwidth]{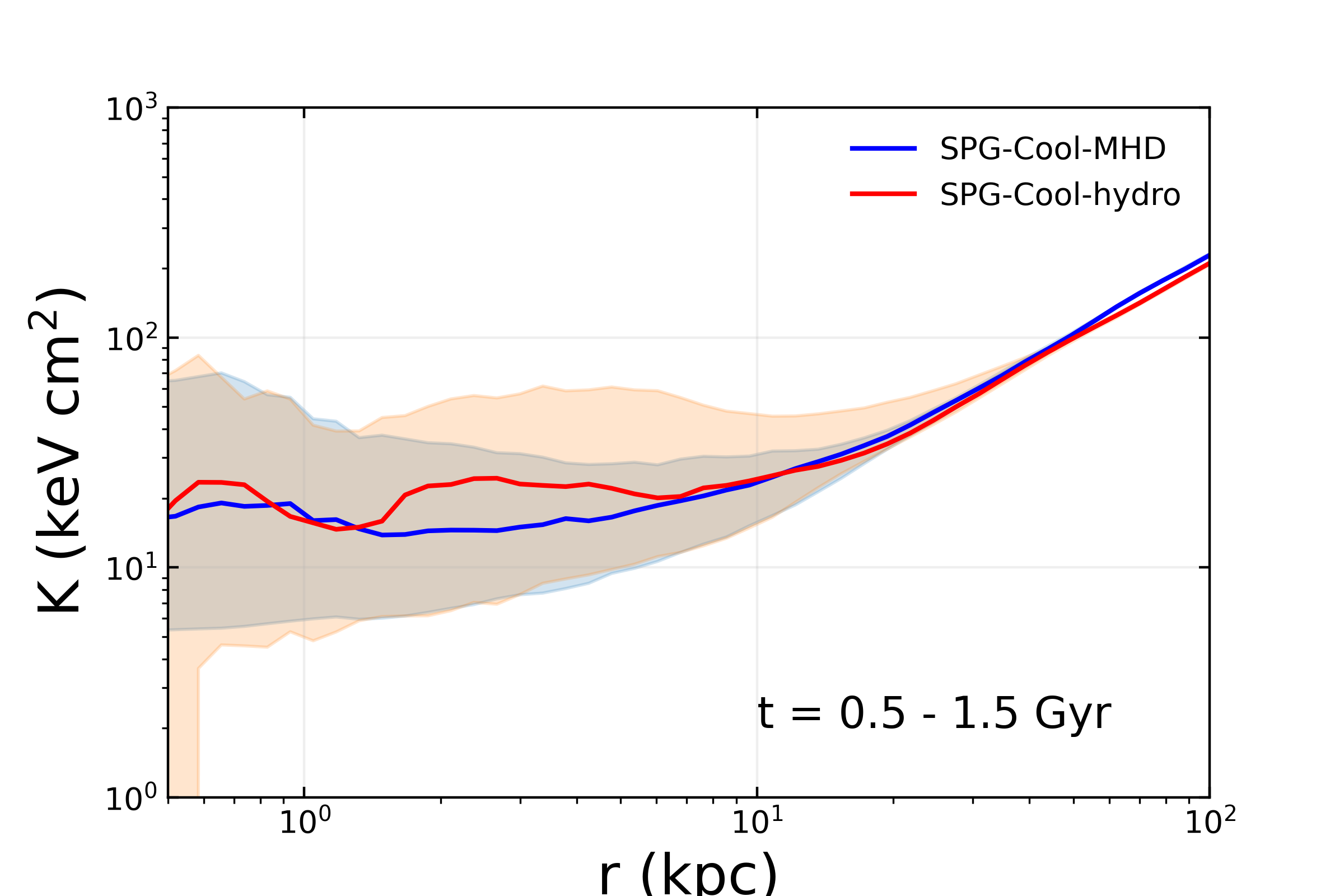}
 \includegraphics[width=0.47\textwidth]{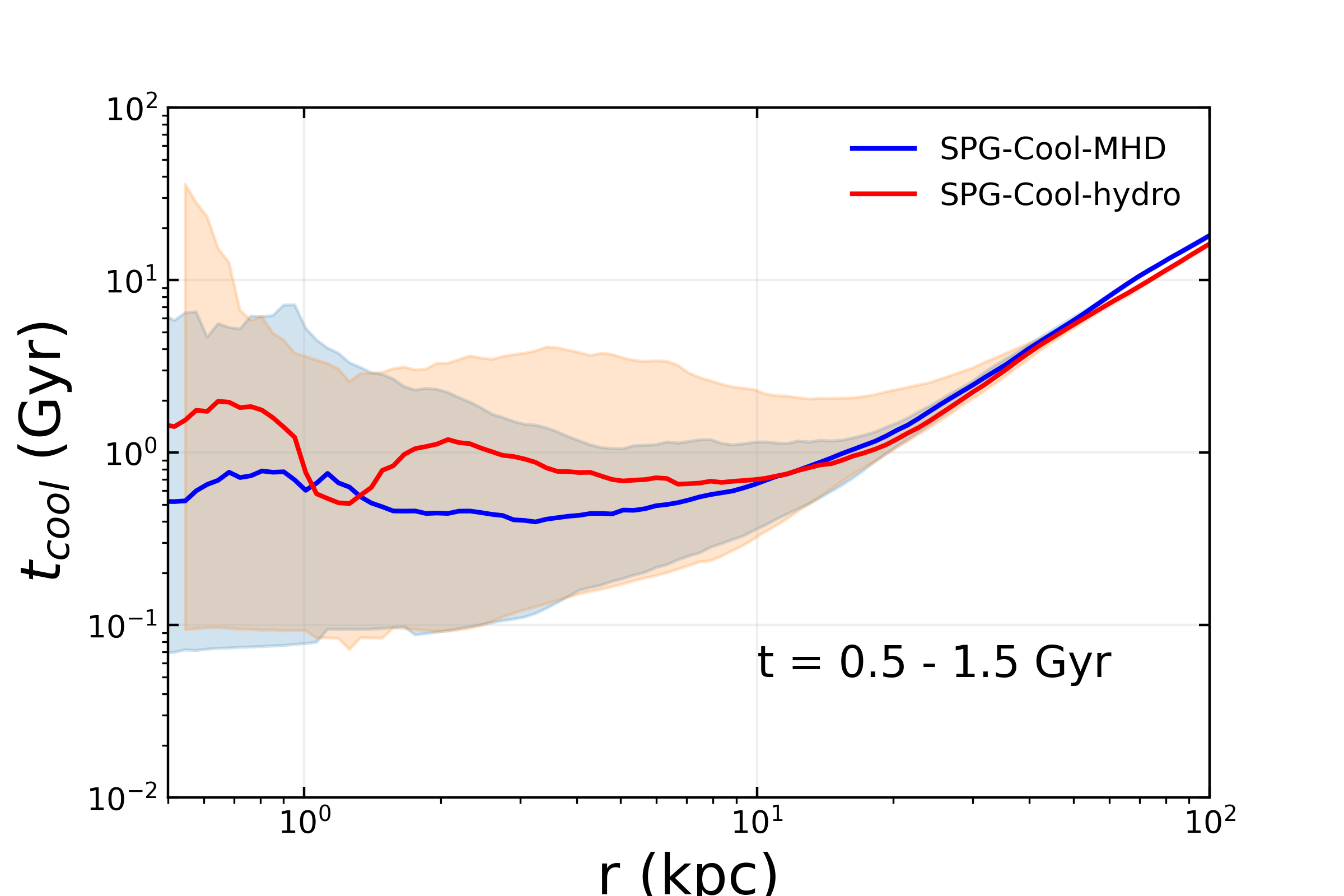}
 \includegraphics[width=0.47\textwidth]{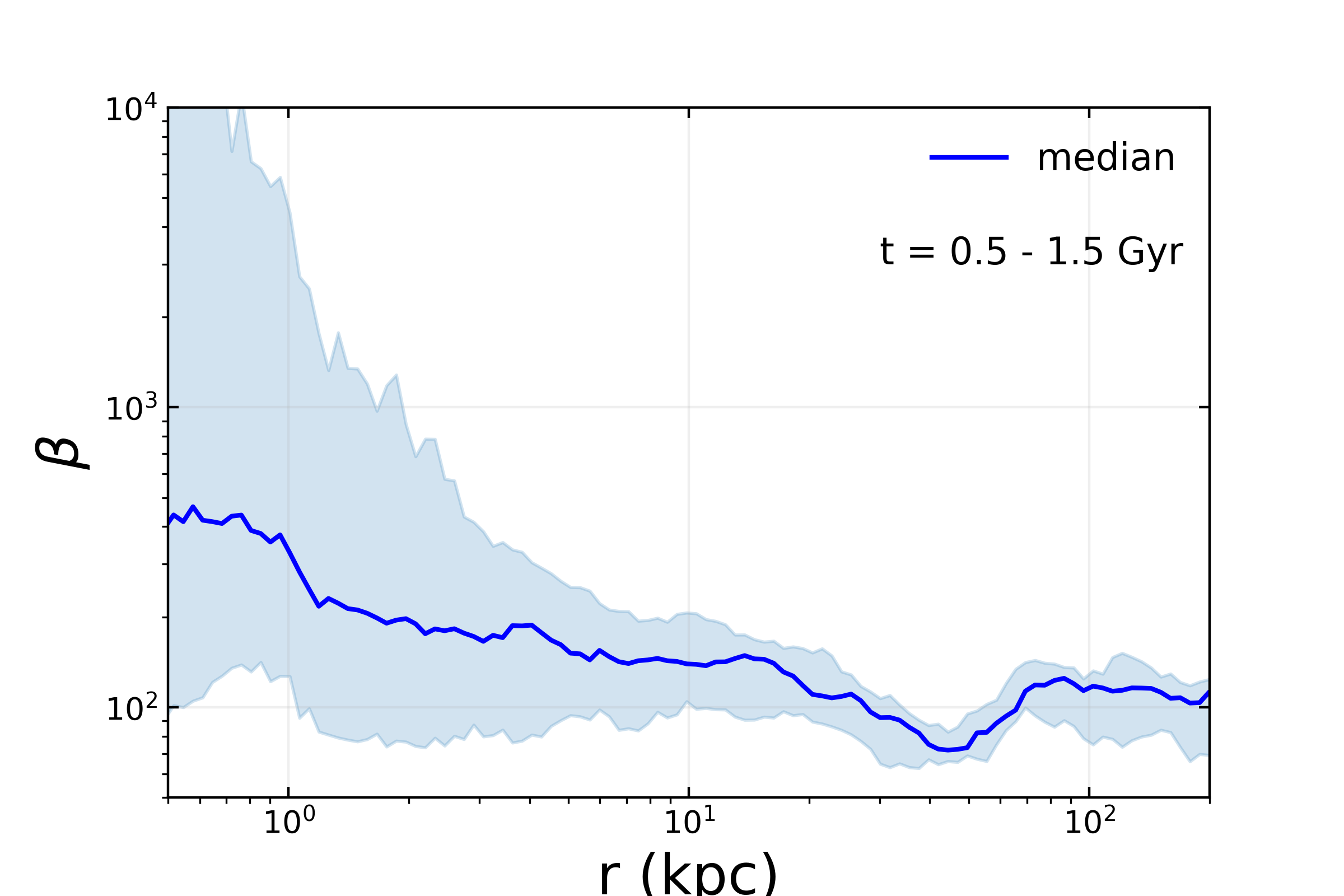}
   \includegraphics[width=3.3in,height=2.2in]{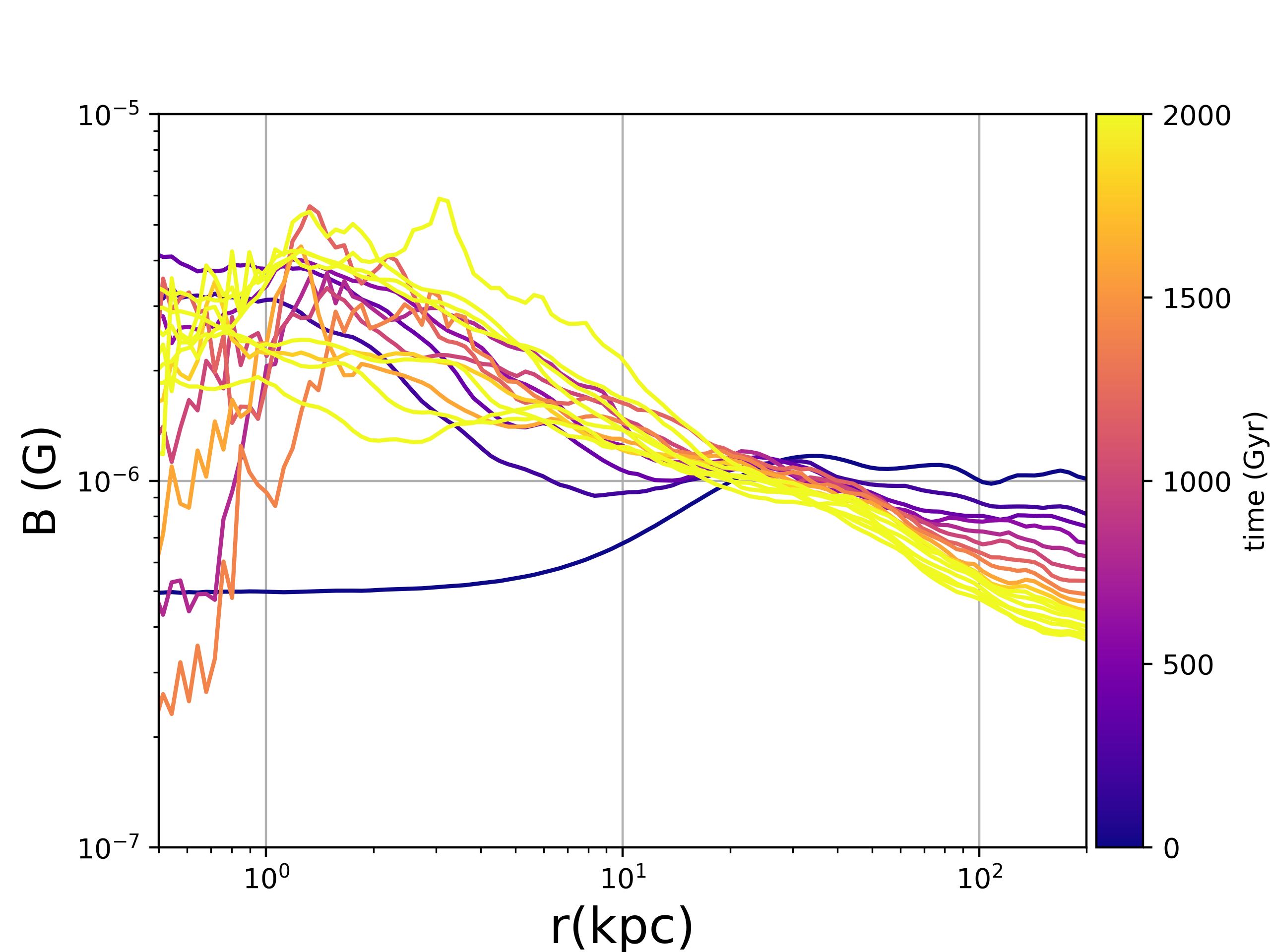}
 \caption{ {\it Top left panel : } Median mass-weighted radial entropy profile for the  X-ray gas ($0.2$ keV$<T<8$ keV) for SPG-Cool-MHD (blue line) and SPG-Cool-hydro (red line) runs. The faded cyan, red and grey region represent the 5th-95th percentile range of radial entropy profile at each radius between t=0.5-1.5 gyr for the SPG-Cool-MHD and SPG-Cool-hydro runs respectively. 
{\it Top right panel :} Median radial $t_{\rm cool}$ for $\Delta t=0.5-1.5$ Gyr for the SPG-Cool-MHD (blue line) and SPG-Cool-hydro (red line) runs. {\it Bottom left panel :} Median plasma-$\beta$ ($=P/[B^2/2\mu_0]$) for the SPG-Cool-MHD run (solid blue line) with the shaded region showing the spread of plasma-$\beta$ between 5th and 95th percentile at each radius. {\it Bottom right panel :} Angle-averaged radial B-field with time for the SPG-Cool-MHD run. The color of the radial profile shows the time between t = 0-2 Gyr with a cadence of 100 Myr.
}
\label{fig:spg_cool_core_radial}
\end{figure*}
Figure \ref{fig:spg_cool_core_radial} analyses the entropy, cooling time, magnetic field strength, and plasma $\beta$ ($\equiv P_{th}/[B^2/2\mu_0]$) for the SPG runs with cooler core using radial (spherically-averaged) profiles. The top left panel shows the median of the mass-weighted radial entropy profile for the X-ray gas ($0.2<T_{\rm keV}<8$) between $t=0.5- 1.5$ Gyr for the SPG-Cool-MHD (blue line) and SPG-Cool-hydro (red line) runs. The cyan and red shaded regions show the 5th-95th percentile range of the entropy at each radius for MHD and hydro runs respectively. Both runs show very similar median entropy and spread at $r>8$~kpc within the $t=0.5-1.5$ Gyr time window. Within $2$kpc$<r<8$ kpc, the median entropy for the hydro run is higher by a factor of 2 compared to the MHD run as AGN jets cause much larger disruption in the CGM in the hydro case. The entropy range within $r \simeq 10$ kpc is consistent with a heated core for both  runs. As with the entropy profile, the SPG-Cool-hydro run shows a higher median $t_{\rm cool}$ (top right panel) within $2$ kpc $<r<8$ kpc compared to the SPG-Cool-MHD run. The $t_{\rm cool}$ profiles show a larger spread at each radius compared to the entropy profiles, because they include all the gas.  

The bottom right panel in Figure \ref{fig:spg_cool_core_radial}  shows spherically-averaged radial profiles magnetic field magnitude with time from $t=0-1.7$ Gyr, with a cadence of 100 Myr for the SPG-Cool-MHD run. This shows that the magnetic field strength monotonically declines with time beyond $r\gtrsim 20$ kpc and strengthens within $r\lesssim20$ kpc. This is similar to the SPG and MPG runs, where we see the magnetic field strength monotonically declining at $r>50$ kpc and rising to a value in the central regions that stays roughly constant with time.  

\subsubsection{SNIa heating and Radiative Cooling}
\begin{figure*}
\centering
 \includegraphics[width=0.45\textwidth]{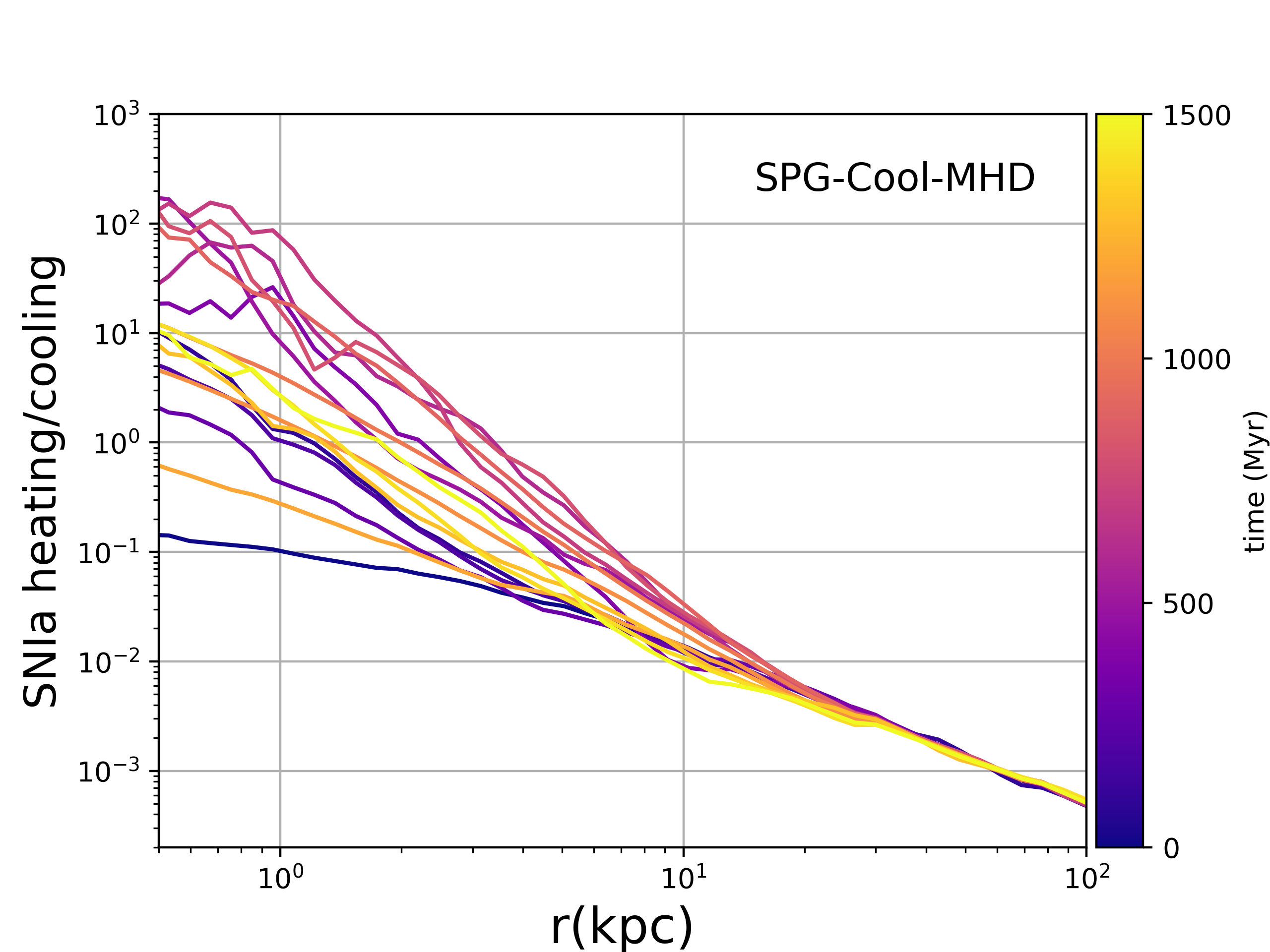}
 \includegraphics[width=0.45\textwidth]{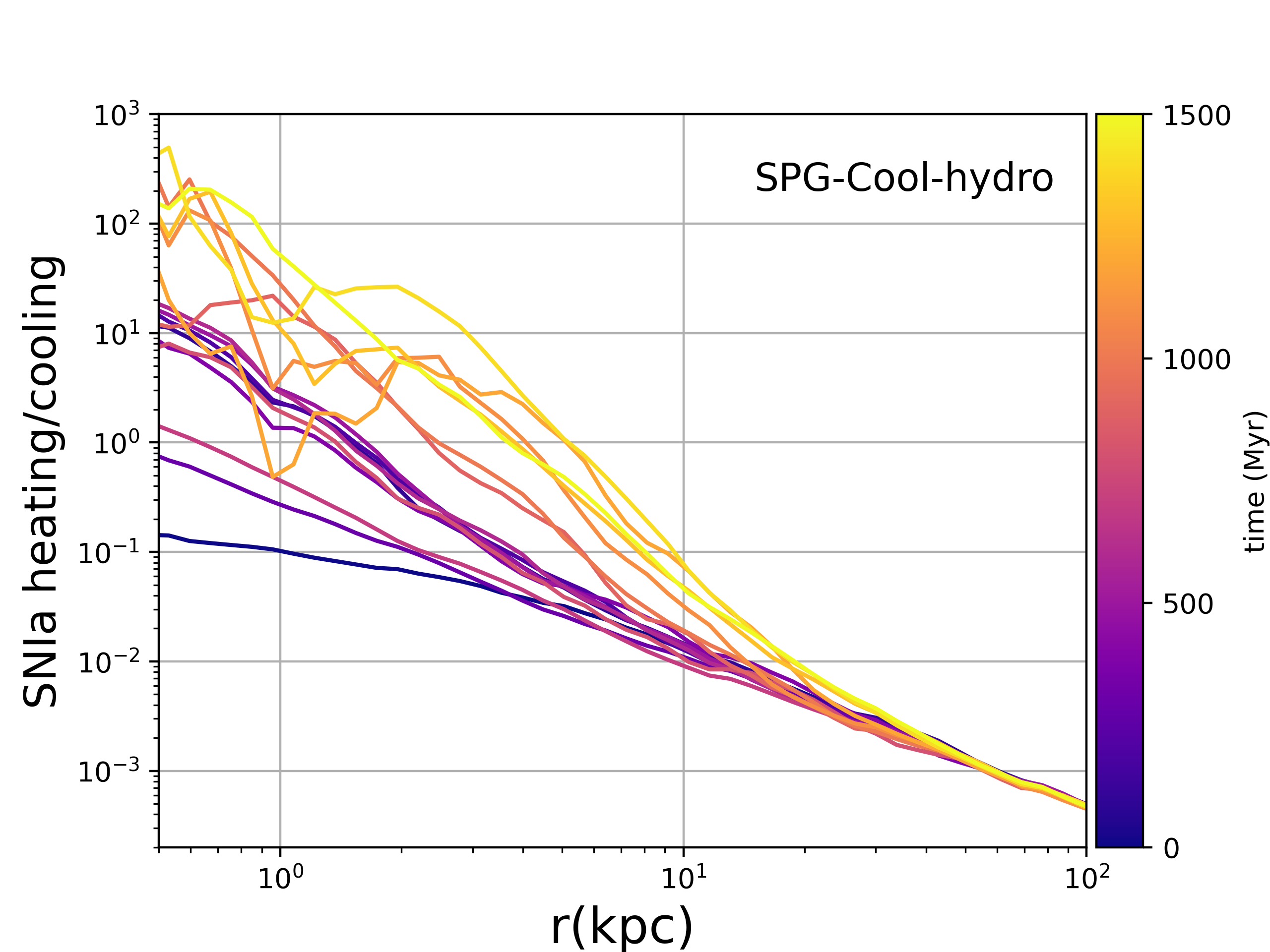} 
 \caption{ SNIa heating to cooling ratio for the SPG-Cool-MHD (left panel) and SPG-Cool-hydro (right panel) runs with a cadence of 100 Myr between $t=0-1.5$ Gyr. The color of the lines represents the time of the galaxy evolution. Dominant cooling at $r>1$ kpc leads to formation of extended cold gas filaments for both the runs. 
}
 \label{fig:spg_cool_ratio}
\end{figure*}
Figure \ref{fig:spg_cool_ratio} shows the evolution of the ratio of SNIa heating to radiative cooling of the CGM for the SPG-Cool-MHD (left panel) and SPG-Cool-hydro (right panel) runs between $t = 0-1.5$ Gyr. Initially, radiative cooling dominates over SNIa heating at all radii. As the galaxy evolves, AGN feedback pushes the gas from the inner $r<5$ kpc outwards, lowering the CGM pressure. This allows SNIa heating to dominate over radiative cooling within $r \simeq 3$ kpc. The AGN jets are not powerful enough to keep the CGM in the hot state for long, however, and as such cooling becomes dominant over SNIa heating again. That leads to the formation of cold gas, which leads to the next AGN outburst. A similar heating-cooling cycle is seen for both MHD and hydro runs for $t=1.5$ Gyr. 

\subsection{Jet Morphology}
\label{sec:morph}
\begin{figure*}
\centering
 \includegraphics[width=0.95\textwidth]{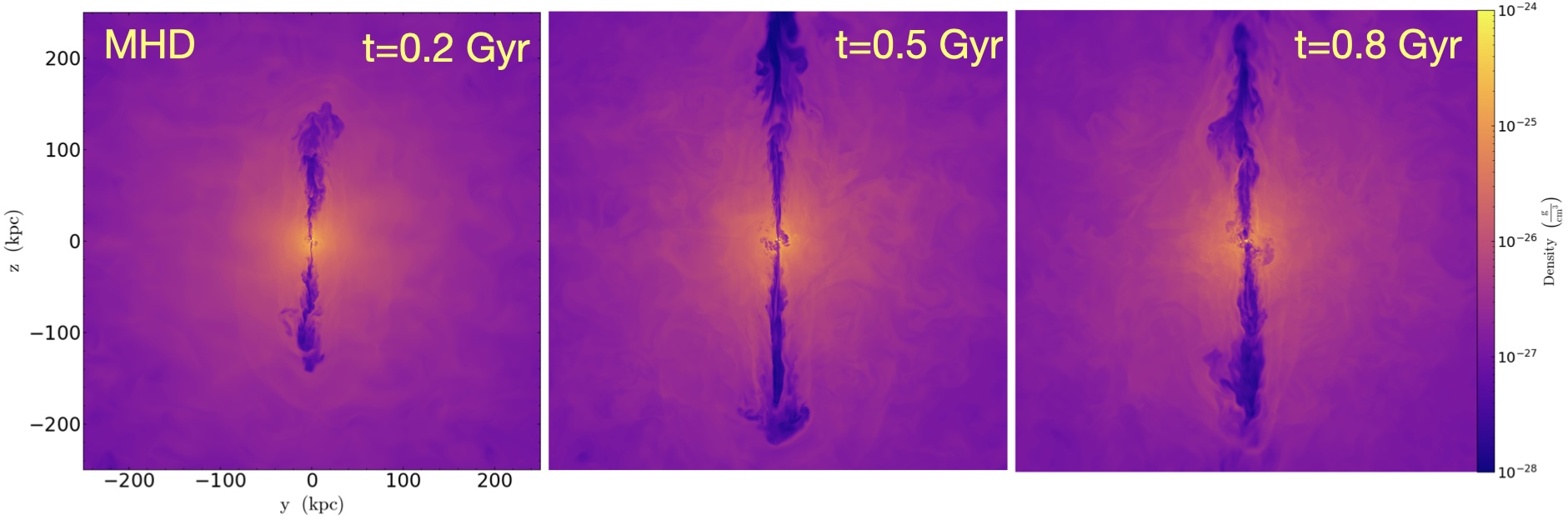}
 \includegraphics[width=0.95\textwidth]{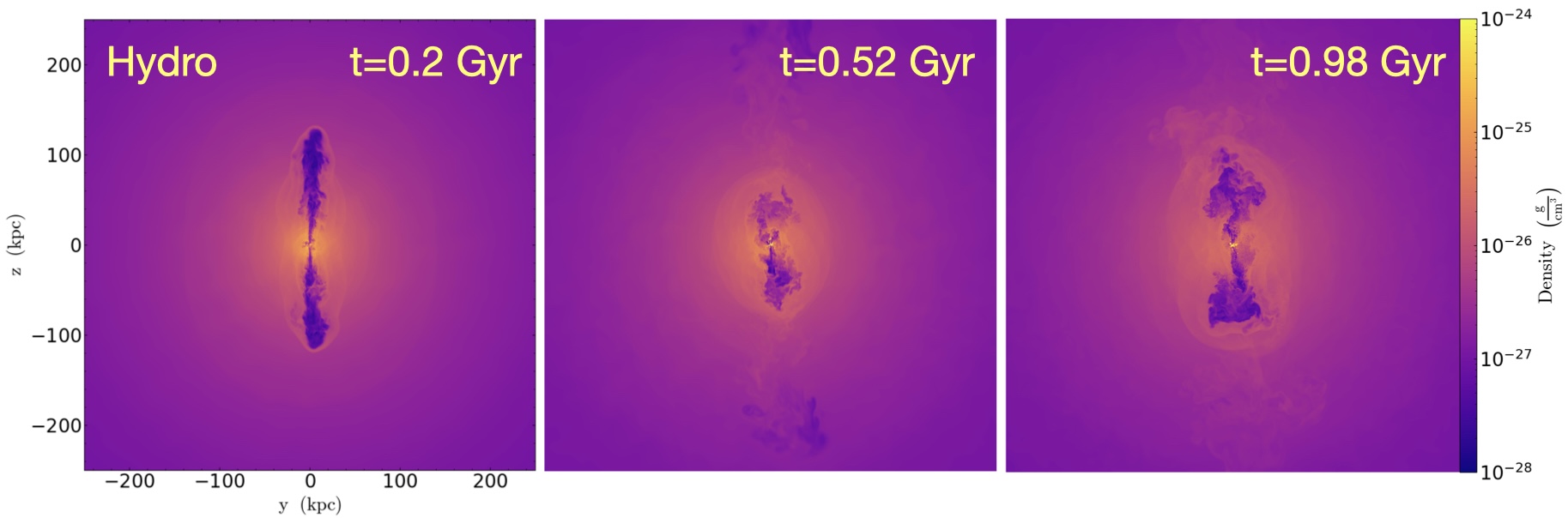}
 \includegraphics[width=0.94\textwidth]{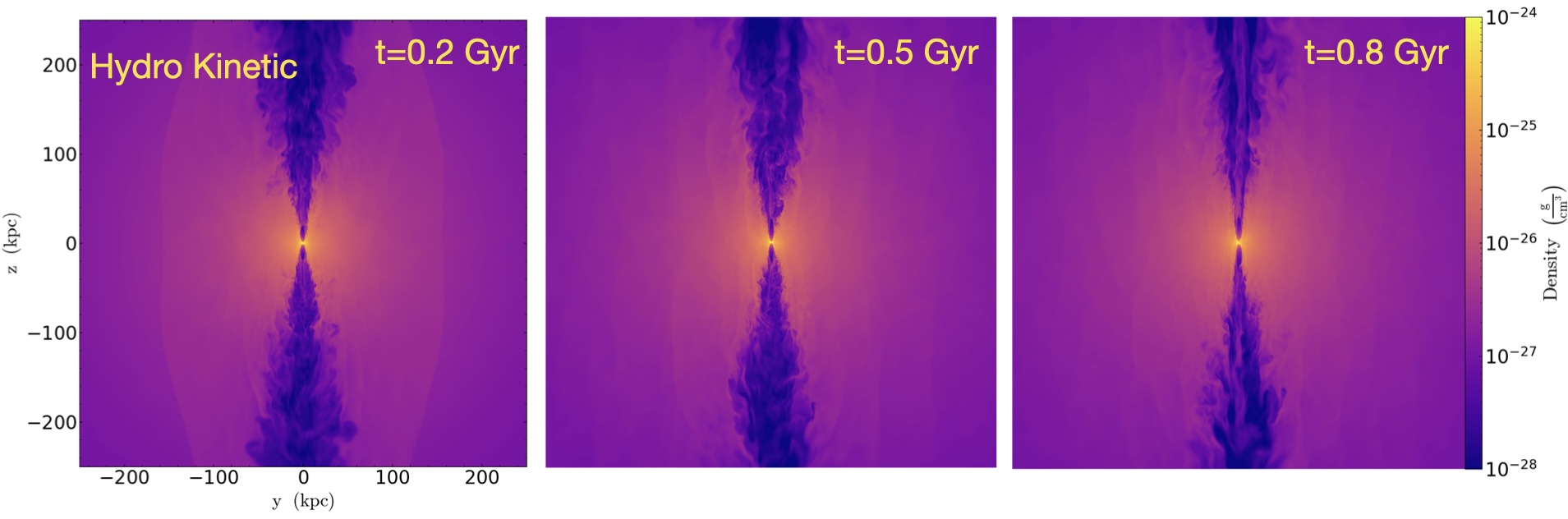}
 \caption{ Density slices showing AGN jets for the MPG-MHD (upper panels), MPG-hydro (middle panels) and MPG-hydro-kinetic (lower panel) runs. The time of the snapshot has been chosen such that AGN is close to peak power. Pure kinetic and MHD jets travel to much larger distances in a collimated state compared to hydro jet with partial thermal energy which tend to inflate cavities much closer ($r\sim30$ kpc) to SMBH which then rises buoyantly to larger radii.
 }
 \label{fig:jet_morph_den}
\end{figure*}  
\begin{figure*}
\centering
 \includegraphics[width=0.95\textwidth]{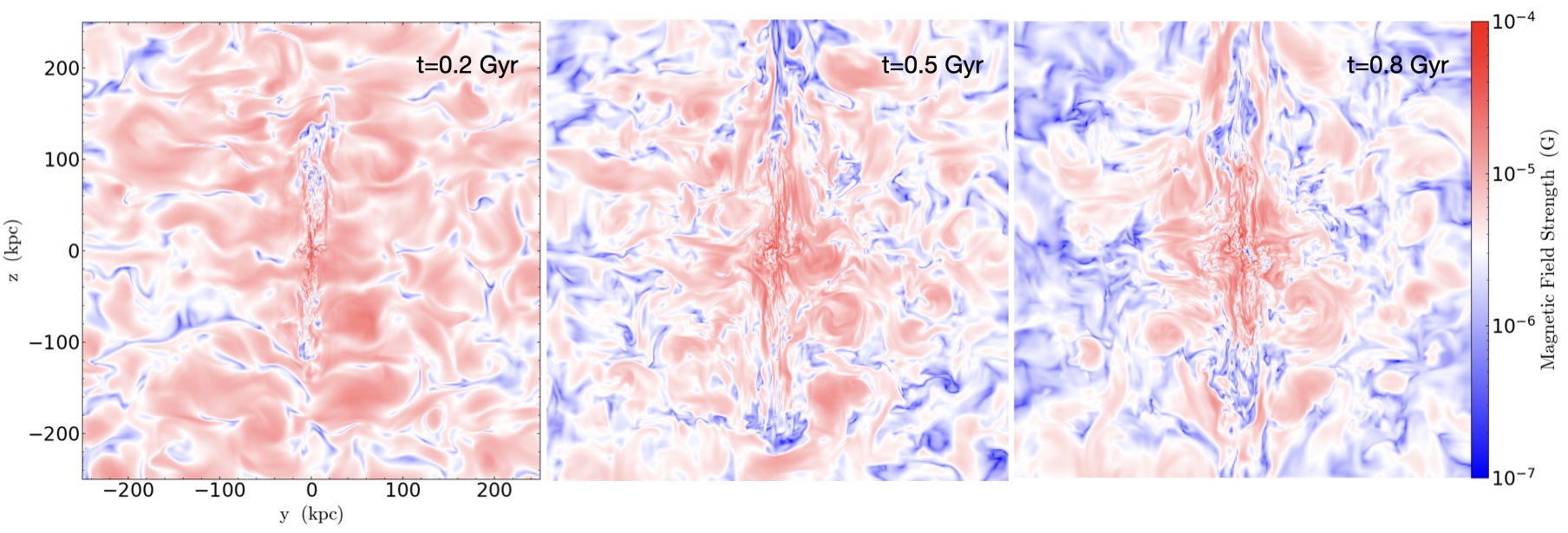}
 \caption{ Magnetic field strength slice at t=0.2 Gyr (left), 0.5 Gyr (middle), and 0.8 Gyr (right) for the MPG-MHD run. Magnetic fields concentrate along the jet axis, while they decline in strength at larger radii.    
}
 \label{fig:jet_mag_field}
\end{figure*}
\begin{figure*}
\centering
 \includegraphics[width=0.95\textwidth]{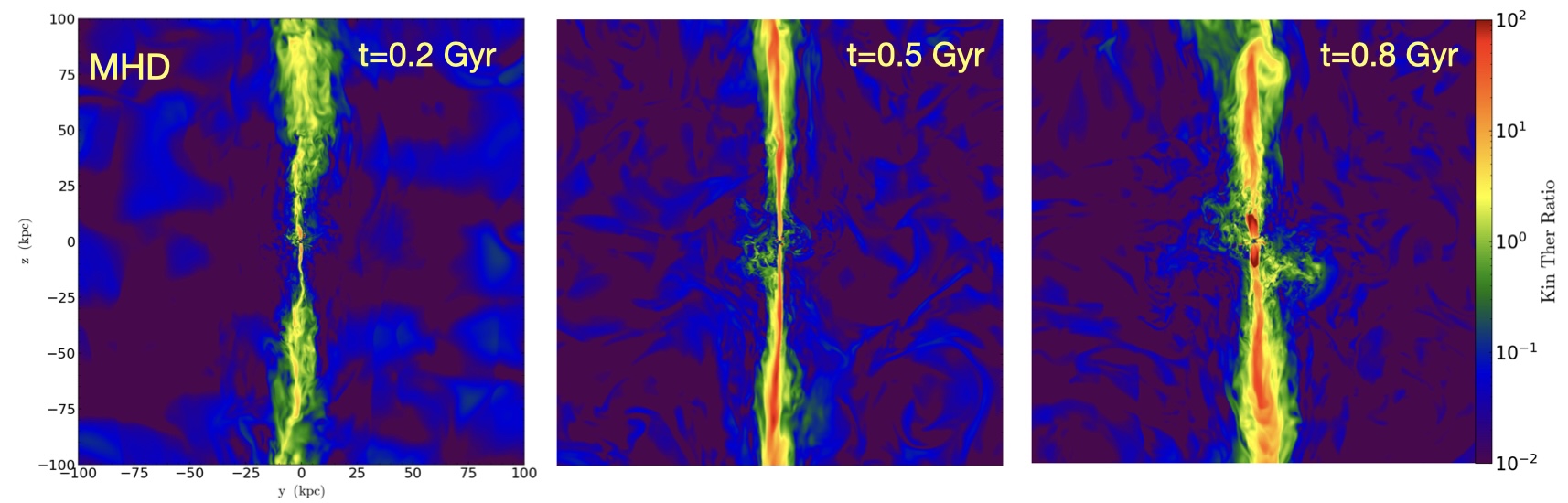}
 \includegraphics[width=0.95\textwidth]{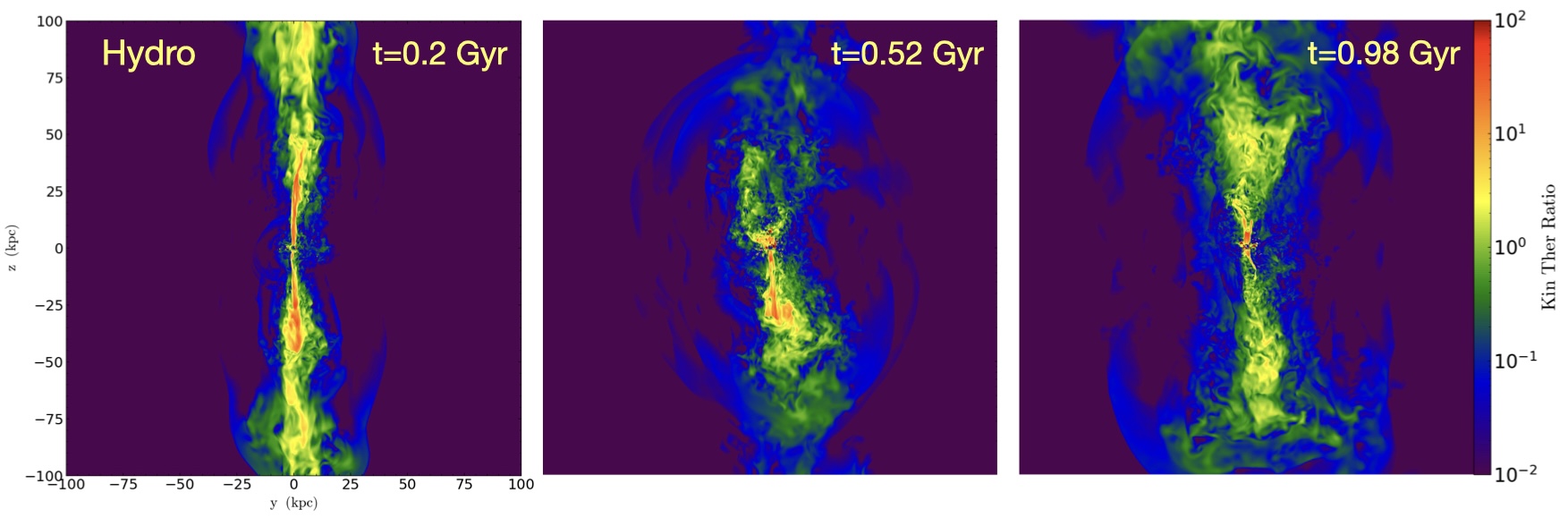}
 \includegraphics[width=0.95\textwidth]{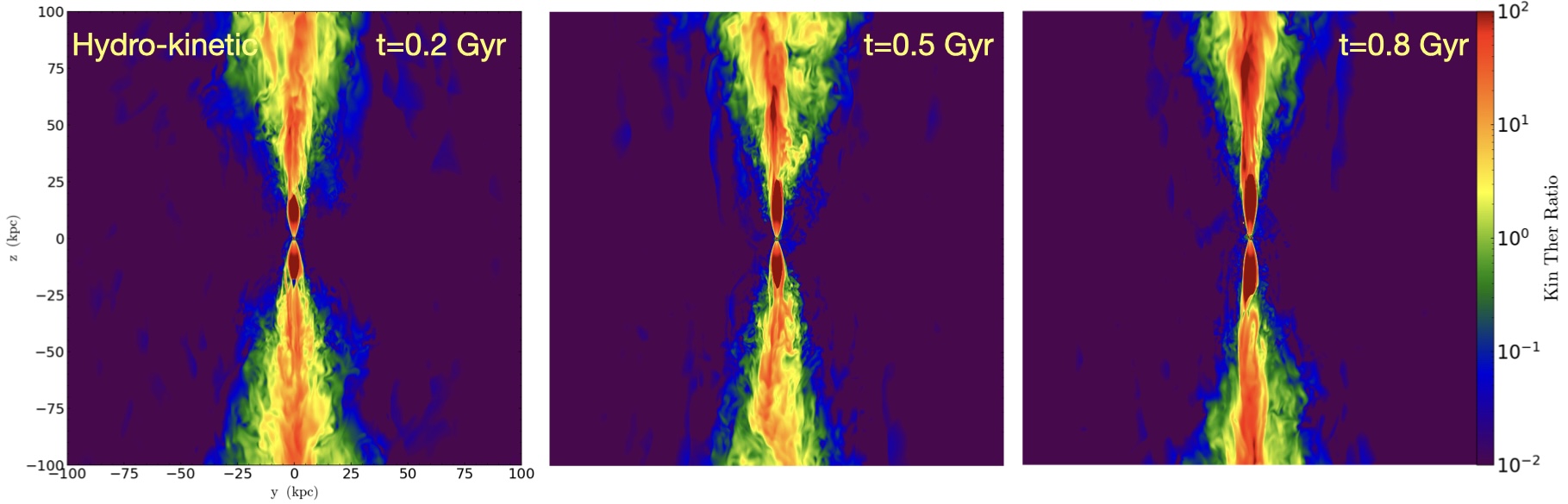}
 \caption{  {\it Upper panels} : Kinetic to thermal energy  slices for the MPG-MHD (upper panels), MPG-hydro (middle panels) and MPG-hydro-kinetic (lower panels) runs at same times as figure \ref{fig:jet_morph_den}. The box size is 200 kpc x 200 kpc. Pure kinetic and MHD jets show kinetically dominated jets to larger radii compared to hydro jets with partial thermal feedback which thermalise within $r<20$ kpc.  }
 \label{fig:jet_morph_ratio}
\end{figure*}
Figure \ref{fig:jet_morph_den} shows the density snapshots for the MPG-MHD (upper panels), MPG-hydro (middle panels) and MPG-hydro-kinetic runs at times when the AGN power is close to its peak. The AGN jets are highly collimated to 100s of kpc for the pure kinetic and MHD runs. Consequently, they travel to larger distances before being disrupted compared to the hydro run. The higher injected kinetic energy in the pure kinetic feedback run leads to greater initial velocity of the jet plasma, allowing the resulting jets to travel much larger distances. In the MHD case, magnetic fields provide additional collimation to the AGN jets thus allowing them to travel larger distances. Figure \ref{fig:jet_mag_field} shows the magnetic field strength in the CGM for the MPG-MHD run at the same times as the density slices 
for the MPG-MHD run in Figure \ref{fig:jet_morph_den}. We see a gradual amplification and alignment of the magnetic field strength along the jet axis as the galaxy evolves. Plots also show a clear decline in magnetic field strength in regions away from the jet axis (also see bottom right plot in Figure \ref{fig:mpg_radial_prof}). The amplified magnetic fields provide  collimation to the jets in addition to contributing to the CGM pressure.

Figure \ref{fig:jet_morph_ratio} shows slices of the kinetic to thermal ratio of the CGM for the MPG-MHD (upper panels), MPG-hydro (middle panels), and MPG-hydro-kinetic (lower panels) runs at the same times as Figure \ref{fig:jet_morph_den}. The kinetic to thermal ratio slices 
show that the highly collimated jets for the pure kinetic and MHD runs are kinetically dominated for much larger distances ($r\sim 100$ kpc) compared to hydro jets with partial thermal feedback, which thermalise at $r\sim 20$ kpc. This difference becomes more pronounced for the late stage AGN activity for the MPG-MHD case as the gradual buildup of magnetic fields along the jet axis creates a channel for jets to travel to larger distances before getting disrupted. On the other hand, the hydro jets with partial thermal feedback likely rises buoyantly to 100s of kpc as the kinetic-to-thermal ratio is approximately 1 due to their thermalisation at smaller radii ($r\sim10$s kpc).

\begin{figure*}
\centering
 \includegraphics[width=0.3\textwidth]{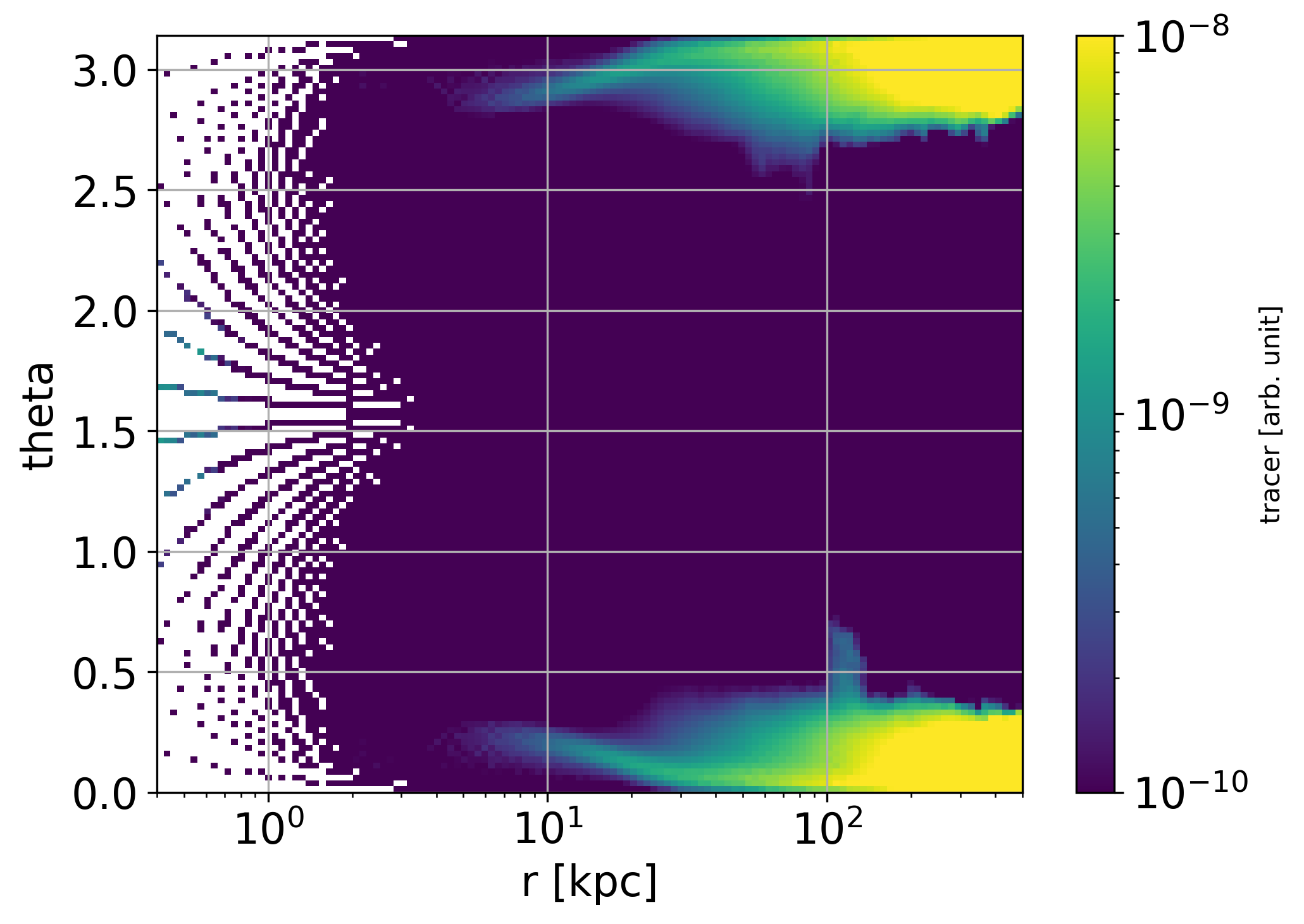}
 \includegraphics[width=0.3\textwidth]{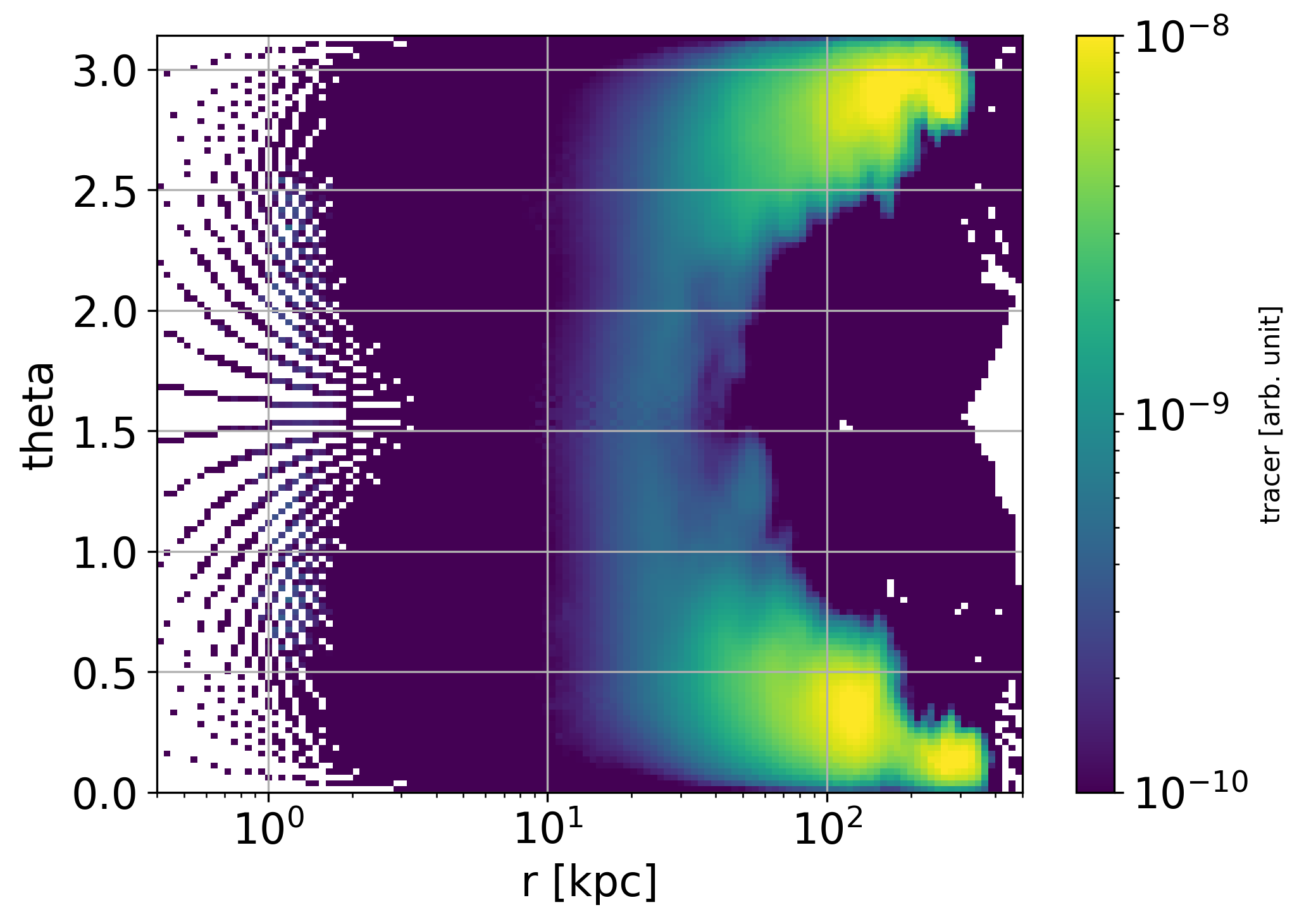}
 \includegraphics[width=0.3\textwidth]{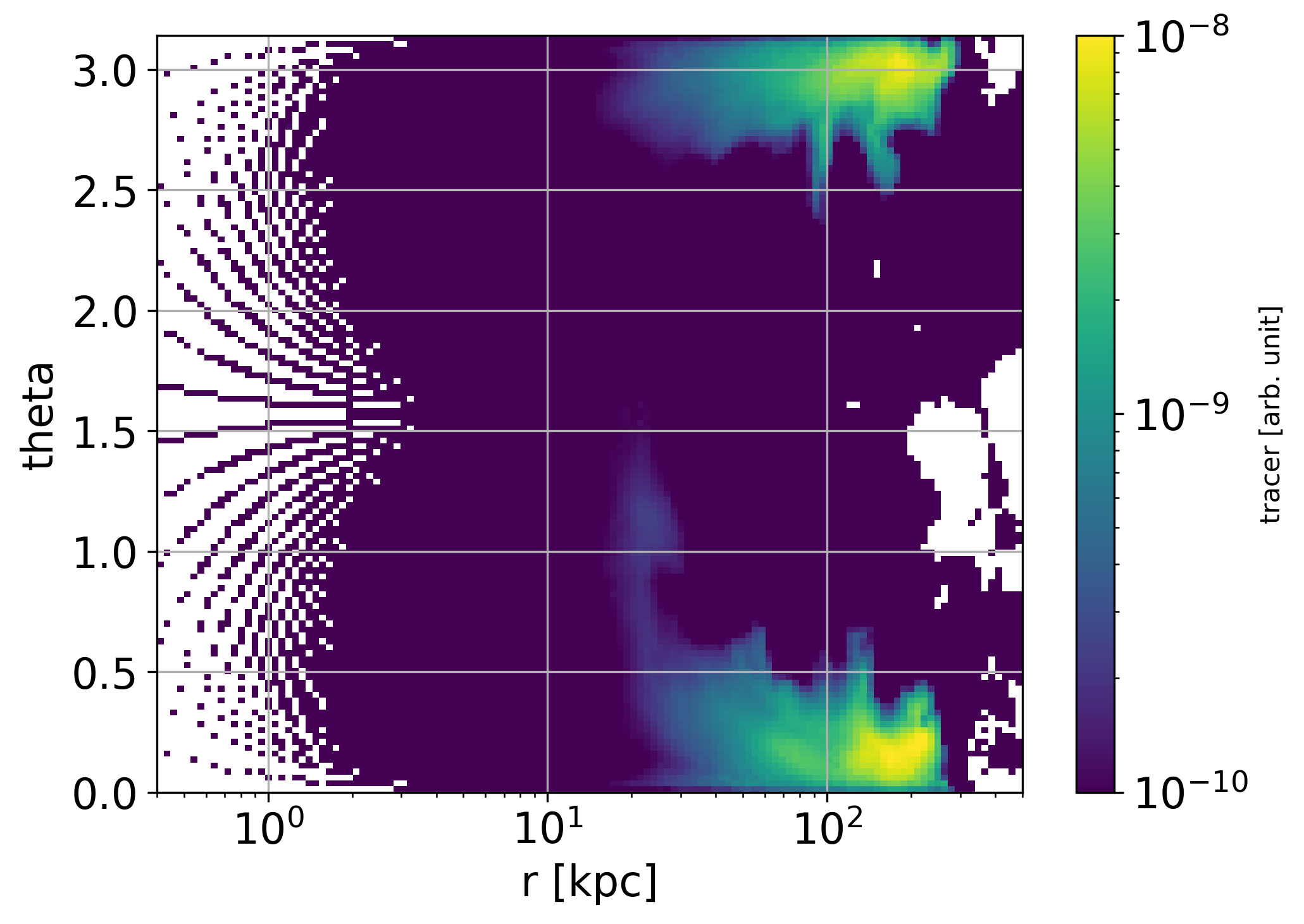}
 \caption{
 Radial - polar angle (r-$\theta$) distribution of the jet tracer fluid for the MPG-hydro-kinetic (left panel), MPG-hydro (middle panel) and MPG-MHD (right panel) runs between t=0.5-1 Gyr. For the pure kinetic AGN feedback hydro run, jet material remains confined along the jet axis with very little mixing beyond the jet cone. AGN jets travel to large distances before thermalising. On the other hand, for the hydro run with kinetic+thermal AGN feedback, the jet material is more isotropically distributed at smaller radii ($r\sim10-30$ kpc) where the jets thermalise, while jet material remains concentrated along the jet axis at larger radii as it rises buoyantly to 100s of kpc. In the case of the MPG-MHD run, the jet material is concentrated along the jet axis at all radii with small spread at lower radii ($r\sim 15$ kpc).
}
 \label{fig:R-theta_tracer}
\end{figure*}

To explore the flow of jet plasma in the CGM, we injected and tracked a passive tracer fluid with the AGN jet material. Figure \ref{fig:R-theta_tracer} shows the radial-polar ($R-\theta$) distribution of the passive tracer averaged over $t = 0.5 - 1.0$ Gyr for all MPG runs. In the case of pure kinetic AGN feedback, the passive tracer remains confined along the jet axis with AGN jets traveling very large distances ($r>$ few 100 kpc) before thermalising and mixing with the ICM. There is little mixing of the jet plasma at smaller radii. 
For the MPG-hydro run, the distribution of passive tracer is more isotropic between $r\sim10-40$ kpc, where AGN jets thermalise. However, after thermalisation the cavities inflated by the jets tend to rise to larger radii along the jet axis. On the other hand, as the jets thermalise at $r>100$ kpc in the MPG-MHD case, the tracers are largely confined along the jet axis out to $r\sim200$ kpc with very little spread beyond the jet cone at smaller radii ($r\sim 10$s kpc) similar to MPG-hydro-kinetic run. However, unlike the MPG-hydro-kinetic run where passive tracer fluid is seen beyond $r\sim 500$ kpc, the tracer fluid for the MPG-MHD run remains confined within $r\sim200$ kpc as the jet power is an order of magnitude smaller.

\subsection{Gas Mass Flow}
\label{sec:config}
In all of our simulations, most of the AGN's energy output ultimately goes into reconfiguration of the circumgalactic gas rather than heating CGM within the galactic halo. 
\begin{figure*}
\centering
\includegraphics[width=0.3\textwidth]{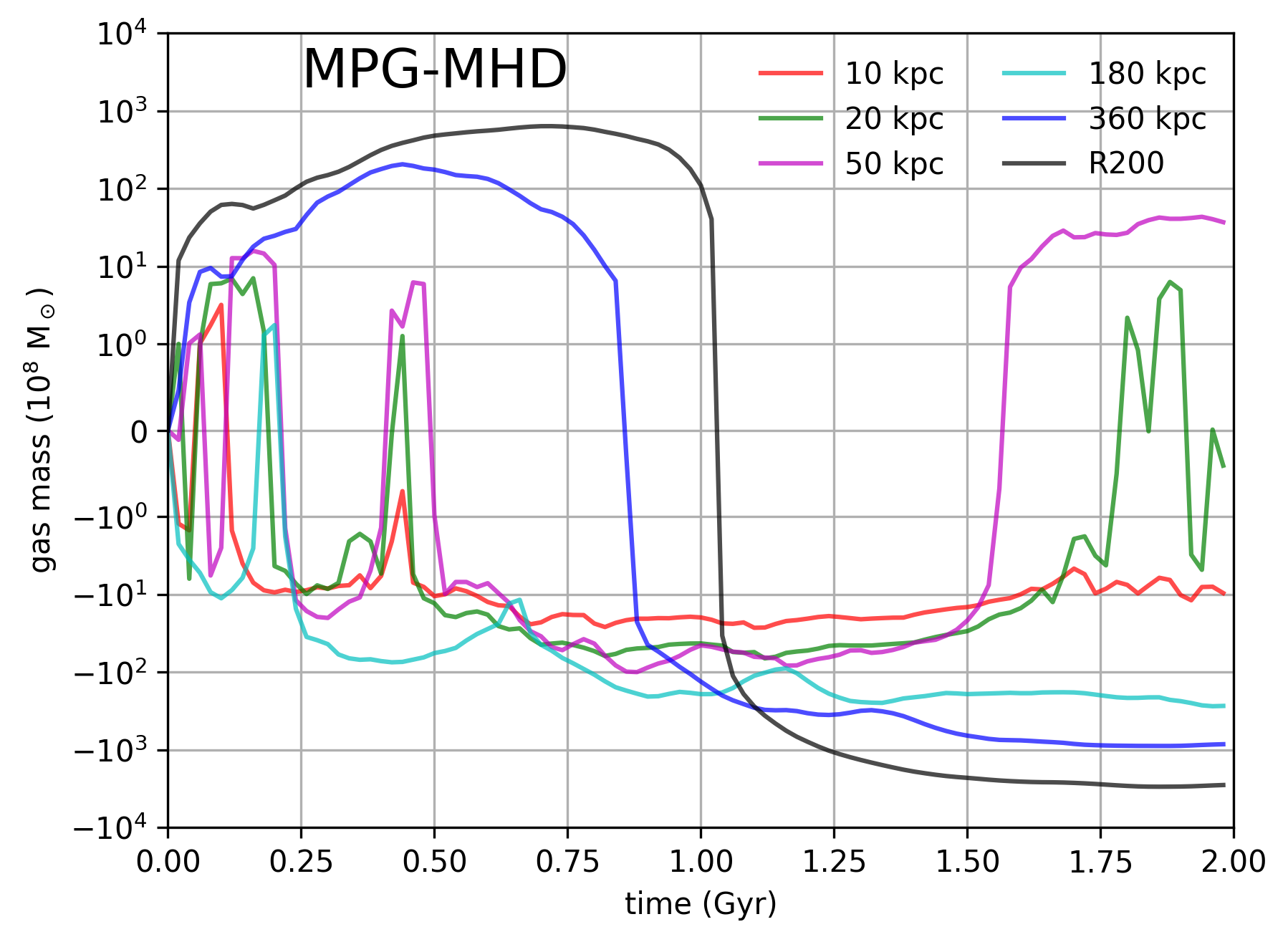}
\includegraphics[width=0.3\textwidth]{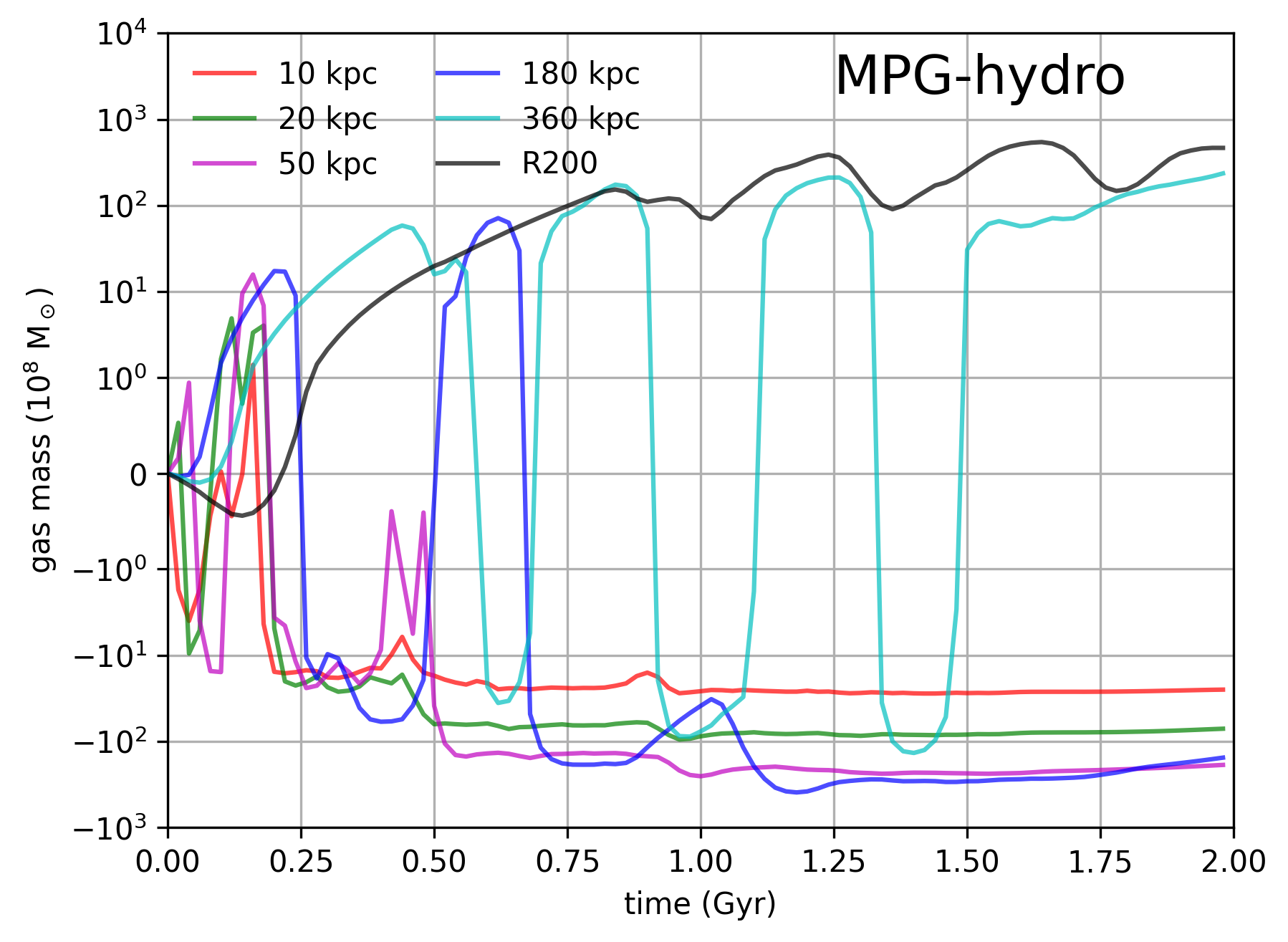}
\includegraphics[width=0.3\textwidth]{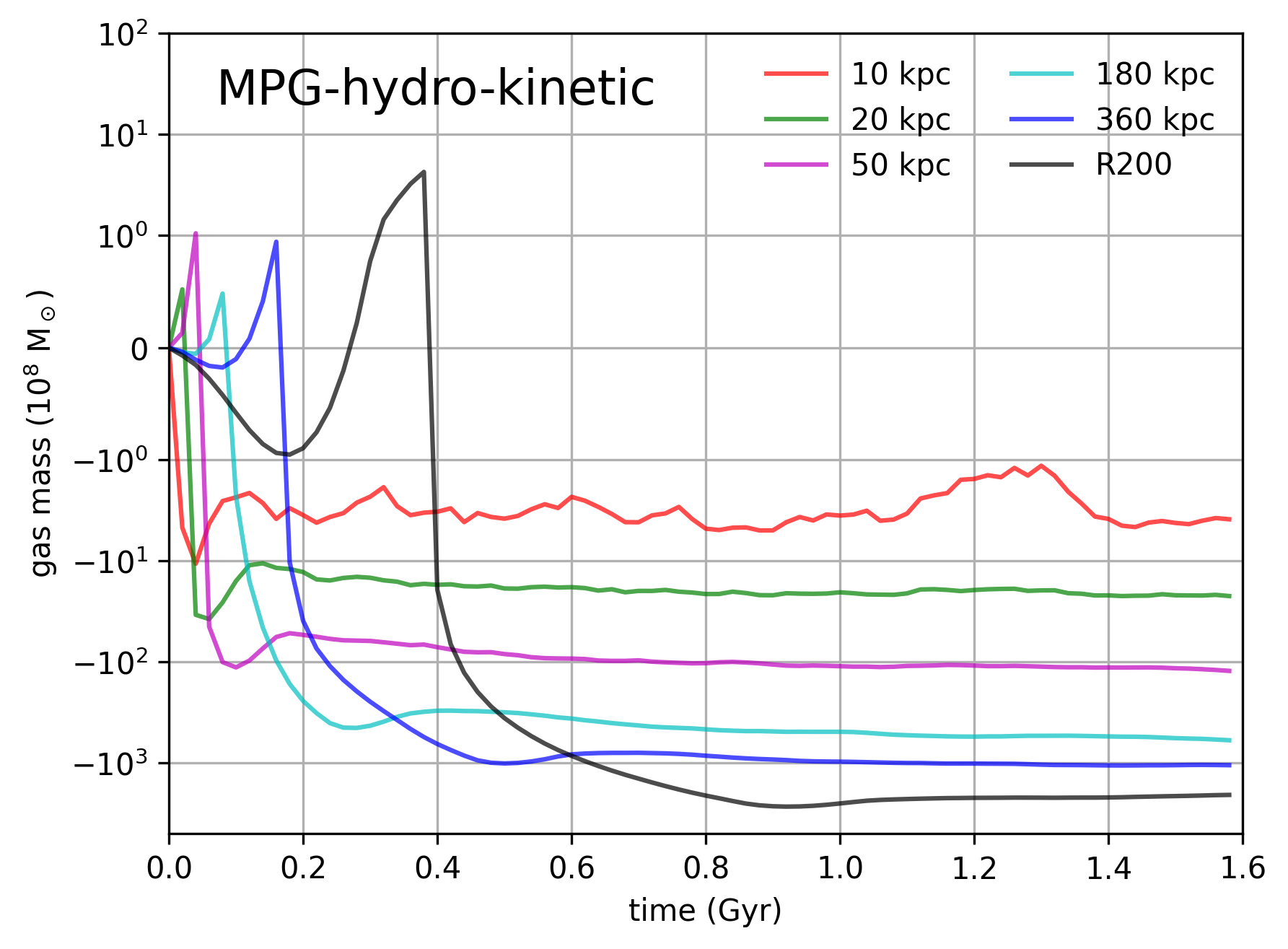}
\caption{Change in the total enclosed gas mass within different radii with time compared to the initial state for the MPG-MHD (left panel), MPG-hydro (middle panel) and MPG-hydro-kinetic (right panel) runs. Note the different behaviour of the total enclosed gas mass within $R_{500}$ ($\sim 360$ kpc) and $R_{200}$ ($\sim730$ kpc) for the runs with collimated jets (MPG-hydro-kinetic and MPG-MHD) and one with jet thermalisation at 10s kpc (MPG-hydro). }
\label{fig:encl_bry}
\end{figure*}

Figure \ref{fig:encl_bry} shows the changes in total enclosed baryon mass within r=10 kpc, 20 kpc, 50 kpc, $R_{2500}$ (180 kpc), $R_{500}$ (360 kpc) and $R_{200}$ (730 kpc) as a function of time for the MPG-MHD (left panel), MPG-hydro (middle panel) and MPG-hydro-kinetic (right panel) runs. For the MPG-hydro-kinetic run, the total baryon mass within all radii shows an initial rise and then a sharp drop by $t\sim 0.5$~Gyr. This is largely because powerful kinetic AGN jets drives large outflows, pushing the gas well beyond $R_{200}$. As a result, there is a net decline of the total baryon mass within $R_{200}$ $\sim10^{11}$ M$_\odot$. However, the decline in total baryon mass is not monotonic and the outflow of gas beyond $R_{200}$ is arrested by $t\sim 1$ Gyr. For the MPG-hydro run with partial thermal feedback, AGN jets are unable to drive very large outflow to several 100 kpc and as such there is no sharp decline in the total baryon mass within $R_{200}$. However, the MPG-hydro run shows an oscillation in the total baryon mass within $R_{500}$ (360 kpc). The total baryon mass within $R_{200}$ shows a monotonic rise with time with some oscillation in baryon mass. In the central regions ($r \lesssim 50$ kpc) the AGN is able to raise $\sim 10^{10}$ M$_\odot$ of gas to larger radii with the total baryon mass not recovering to its initial state after $t\sim0.5$ Gyr. For the MPG-MHD run, the total enclosed gas mass within $R_{200}$ shows a sharp decline after $t\sim1$ Gyr. This decline in total baryon mass within $R_{200}$ is similar to the MPG-hydro-kinetic run albeit with a delay. Within the central $r=50$ kpc the enclosed baryon shows a rise and fall that coincides with the jet cycle.

\section{Discussion}
\label{sec:disc}

\subsection{Kinetic vs. Thermal Feedback}
\label{sec:kinetic}

\begin{figure*}
\centering
 \includegraphics[width=0.47\textwidth]{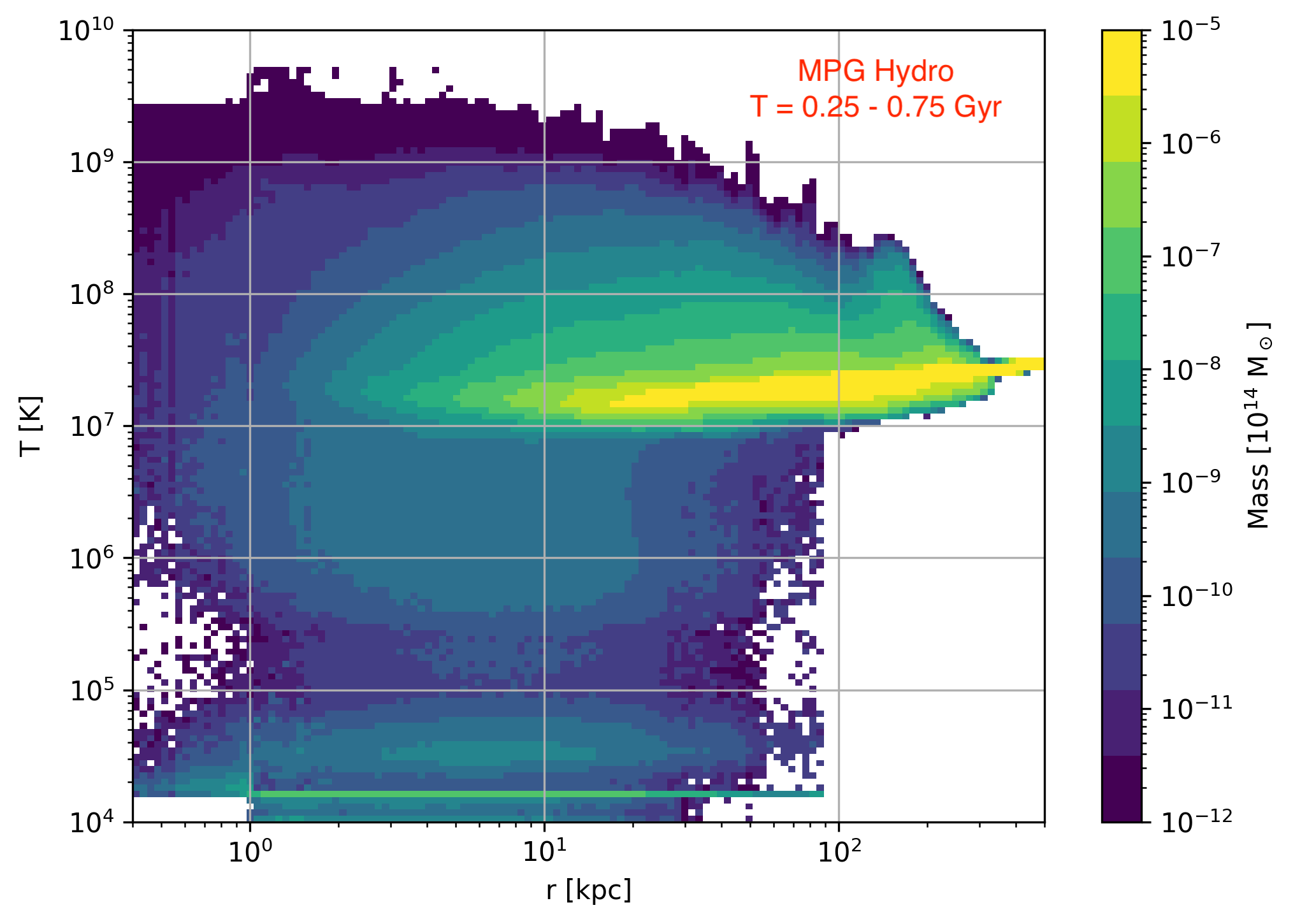}
 \includegraphics[width=0.47\textwidth]{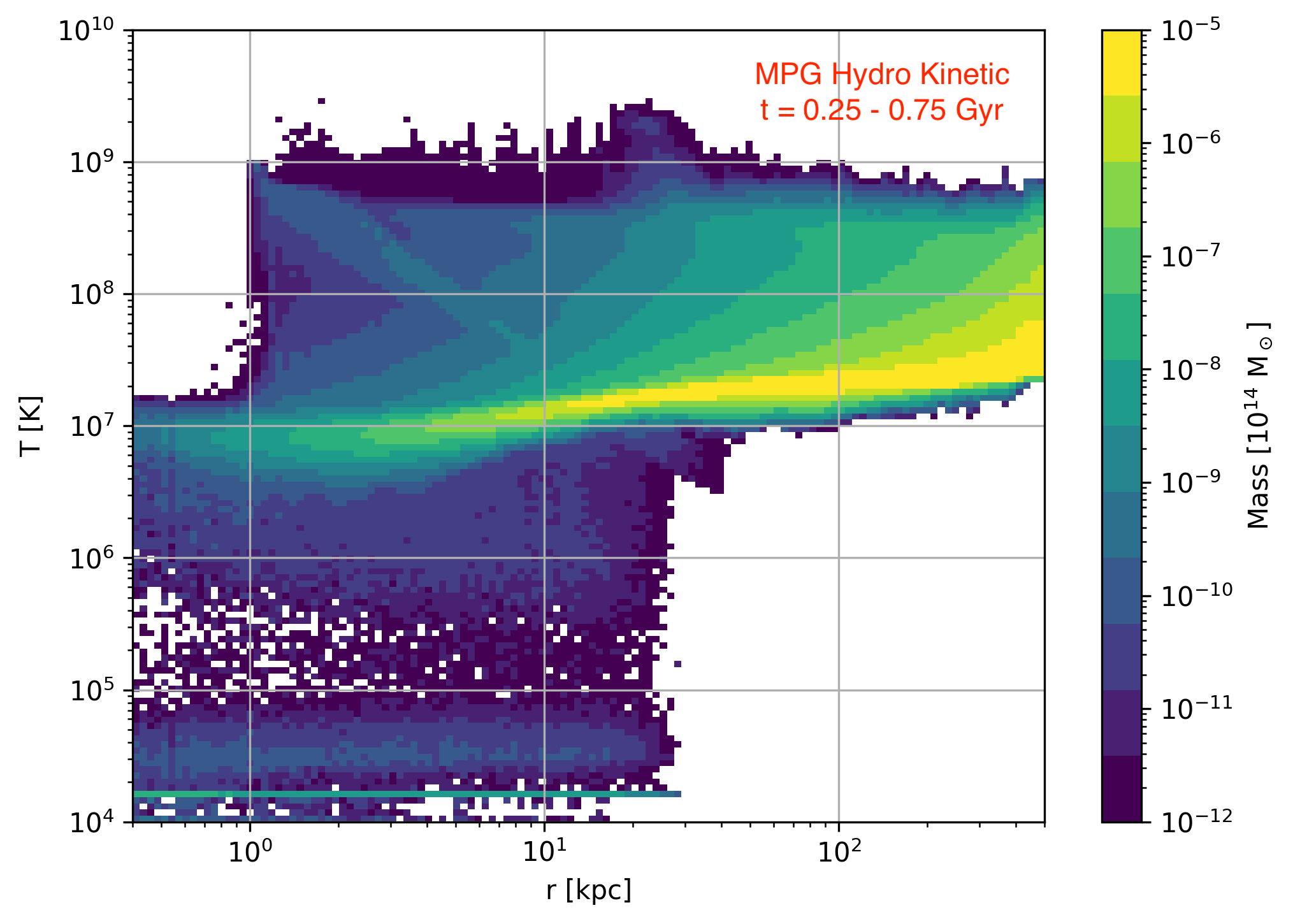}
 \caption{ Average temperature-radius distribution of the CGM for the MPG-hydro (left panel) and the MPG-hydro-kinetic (right panel) runs between $t=0.25-0.75$ Gyr. The colour shows the mass of gas cells. Note the apparent lack of cold gas below $r<1$ kpc with most of the cold gas accumulating between $3$ kpc $<r<15$ kpc for the MPG-hydro run. On the other hand, for the purely kinetic AGN feedback run, the cold gas mass peaks below $r<1$ kpc.   
}
 \label{fig:phase_sp}
\end{figure*}

Thermal AGN feedback in our simulations impact the formation and distribution of cold gas ($T<10^5$ K) in the CGM. The left panel of Figure \ref{fig:phase_sp} shows the temperature distribution of the CGM as a function of radius for the MPG-hydro run, with colour representing the average mass at that radius between $t=0.25-0.75$ Gyr. There appears to be a lack of cold gas ($T<10^5$ K) within $r\simeq1$ kpc and most of the cold gas accumulates between $3$ kpc $<r<15$ kpc. This is because thermal AGN feedback within $r \simeq 1$ kpc does not allow a large amount of cold gas to accumulate within $r \approx 1$ kpc. Cold gas clumps extend up to $r\sim80$ kpc.

The right panel of Figure \ref{fig:phase_sp} shows the temperature distribution of the CGM as a function of radius between $t=0.25-0.75$ Gyr for the MPG-hydro-kinetic run. Unlike the thermal+kinetic MPG-hydro case, we do not see any extended cold gas ($T<10^5$ K) beyond $r\sim30$ kpc. The pure kinetic AGN feedback run also shows that the amount of cold gas peaks within $r \simeq 1$ kpc, showing that the lack of thermal AGN feedback within $r \simeq 1$ kpc allows for cold gas to accumulate closer to the central SMBH.   

\subsection{Magnetic Field and Cold Gas}

\label{sec:mag_disc}
\begin{figure}
\centering
 \includegraphics[width=0.47\textwidth]{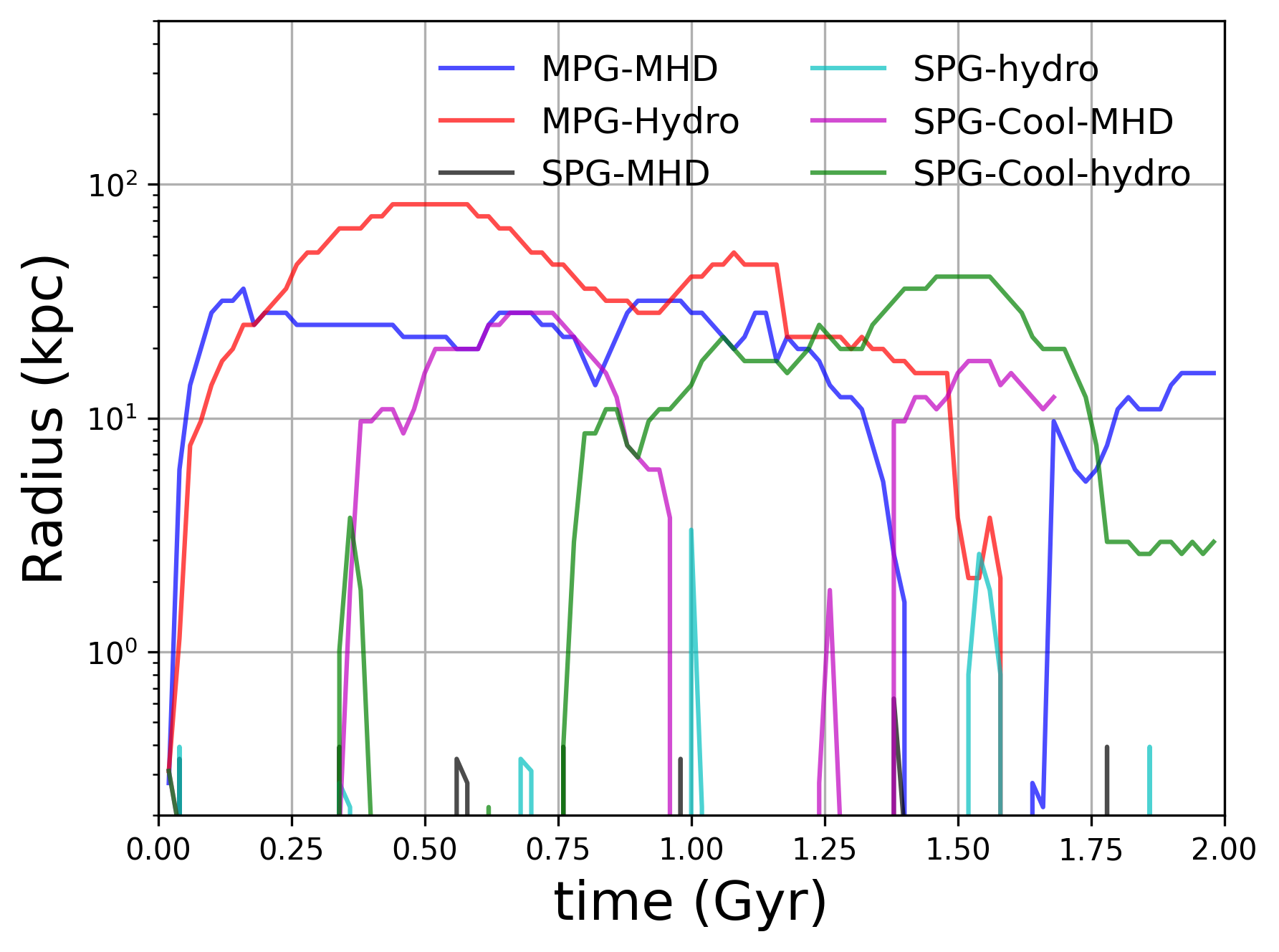}
 \caption{ Radial extent (radii of the most distant cold gas cell ($T<10^5$ K) from the central SMBH) of the cold gas clumps with time for all the runs  
 Cold gas formation remain concentrated within $r\sim 1$ kpc for the SPG-MHD and SPG-hydro runs while MPG and SPG Cool-core MHD and hydro runs show extended cold gas filaments.  
}
 \label{fig:rad_ext}
\end{figure}

As discussed in Section \ref{sec:hyd_spg}, we do not see any significant difference in the cold gas mass or the radial extent of the cold gas clumps in the single phase galaxy MHD and hydro runs (see Figure \ref{fig:rad_ext}). However, the total number of AGN cycles within $t=0-2$ Gyr is higher for the SPG-MHD run compared to the SPG-hydro run. This is likely because magnetic fields prevents the low entropy gas in the central $r \simeq 5$ kpc from completely being evacuated due to AGN activity, thus allowing the cooling phase to start on a short interval. 

For the MPG-MHD and MPG-hydro runs, the amount of cold gas is similar but the cumulative stellar mass for the MHD case is higher by a factor of 2 compared to the hydro run. Figure \ref{fig:rad_ext} shows that the radial extent of the cold gas clumps (farthest distance of the cold gas ($T<10^5$ K) cells from the central SMBH) for the MPG-hydro run is much larger ($\sim 75$ kpc) compared to the MPG-MHD run ($\sim 30$ kpc). This is likely because AGN jets are able to evacuate low entropy gas from smaller radii to larger radii in the hydro case as inferred from the sharp drop in the total gas mass within $r=20$ kpc seen in middle panel of Figure \ref{fig:encl_bry}, which leads to the formation of cold gas clumps at larger radii. On the other hand, in MPG-MHD run, magnetic field prevents the evacuation of large amounts of low entropy gas by the AGN jets from smaller radii to larger radii as seen in left panel of Figure \ref{fig:encl_bry}. As such cold gas formation remains constrained for MHD run.  

The presence of magnetic fields seems to have a significant impact on the evolution of the SPG-Cool halos. Figure \ref{fig:rad_ext} shows that both SPG-Cool-MHD and SPG-Cool-hydro runs form extended cold gas ($r\sim 20$ kpc). However, in the hydro case the amount of cold gas is higher by an order of magnitude compared to the MHD run. As such, the total star formation in the hydro case is significantly higher compared to the MHD case.   
Similar to the SPG-MHD and MPG-MHD runs, the magnetic field prevents the CGM from being evacuated from the central $r \simeq 10$ kpc, as inferred from the X-ray luminosity shown in Figure \ref{fig:spg_cool_time}.

\subsection{The Limits of AGN Feedback}
\label{sec:limits}
In Section \ref{sec:spg_cool} we presented the findings of our numerical experiments for the SPG-Cool galaxy halo, where the halo gravitational potential and central stellar density profile have been modeled as our SPG system while the baryon profile was closer to the MPG system, with higher central density and greater pressure (and thus shorter cooling times). The main goal of these experiments is to explore whether stellar and AGN feedback processes alone are able to reconfigure the CGM with high initial pressure and density (MPG-like) state to a SPG-like state, which requires a substantial amount of energy to raise the CGM plasma within the galactic gravitational potential. More precisely, in the gravitational potential for the single-phase galaxy the total binding energy of the SPG-like CGM within the virial radius is $\sim 1.7\times10^{61}$ ergs, whereas for the MPG-like CGM it is $\sim 2.0\times10^{61}$ ergs -- a difference of $\sim 3\times10^{60}$ ergs.  This means that it would take $\sim 3\times10^{60}$ ergs of \textit{net} energy -- i.e., energy injection above the net radiative loss from X-ray emission -- from one or more sources to reconfigure the CGM into a single-phase state. 

As the results from these experiments showed, our AGN and stellar feedback algorithm cannot prevent formation of extended multiphase cold gas clumps and filaments similar to a multiphase galaxy. The average $P_{\rm jet}$ for the SPG-Cool-MHD simulation from $t=0-1.7$ Gyr is $\sim 1.2\times10^{43}$ erg/s. During the same period, the X-ray luminosity in the same calculation averages $\sim 4 \times 10^{42}$ erg/s, which translates into $\sim 6.4\times10^{59}$ ergs of net injection of AGN energy into the CGM.
The total AGN energy injection for 1.7 Gyr is over an order of magnitude smaller than the total CGM binding energy of the SPG-Cool galaxy, which is $\approx 2\times 10^{61}$ ergs, and a factor of roughly five less than the energy it would take to reconfigure the CGM in the galaxy from the MPG-like state to the SPG-like state. This means that purely from the standpoint of net energy injection (and ignoring the details of the \textit{location} of that energy injection), it is impossible for feedback to put the CGM in a galaxy of this mass into a state where multiphase gas is only possible at the center. 

On a smaller scale, the AGN energy leads to a significant outflow of gas ($\sim10^9$ M$_\odot$) from the central $r \simeq 30$ kpc as AGN activity overheats and/or raises the CGM in the SPG-Cool-hydro case. While the thermal energy deposition accounting for AGN and stellar feedback is spherical in our simulations, the kinetic and magnetic energy deposition from the AGN jet is bipolar and as such a major fraction of AGN energy is channeled to large radii before it thermalises. This results in AGN heating being insufficient to overpower cooling at any point of time, and as such significant and spatially extended net cooling is seen for both the SPG-Cool runs.

\subsection{Comparisons with our previous works}
\label{sec:comparison}

In the work presented here, we extend our previous efforts in \citet{prasad2020} to explore the role of additional physics including thermal AGN feedback, magnetic fields, and magnetised AGN feedback. In addition, we have used different parameters for kinetic AGN feedback based on our understanding of the limitations of the AGN feedback implementation in \citet{prasad2020}. 

In \citet{Prasad2022}, we found that the choice of accretion efficiency $\epsilon=10^{-4}$ resulted in momentum-heavy (higher momentum to kinetic energy ratio) jets. That allowed for AGN jets to create large atmospheric circulation, resulting in galactic entropy profiles at scales of a few kpc that are elevated compared to observations of the corresponding galaxies. To overcome that limitation we have raised the accretion efficiency, $\epsilon$, by an order of magnitude to $10^{-3}$ which serves to make the simulated jets lighter and faster. Another important difference from our earlier work has been the accretion time of the gas within the accretion zone, which has been raised from $t_{acc}=1$ Myr to $t_{acc}=10$ Myr. This is because cold gas within the accretion zone $r\lesssim1$ kpc does not free-fall onto the SMBH but has angular momentum which delays its infall into the SMBH. 

All but one (MPG-hydro-kinetic run) of our runs have thermal feedback where mass and energy is injected uniformly proportional to the energy fraction within the central $r=1$ kpc. This is unlike our earlier works, where only thermal energy was injected in the bipolar jet source region. Furthermore, our runs use a higher thermal fraction of 0.25 compared to 0.1 in our earlier works. The use of a higher thermal energy fraction with spherical deposition of mass and energy within $r=1$ kpc has resulted in overheating of the CGM within $r\simeq 5$ kpc in runs with partial thermal feedback.

In \citet{grete2025}, we analysed the MHD and hydro AGN feedback in galaxy clusters ($M_{200} \sim 6.7\times10^{14}$ M$_\odot$). 
Magnetic field strengths for the SPG-MHD, MPG-MHD and SPG-Cool-MHD runs show similar behaviour to the fiducial cluster run where the intracluster medium (ICM) was initialised with a similar $1~\mu G$ magnetic field. The fiducial cluster run also shows that the ICM magnetic field strength declines at large radii ($r>100$ kpc) with time and rises to a saturation level within the core ($r<30$ kpc). The buildup of magnetic fields in the core allows for collimation of jets for the fiducial cluster run. However, jets in the cluster simulations thermalise within the cluster core (at $r\lesssim30$ kpc) unlike the SPG-MHD and MPG-MHD runs studied here, where the passive tracer fluid shows a more isotropic distribution within the central $r \simeq 20$ kpc and the jet cavities rising along the jet axis to much larger radii. Because giant elliptical galaxies have shallower potential wells compared to galaxy clusters, collimation of jets allows the AGN jets to travel to larger radii ($r>100$ kpc) before thermalising.   

\subsection{Comparison with other works}
\label{sec:analog}
The analysis presented in \citet{voit15N, voit2020} has motivated multiple simulations exploring the interplay between radiative cooling, stellar feedback and AGN feedback in the evolution of massive elliptical galaxies. \citet{wang2019} performed hydrodynamic simulations focused on galaxies similar to our single phase and multiphase galaxies. Similar to \citet{prasad2020}, they also find that in single phase galaxy AGN feedback is able to maintain a steady hot CGM with a small amount of centrally-concentrated cold gas. A similar mass multiphase galaxy forms a large amount of extended cold gas. Our simulations are largely in line with these findings for the single phase and multiphase galaxies despite having very different numerics and physics modules. Our results deviate from \citet{prasad2020} in the central $r \simeq 3$ kpc, however, as we see a flattening of the entropy profile for both SPG-hydro and MPG-hydro simulations.  We speculate that this is due to the strong thermal feedback in our simulations within $r\simeq 1$ kpc. Our MHD runs also show similar behaviour to the hydro runs with partial thermal AGN feedback in the radial profile. The spherical heat injection by AGN feedback in addition to the thermal feedback due to Type Ia supernova feedback leads to overheating within the central $r \simeq 3$ kpc. Recent work by \citet{mohapatra2025} suggests that modeling the Type Ia supernova feedback via discrete explosions rather than uniformly distributed heating helps to maintain the low central entropy within $r \simeq 5$ kpc although these simulations were run for shorter time ($\sim$ few 100 Myr) and lacked AGN feedback.   

 For our simulations, we have kept the thermal energy fraction at $25\%$ of the total injected energy for the AGN for all the runs except for MPG-hydro-kinetic run. This is because even tough AGN jets thermalise as they travel through the dense CGM, they still remain kinetically dominated at $r\sim1$ kpc. Similar studies by \citet{yu2021} have equipartitioned the AGN energy in kinetic and thermal part, arguing the AGN jets thermalise close to SMBH through shock heating. Simulations of galaxy cluster halos by \citet{li2014a,li2014b} find that choice of AGN energy partition between kinetic and thermal fraction does not significantly affect the AGN-ICM interaction. However, we find that the partitioning of AGN energy in kinetic and thermal fraction has significant impact of AGN-CGM interaction. One of the possible reason might be that the galaxy clusters being larger halos have a much deeper potential well and higher ICM density and pressure compared to giant elliptical galaxies. Other possibility might be the way thermal energy feedback is implemented in our simulations. \citet{li2014a} injects the thermal energy within the bipolar jet source region while in our simulations, AGN thermal energy is injected within a spherical volume within $r\sim 1$ kpc around the SMBH.

\cblu{}

\section{Conclusions}
\label{sec:conc}
This paper has investigated how AGN feedback might achieve self-regulation in the presence and absence of magnetic fields in halos of mass $\sim 10^{13.5} \, M_\odot$. Galaxies with three distinct initial configurations were simulated. One of them (the single phase galaxy/SPG) starts with a lower density, higher entropy atmosphere that has a deeper potential well at the center ($\sigma_v \approx 280 \, {\rm km \, s^{-1}}$). The other (the multiphase galaxy/MPG) has a higher density, lower entropy atmosphere with a shallower potential well at the center ($\sigma_v \approx 230 \, {\rm km \, s^{-1}}$). The third galaxy setup (SPG-Cool) has been initialised with the SPG gravitational potential and MPG-like baryon profile. 

The following points summarize our findings:
\begin{itemize}
\item Our kinetic feedback algorithm is highly efficient in controlling the cooling flow in the CGM while maintaining the entropy profile within observed ranges. However, splitting the AGN power into a kinetic and thermal part results in an excess entropy bump within $r \simeq 15$ kpc compared to observed entropy profiles for both single phase and multiphase galaxies.

\item Magnetic fields do not seem to cause major differences in the way AGN jets interact with the CGM for SPG runs, while MHD runs tend to have more collimated jets. MHD jets thermalise at larger radii ($r\sim50$ kpc) compared to the hydro runs with kinetic+thermal AGN feedback, where AGN jets tend to thermalise at $r \lesssim 10$ kpc.

\item Magnetic fields prevent major disruption of the gas in the core ($r \lesssim 10$ kpc). For the MHD runs, the total gas mass within the central $r \simeq 10$ kpc recovers to its pre-AGN state within a few 100 Myr. On the other hand, AGN jets cause major disruption to the total gas mass within $r \simeq 10$ kpc for hydro runs with kinetic+thermal AGN feedback by evacuating significant amount of gas ($\sim 10^9$ M$_\odot$) from within $r \simeq 10$ kpc.

\item All our MHD runs (with kinetic+thermal+magnetic AGN feedback) show magnetic field amplification and alignment along the jet axis closer to the center ($r \lesssim 30$ kpc) while there is a decline in the magnetic field strength at larger radii ($r>30$ kpc) away from the jet axis.

\item The SPG-Cool runs show that AGN feedback in our simulations are unable to completely disrupt the high CGM density and pressure. Higher CGM pressure leads to the formation of extended cold gas, despite the galaxy having $\sigma_v > 240$ km/s during late stages of evolution ($t\gtrsim1$ Gyr).
\end{itemize}

The results presented in this paper illuminate the role that different aspects of AGN feedback model have in the evolution of the most massive galaxies of the universe. Models that have thermal AGN feedback deviate from the observed entropy profiles as they tend to overheat the CGM in the central regions. Future numerical studies of AGN feedback should keep the thermal energy fraction to no more than $\sim 5\%$ of the total AGN energy and avoid spherically symmetric thermal energy injection in the domain. While the presence of magnetic fields does not lead to radically different evolution of these massive halos in our simulations, their role could be critical if cosmic rays are introduced into the simulations.  
\section*{Acknowledgements}
DP is supported by Royal Society grant no. RF-ERE-210263 (FvdV as PI). PG acknowledges funding by the Deutsche Forschungsgemeinschaft (DFG, German Research Foundation) – 555983577. BWO acknowledges further support from NASA ATP grants NNX15AP39G and 80NSSC18K1105 and NSF grant AST-1908109. MF acknowledge funding by the Deutsche Forschungsgemeinschaft
(DFG, German Research Foundation) under Germany’s Excellence Strategy –
EXC 2121 “Quantum Universe” – 390833306 and project number 443220636
(DFG research unit FOR 5195: "Relativistic Jets in Active Galaxies"). FvdV is supported by a Royal Society University Research Fellowship (URF\textbackslash R1\textbackslash191703 and URF\textbackslash R\textbackslash241005). This work used the INCITE computational grant number AST-146, as well as the resources of the Super Computing Wales (project no.- scw2065). Computations described in this work were performed using the publicly-available AthenaPK (\citealt{grete2022}) and YT \citep{YT} codes, which are the products of the collaborative effort of many independent  scientists  from  numerous  institutions  around  the world.

\section*{Data Availability}
The data underlying this article will be shared on reasonable request to the corresponding author.

\bibliographystyle{mnras}
\bibliography{reference} 
\label{lastpage}

\end{document}